\documentclass[useAMS,usenatbib]{mn2e} 
\pdfoutput=1 
\usepackage{journals}
\usepackage[pdftex]{graphicx,color} 
\usepackage[latin1]{inputenc}
\usepackage{graphics} 
\usepackage{amsfonts} 
\usepackage{amsmath}
\usepackage{multicol} 
\usepackage{layout} 
\usepackage{amssymb}
\usepackage[a4paper,colorlinks=true,pdfstartview=FitV,
linkcolor=red,citecolor=blue,urlcolor=magenta]{hyperref}

\title[Halo formation times and concentrations]
      {Formation times, mass growth histories and concentrations of dark matter haloes}

\author[C. Giocoli et al. 2011]
{\parbox{\textwidth}{Carlo Giocoli$^{1,2}$\thanks{E-mail:
 \href{mailto:cgiocoli@oabo.inaf.it} {cgiocoli@oabo.inaf.it}},
 Giuseppe Tormen$^{3}$, Ravi K. Sheth$^{4,5}$} \\ \\ 
 $^{1}$ INAF - Osservatorio Astronomico di Bologna, via Ranzani 1, 
40127, Bologna, Italy \\ 
 $^{2}$ INFN - Sezione di Bologna, viale Berti Pichat 6/2, 40127,
 Bologna, Italy \\
 $^{3}$ Dipartimento di Astronomia, Universit\`a di Padova, vicolo
dell'Osservatorio 3, 35122, Padova, Italy \\ 
 $^{4}$ The Abdus Salam International Center for Theoretical Physics, 
Strada Costiera 11, 34151 Trieste, Italy \\
 $^{5}$ Center for Particle Cosmology, University of Pennsylvania, 
       209 S 33rd St, Philadelphia, PA 19104, USA}

\begin{document}
\date{}
\maketitle
\label{firstpage}
\pagerange{\pageref{firstpage}--\pageref{lastpage}} \pubyear{2011}

\begin{abstract}
  We develop a  simple model for estimating the  mass growth histories
  of dark matter halos.  The model  is based on a fit to the formation
  time distribution,  where formation is defined as  the earliest time
  that the main  branch of the merger tree contains  a fraction $f$ of
  the final mass $M$.  Our  analysis exploits the fact that the median
  formation time as a function of $f$ is the same as the median of the
  main  progenitor mass  distribution  as a  function  of time.   When
  coupled with previous work showing that the concentration $c$ of the
  final halo  is related to  the formation time $t_f$  associated with
  $f\sim  0.04$,   our  approach  provides  a   simple  algorithm  for
  estimating  how  the  distribution  of halo  concentrations  may  be
  expected to  depend on mass,  redshift and the expansion  history of
  the background  cosmology.  We  also show that  one can  predict $\log_{10}c$
  with a  precision of about  0.13 and 0.09~dex  if only its  mass, or
  both mass  and $t_f$  are known.  And,  conversely, one  can predict
  $\log_{10}t_f$  from mass  or  $c$ with  a  precision of  0.12 and  0.09~dex,
  approximately independent of $f$.   Adding the mass to the $c$-based
  estimate  does  not result  in  further  improvement.  These  latter
  results may be  useful for studies which seek to  compare the age of
  the stars in the central galaxy in a halo with the time the core was
  first assembled.
\end{abstract}

\section{Introduction}

One main focus  of modern observational cosmology is  the study of the
uabundance and structural properties of galaxy clusters: typically, the
abundance is expressed  as a function of mass  $M$, and the structural
property of most interest is the central concentration $c$ of the mass
density profile.  There are  various ongoing and planned surveys which
are    capable    of    finding    thousands   of    clusters    using
well-controlled/calibrated  selection criteria, so  a number  of large
homogeneous samples of  galaxy clusters will be available  in the near
future  \citep{euclidredbook}.   Such samples  have  the potential  to
strongly constrain the fundamental  parameters of a given cosmological
model  (e.g.   $\Lambda$CDM).   This  is  in  large  part  because  of
numerical simulations \cite[e.g.]{navarro97} which have shown that, at
fixed  mass,  halos  are  less  concentrated if  the  matter  density,
$\Omega_m$, is lower.  In addition,  for a given cosmology, there is a
reasonably tight monotonic correlation between  $M$ and $c$,
although this $M-c$ relation depends  on how $c$  is estimated. 
Recently \citet{prada11} report that, if $c$ is determined by 
fitting to the circular velocity rather than the density profile, then 
the $c-M$ relation is no longer monotonic.
Provided one  knows which method has been used to  estimate $c$,
measuring the $M-c$ relation at a range of redshifts  provides important
constraints on the background cosmology 
\citep{eke01,dolag04,neto07,maccio08}.)   The tightness  of these 
constraints  rests  on   accurately  estimating  the  mass  and
concentration of each cluster. 

In all cases,  these relations are thought to be  a consequence of the
fact  that the  structure  of a  halo,  and its  concentration $c$  in
particular,      depend       upon      its      assembly      history
\citep{navarro96,bullock01a,wechsler02,zhao03b,zhao09,giocoli10}.
Roughly  speaking, $c$  is  related  to the  ratio  of the  background
density at the time at which the  mass within the core of the halo was
first  assembled to  that  at  which the  halo  was identified  (e.g.,
today).    Lower   formation   redshifts   typically   imply   smaller
concentrations,  but  the  exact  dependence  is  a  function  of  the
expansion history of the background cosmology.

Recently \citet{zhao09}  have provided  a prescription for  relating a
halo's concentration to its  mass assembly history.  This prescription
requires knowledge  of the time at  which a halo had  assembled 4\% of
its mass:
\begin{equation}
 c_{vir} = 4 \,\left[1 + \left(\frac{t_1}{3.75t_{0.04}}\right)^{8.4}\right]^{1/8}
 \label{czhao09}
\end{equation}
where  $t_f$  denotes the  time  at which  the  halo  had assembled  a
fraction $f$  of its  mass (in what  follows we will  use indistinctly
both $t_1$ and $t_0$ to denote the present day time, at which the halo
had assembled all its mass).   If this remarkable relation were exact,
then it  would imply  that the $M-c$  relation, and its  dependence on
cosmology,  comes entirely  from the  fact that  $t_{0.04}/t_1$  has a
distribution  which depends on  halo mass  (recall that  massive halos
assemble their  mass later,  so they typically  have larger  values of
$t_{0.04}/t_1$).  This mass dependence itself depends on cosmology, as
does the conversion between time and the observable, redshift. 

Unfortunately,  estimating  the   mass  and  cosmology  dependence  of
$t_f/t_1$  is not  straightforward.  Analytic  arguments for  this are
only available for $f>1/2$ \citep{lacey93,nusser99}, and these are not
particularly  accurate \citep{giocoli07a}.   So the  main goal  of the
present work is to provide a simple prescription that is accurate even
when $f\ll 1$.

Although the close relation between $c$ and assembly history is one of
the main motivations  of the present work, we note  that there is also
interest  in using  the  distribution of  galaxy velocity  dispersions
\citep{sheth03a} and the weak lensing signal to constrain cosmological
parameters  \citep{fu08,schrabback10}.   Both  these  observables  are
expected  to be  related  to  halo mass  and  concentration, so  these
studies will also benefit from a better understanding of the formation
history of dark matter haloes.

In  addition,  there  is  considerable  interest  in  quantifying  the
difference between the time when the  stars in the central galaxy in a
halo formed and the time when  those stars were first assembled into a
single object.  Whereas crude estimates of the stellar age can be made
from the observed colors, there are currently no estimates of the halo
assembly time.  A by-product of our study of the joint distribution of
halo mass,  concentration and assembly  history is an estimate  of how
precisely the  mass and concentration  of a halo predict  its assembly
time.

This  paper  is  organized  as  follows.   In  Section~\ref{sims1}  we
describe the close connection  between the formation time distribution
and  the mass  of the  main progenitor  on which  our model  is built.
Section~\ref{sims2}  describes our  simulation-based  estimate of  the
formation  time distribution  for arbitrary  $f$.  It  also quantifies
correlations  between halo  mass, concentration  and  $t_f$, providing
estimates  of how precisely  the concentration  can be  estimated from
knowledge of  the mass and assembly  history, as well as  how well the
mass  and concentration  of a  halo  predict its  formation time.   In
Section~\ref{mcs}  we  describe how  to  use  our  fitting formula  to
generate  Monte-Carlo merger  histories  from which  to estimate  halo
concentrations.  Section~\ref{forecasts}  quantifies how halo assembly
histories  depend  on cosmological  parameters,  and  a final  section
summarizes.

\section{Formation and the mass of the main progenitor}
\label{sims1}
The mass which makes up a halo  of mass $M_0$ at $z_0$ will be in many
smaller pieces at $z>z_0$.  We  will follow common practice in calling
these pieces  the `progenitors', and  we will use  $N(m,z|M_0,z_0)$ to
denote the  `mass function' of  progenitors: i.e., the mean  number of
progenitors  of $M_0$ that,  at $z>z_0$,  had mass  $m$.  We  will use
$m_1(z)$ to denote the most  massive of the set of progenitors present
at $z$.  This `most massive  progenitor' may be different from what we
will call  `the main progenitor',  $m_{\rm MP}(z)$, which is  the most
massive progenitor of the most  massive progenitor of the most massive
progenitor ... and so on  back to some high redshift $z>z_0$.  Indeed,
it  is almost  certainly  true that  $m_1(z)\ge  m_{\rm MP}(z)$,  with
equality  only guaranteed  when  $m_1(z) >  M_0/2$.   We quantify  the
differences between these two quantities in the Appendix.  But because
it is  $m_{\rm MP}$ which  is relevant to  equation~(\ref{czhao09}) we
will use it in what follows.

The formation time  of a halo, $\mathbf{p_f(>z_f|M_0,z_0)}$, is usually 
defined  as the earliest time
when  $m_{\rm MP}(z)  >  fM_0$.   (Note that  this  is different  from
defining  formation as  the  earliest time  that  a single  progenitor
contains a fraction  $f$ of the final mass,  which would correspond to
requiring $m_1(z) >  fM_0$.  We discuss this choice  in the Appendix.)
If  $p_{\rm  MP}(m,z|M_0,z_0)$ denotes  the  probability that  
$m_{\rm MP}(z) = m$, then
\begin{equation}
 p_f(>z_f|M_0,z_0) = p_{\rm MP}(>fM_0,z_f|M_0,z_0),
 \label{zfmmp}
\end{equation}
where the subscript  $f$ on the left hand side is  a reminder that the
shape  of  the  distribution  depends   on  $f$.   This  is  a  simple
generalization,  to  arbitrary  values  of  $f$, of  the  argument  in
\citet{lacey93} who  assumed that $f = 1/2$,  and \citet{nusser99} who
considered $f > 1/2$.

When $f\ge 1/2$  then $m_{\rm MP}=m_1$, since a  halo cannot have more
than  one progenitor  with mass  greater  than $fM_0$.   As a  result,
$p_{\rm  MP}(m,z|M_0,z_0)$ is  simply related  to the  progenitor mass
function $N(m,z|M_0,z_0)$.  In particular,
\begin{equation}
 p_{\rm MP}(>fM_0,z_f|M_0,z_0) = N(>fM_0,z_f|M_0,z_0) \quad {\rm for}\ f\ge 1/2.
\end{equation}
This    is   fortunate,   since    simple   analytic    estimates   of
$N(>fM_0,z_f|M_0,z_0)$ are  available.  Unfortunately, this simplicity
is lost  when $f<1/2$,  and so, although  equation~(\ref{zfmmp}) still
holds,    there    is     no    analytic    estimate    for    $p_{\rm
  MP}(>fM_0,z_f|M_0,z_0)$ in this case.

In  what follows,  we  will simply  measure $p_f(>z_f|M_0,z_0)$,  with
$z_f$   determined   by  requiring   $m_{\rm   MP}(z_f)\ge  fM_0$   in
simulations, and use it to  determine the MP distribution at any given
$z$.   Namely, whereas  the usual  procedure (when  $f\ge 1/2$)  is to
estimate the formation time  distribution by differentiating the right
hand  side of equation~(\ref{zfmmp})  with respect  to $z_f$,  we will
instead estimate the MP distribution (when $f<1/2$) by differentiating
the left-hand side with respect to $f$ (at fixed $z$).  From this, one
can derive an estimate of the  mean value of $m_{\rm MP}$ at each $z$.
But  note  that  this  mean  value  can  be  got  more  directly  from
equation~(\ref{zfmmp}) by using the fact that
\begin{equation}
 \int_0^{M_0} {\rm d}m_{\rm MP}\, p_{\rm MP}(>m_{\rm MP},z|M_0,z_0)
=  \langle m_{\rm MP}(z)|M_0,z_0\rangle .
 \label{meanmmp}
\end{equation}  
Since the median $\tilde{m}_{\rm MP}$ at $z$ is easy to estimate -- it
is that  $f$ at which equation~(\ref{zfmmp}) equals  1/2 -- comparison
of the  median and the  mean provide a  simple estimate of  how skewed
$p_{\rm MP}(m_{\rm MP},z|M_0,z_0)$ is.

In what  follows, we  will be more  interested in the  median relation
than in  the mean.  This is  because the mean formation  redshift as a
function  of  the  required  mass  of the  main  progenitor,  $\langle
z_f|f\rangle$, traces  out a different  curve than does the  mean main
progenitor  mass  as  a  function  of  redshift  $\langle  f|z\rangle$
\citep{nusser99}.  In contrast, the median  $z_f$ as a function of $f$
traces out the same curve as does the median $f$ as a function of $z$:
this   convenient   property    of   the   median   relations,   which
equation~(\ref{zfmmp}) makes obvious, has not been highlighted before.

\begin{figure*}
\begin{center}
\includegraphics[width=7.5cm]{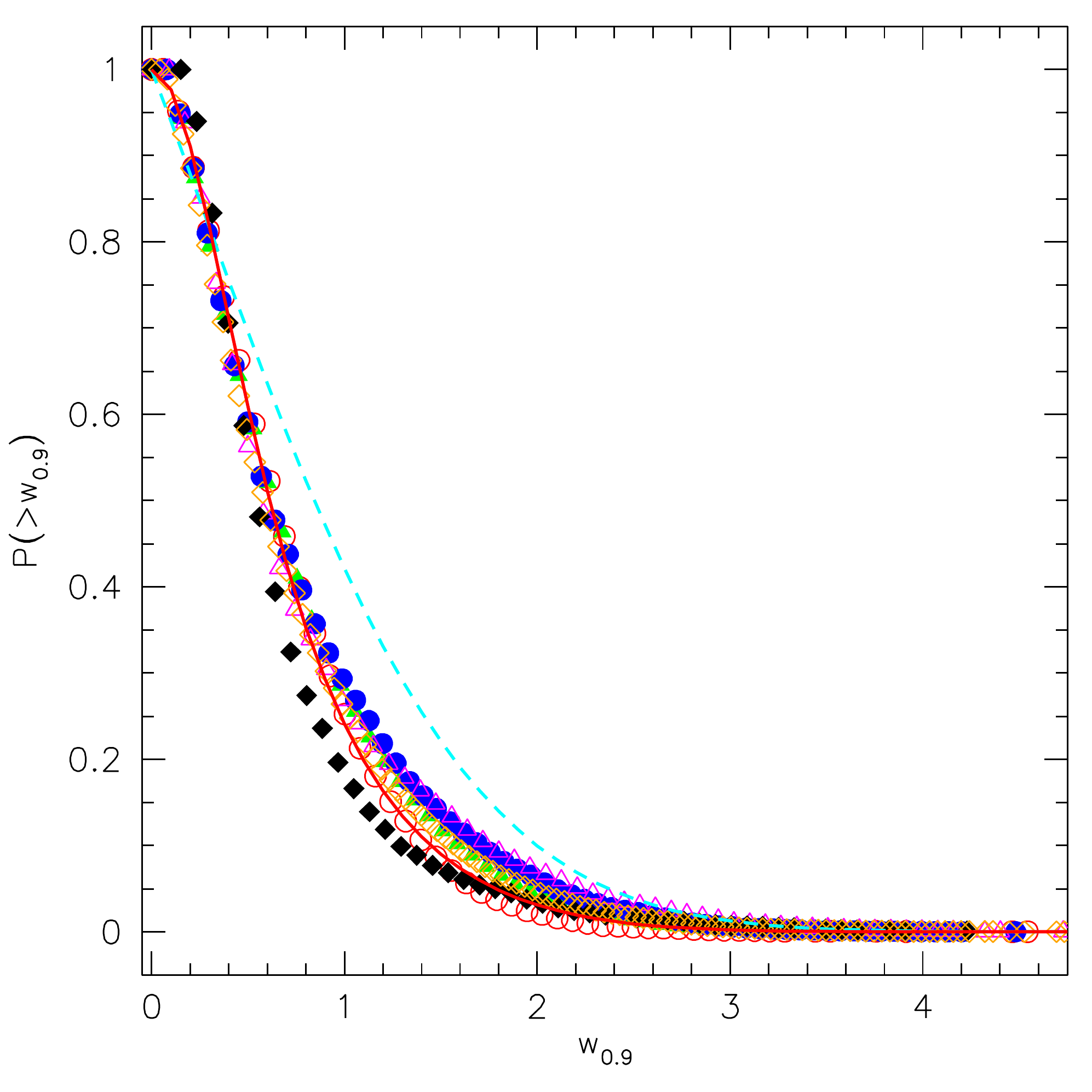}
\includegraphics[width=7.5cm]{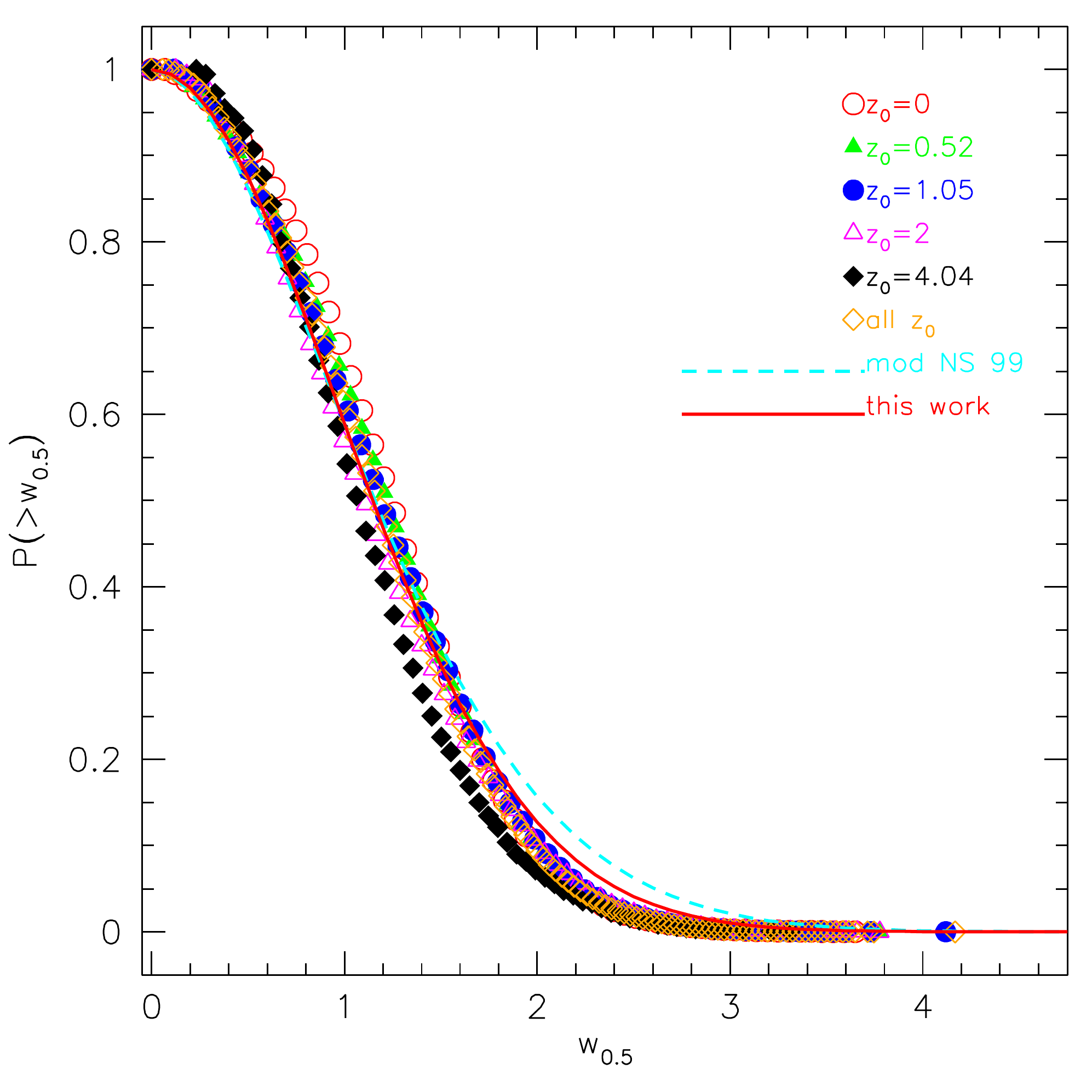}
\includegraphics[width=7.5cm]{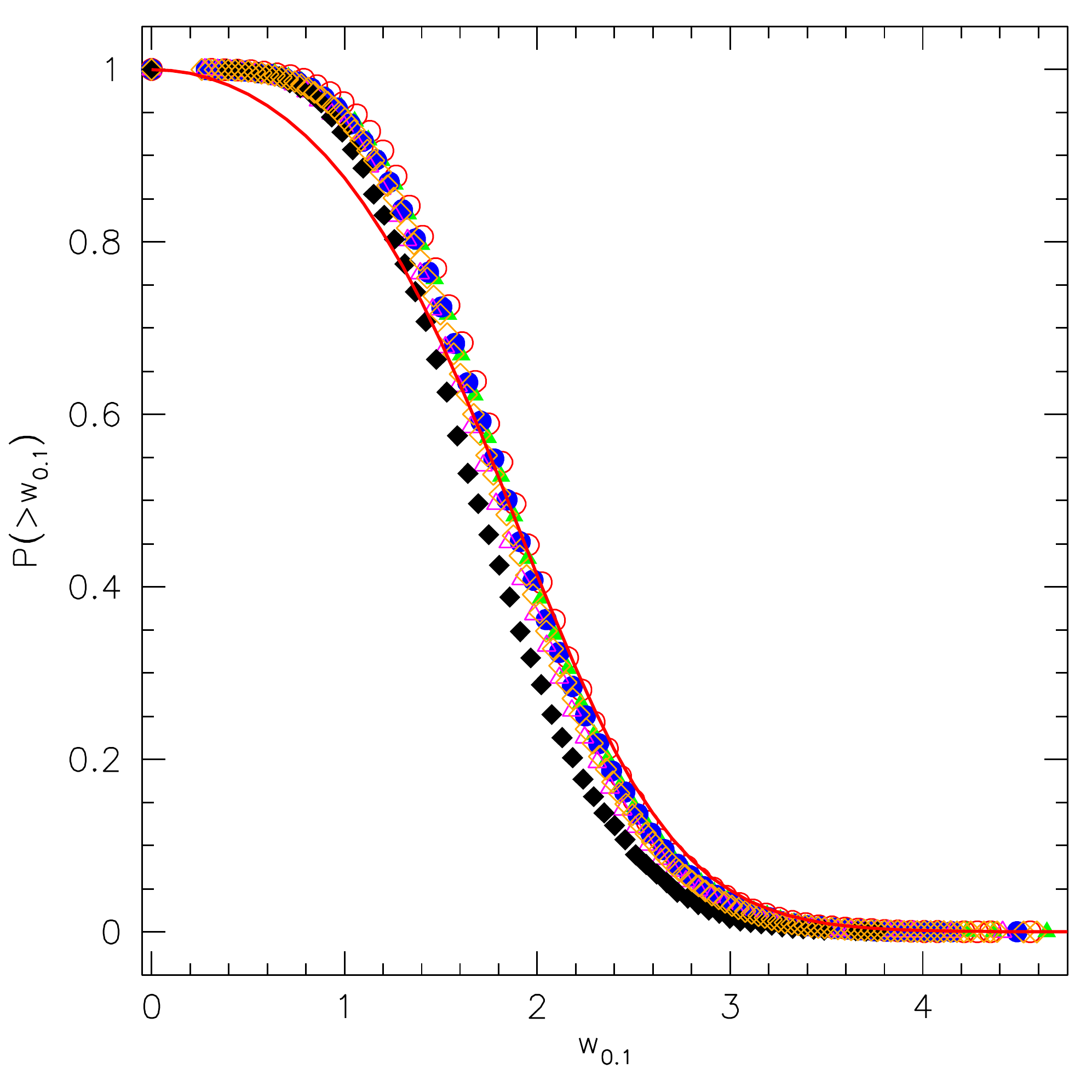}
\includegraphics[width=7.5cm]{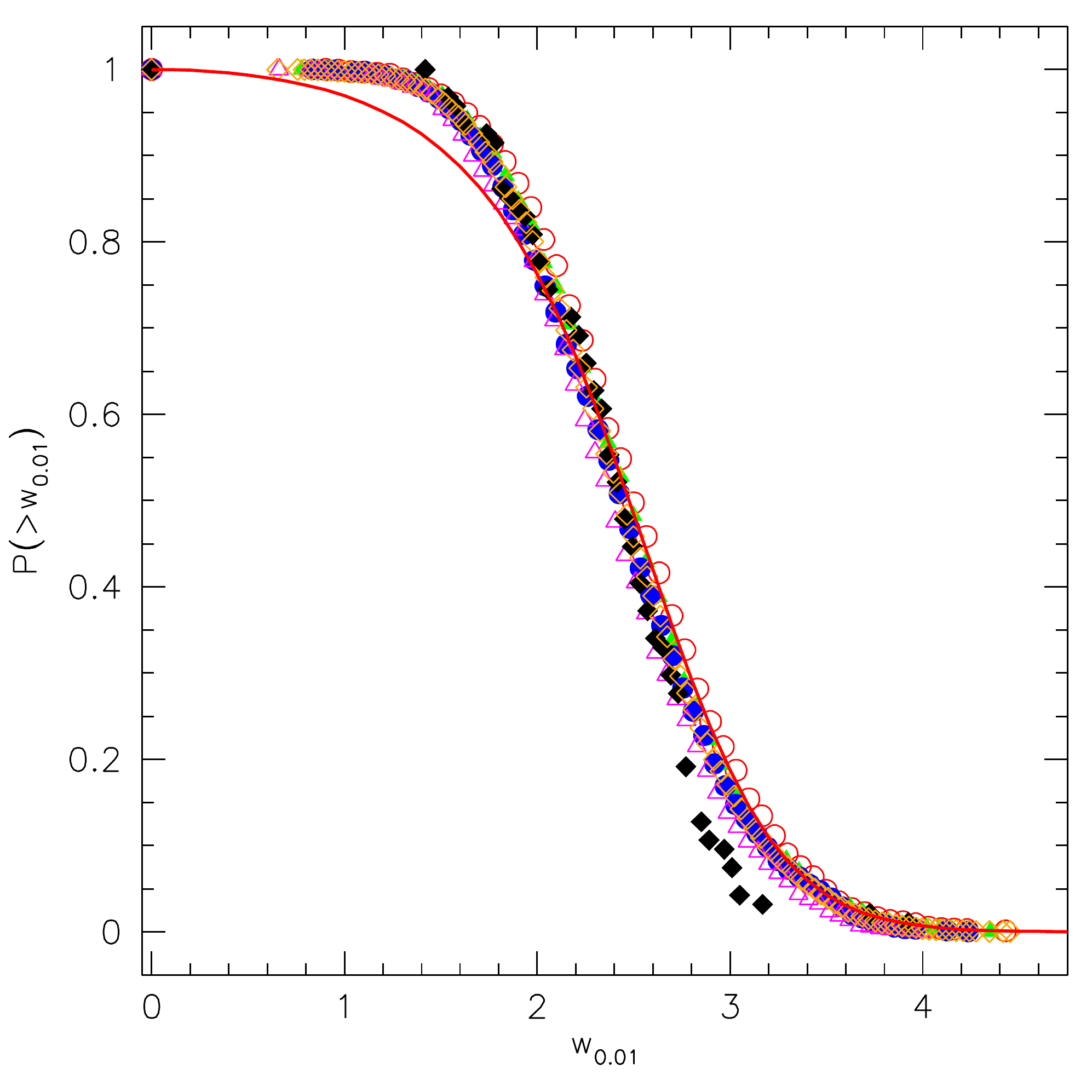}
\caption{Dependence of  rescaled formation time  distribution $p(w_f)$
  on the required  fraction $f$ of the final halo  mass that must have
  been  assembled.  Different   symbols  show  results  for  different
  redshifts  $z_0$ at  which the  hosts  were identified  in the  GIF2
  simulation.   The   dashed  curve  shows  equation~(\ref{eqnusser}),
  modified  following \citet{giocoli07a}.  The  solid curve  shows our
  equation~(\ref{eqmodel1}).
  \label{fPformation}}
\end{center}
\end{figure*}

\subsection{Scaled units}
In  principle, the  distribution $p_f(>z_f|M_0,z_0)$  depends  on $f$,
$M_0$ and  $z_0$.  However,  the analytic solution  for $f\ge  1/2$ is
actually a function of the scaled variable
\begin{equation}
 w_f = \dfrac{\delta_c(z_f) - \delta_c(z_0)}{\sqrt{S(f M_0) - S(M_0)}}\,,
 \label{wf}
\end{equation}
where $\delta_c(z)$ is the  initial overdensity required for spherical
collapse at $z$, extrapolated using linear theory to the present time
and $S(M)$ is the variance in the linear fluctuation field 
when smoothed with a top-hat filter of scale
 $R=\left( 3 M/4 \pi \bar{\rho}\right)^{1/3}$, where 
$\bar{\rho}$ is the comoving density of the background.
When   $z_f$  is   rescaled   in   this  way,   then   the  shape   of
$p(>w_f|M_0,z_0)$ is essentially the same  for all values of $M_0$ and
$z_0$,  although the  dependence on  $f$ is  still strong.   
E.g., for $f\ge 1/2$, the expected shape of this distribution is
\begin{equation}
 p_f(w_f) = \left(\dfrac{1}{f} -1 \right) \mathrm{erfc}\left( \dfrac{w_f}{\sqrt{2}}\right) + 
\left( 2 - \frac{1}{f} \right) \sqrt{\dfrac{2}{\pi}} \mathrm{exp}\left( - \dfrac{w_f^2}{2} \right)\,
\label{eqnusser}
\end{equation}
\citep{lacey93,  nusser99}.  In  what  follows, we  show  that in  
our simulations this remains true for smaller values of $f$ as well.  
Recent work \citep{paranjape11b,musso12} suggests that $w_f$ may not 
be the ideal choice for the scaling variable -- this may account for 
some of the small departures from self-similarity that are evident 
in the following plots.
This means that we do not need  to fit a different functional form for each
value of $f$,  $M_0$ and $z_0$.  Rather we need  only fit to $p(w_f)$,
which is a  function of $f$ and $w_f$ only.   The expression above can
become  negative at  $f<  1/2$, so  it is  not  a viable  model for  a
probability distribution in this  regime.  Therefore, our main goal is
to provide a fitting formula which works even at small $f$.

Although we  can use  this fitting formula  for the  full distribution
$p(w_f)$ to  estimate the mean  $\langle w_f\rangle$, we will  be more
interested in  its median $\tilde{w_f}$.   This is because  the median
formation redshift for fixed $f$ satisfies
\begin{equation}
 \delta_c(z_f) = \delta_c(z_0) + \tilde{w_f}\, \sqrt{S(fM_0)-S(M_0)}
 \label{medzf}
\end{equation}
whereas the median main progenitor mass at fixed $z>z_0$ satisfies
\begin{equation}
 S(\tilde{f}M_0) = S(M_0) + \frac{[\delta_c(z)-\delta_c(z_0)]^2}{\tilde{w_f}^2}.
 \label{medf}
\end{equation}
(Because  both $S(fM_0)$ and  $\tilde{w_f}$ depend  on $f$,  this last
expression must  be solved to yield $\tilde{f}$.)   In practice, these
both trace out the same relation, so for the purposes of making plots,
it suffices to rewrite  the expression for the
median formation redshift as:
\begin{equation}
 \frac{\delta_c(z_f)}{\delta_c(z_0)} =
  1 + \tilde{w_f}\, \sqrt{\frac{S(M_0)}{\delta^2_c(z_0)}} 
  \,\sqrt{\frac{S(fM_0)}{S(M_0)} - 1}\,.
 \label{medzf}
\end{equation}
To an excellent approximation, the left hand side is just the ratio of
linear  theory growth  factors at  the  two redshifts.   E.g., for  an
Einstein-de  Sitter  model  (itself  an  excellent  approximation  for
redshifts  greater than  about 2),  this ratio  would be  $(a_0/a_f) =
(t_0/t_f)^{2/3}$.  And, for a power-law spectrum the final term on the
right hand  side is  just a function  of $f$.  Since  $\tilde{w_f}$ is
also  a function  of $f$  only, the  only mass  dependence  comes from
$S(M_0)/\delta^2_c(z_0)$.   If  we  fix   the  mass,  then  this  term
decreases as  $z_0$ increases, so  the distribution of  $(t_f/t_0)$ is
expected to shift towards unity as the redshift at which the halos are
identified increases.

\section{A fitting formula for the formation time distribution}
\label{sims2}
In  what   follows,  we  will   use  the  GIF2   numerical  simulation
\citep{gao04,giocoli08b}   to  calibrate   our  fitting   formula  for
$p_f(w_f)$.

\subsection{The simulation}
The GIF2  simulation data we  analyze below are publicly  available at
\href{http://www.mpa-garching.mpg.de/Virgo}{http://www.mpa-garching.mpg.de/Virgo}.
The simulation  itself is described  in some detail  by \citet{gao04}.
It   represents   a   flat   $\mathrm{\Lambda}$   cold   dark   matter
($\mathrm{\Lambda}$CDM)   universe    with   parameters   ($\Omega_m$,
$\sigma_8$, $h$,  $\Omega_b h^2$ ) = ($0.3$,  $0.9$, $0.7$, $0.0196$).
The  initial density  fluctuation spectrum  had an  index  $n=1$, with
transfer function produced  by \texttt{cmbfast} \citep{seljak96}.  The
simulation followed  the evolution of $400^3$ particles  in a periodic
cube $110\,\mathrm{Mpc}/h$ on a  side. The individual particle mass is
$1.73 \times  10^9\,M_{\odot}/h$, allowing  us to study  the formation
histories  of a large  sample of  haloes spanning  a wide  mass range.
Particle  positions and  velocities were  stored at  53  output times,
mainly logarithmically  spaced between $1  +z= 20$ and $1$.   Halo and
subhalo merger trees were  constructed from these outputs as described
by \citet{tormen04,giocoli07a,giocoli08b}.

\begin{figure}
\begin{center}
\includegraphics[width=\hsize]{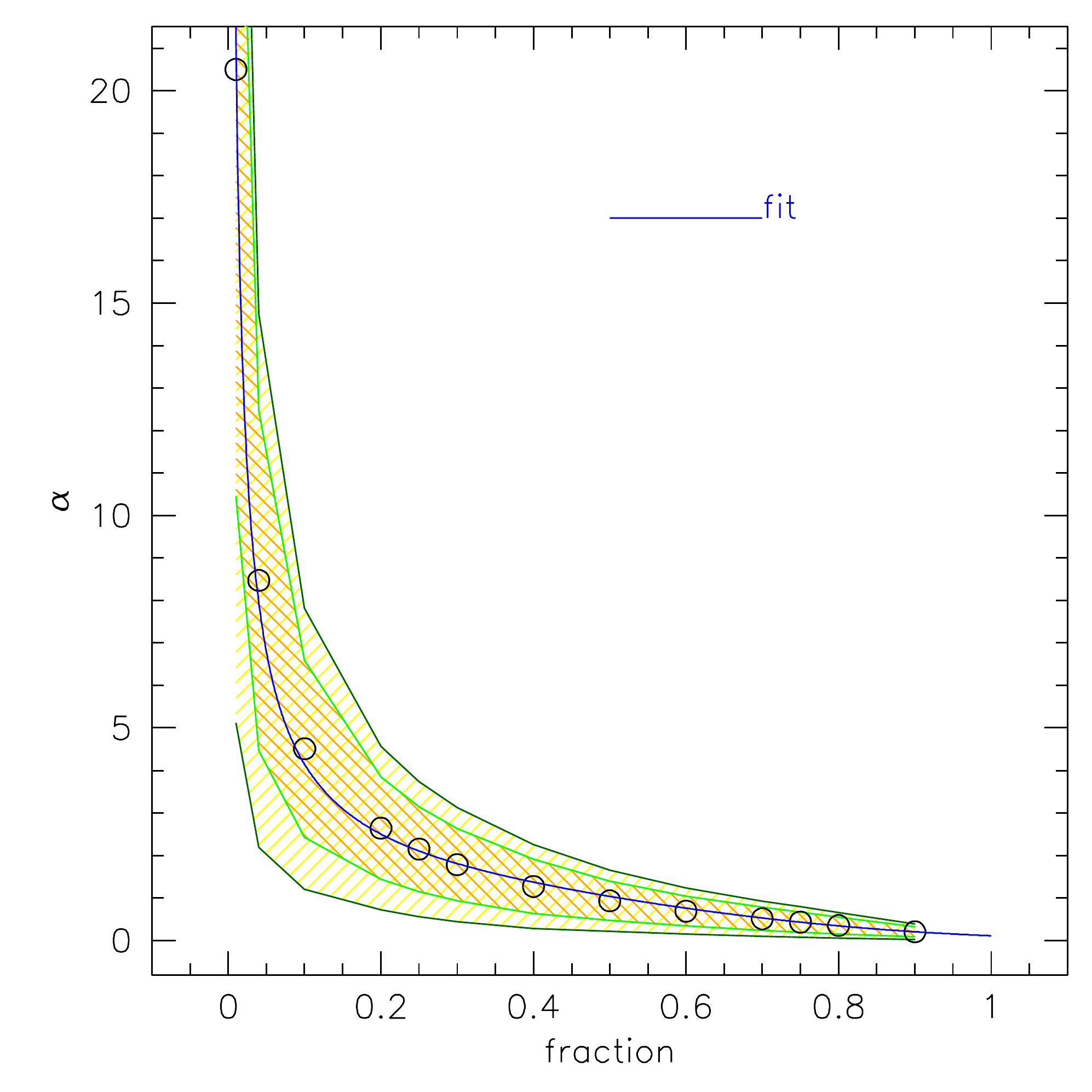}
\caption{Dependence of $\alpha_f$, the free parameter in 
  equation~(\ref{eqmodel1}), on the mass fraction 
  $f = m_{\rm MP}/M_0$ required for formation.  
  \label{fmyfit}}
\end{center}
\end{figure}

At each  simulation snapshot haloes were identified  using a spherical
overdensity criterium.  At each  snapshot the density of each particle
is assumed to be inversely proportional to the cube of the distance to
the tenth nearest  neighbor.  We take as centre of  the first halo the
position  of the densest  particle. We  then grow  a sphere  of matter
around this centre,  and stop when the mean  density within the sphere
falls below the virial value appropriate for the cosmological model at
that redshift; we  used the fitting formulae of  \citet{eke96} for the
virial density.  All the particles  inside this sphere are assigned to
the new forming halo and are removed from the list.  Starting from the
second densest available particle we  build the second halo, and so on
until all the  particles have been scanned. Particles  not assigned to
any collapsed objects are defined as field particles.

For   each    halo   more   massive    than   $10^{11.5}\,M_{\odot}/h$
(i.e. containing at least $\sim 200$ particles) in the five snapshots,
corresponding to $z_0=0$,  $0.52$, $1.05$, $2$ and $4.04$  we build up
the merger history tree as  follows.  Starting from a halo at $z=z_0$,
we  define  its  progenitors  at  the  previous output,  $z  =  z_0  +
\mathrm{d}z_1$, as all haloes containing at least one particle that at
$z = z_0$ will belong to that halo.  The ``main progenitor'' at $z_0 +
\mathrm{d}z_1$ is defined  as the one which provides  the most mass to
the halo  at $z_0$.  Then we  repeat the same  procedure, now starting
with the main  progenitor at $z = z_0+\mathrm{d}z_1$  and identify its
main progenitor at  $z = z_0 + \mathrm{d}z_1  + \mathrm{d}z_2$, and so
on backward in time.  Since we always follow the main progenitor halo,
the resulting merger  tree consists of a main  trunk, which traces the
main progenitor back in  time, $m_{\rm MP}(z)$, and of ``satellites'',
which are all the progenitors  which, at any time, merge directly onto
the main progenitor.

In what follows, to guarantee a well-defined sample of relaxed haloes,
we excluded  all halos  for which $m_{\rm  MP}(z)$ exceeded  the final
host halo mass  by more than $10\%$, as well as  all halos which were,
in fact, unbound.

\subsection{The fitting formula}
Figure~\ref{fPformation} shows the cumulative distribution of rescaled
formation redshifts.   Each panel shows results for  a different value
of  $f$, as  indicated in  the axis  labels $f  = 0.9,  0.5,  0.1$ and
$0.01$.  In each panel, symbols  show the measurements for haloes more
massive than $10^{11.5}\,h^{-1}M_{\odot}$ identified at five different
redshifts,  approximately $z_0  =  0,  1/2, 1,  2,$  and 4.   Clearly,
scaling  $z_f$  to  $w_f$  produces  curves  which  are  approximately
independent of redshift.

The  solid  line shows the result of fitting
\begin{equation}
 P(>w_f) = \frac{\alpha_f}{\mathrm{e}^{w_f^2/2}+\alpha_f-1}
 \label{eqmodel1}
\end{equation}
to the  measurements.  This returns  a value $\alpha_f$ for  each $f$,
which   is  shown   by  the   symbols  in   Figure~\ref{fmyfit}.   The
$\alpha_f(f)$ relation is well described by
\begin{equation}
 \alpha_f = 0.815\, \mathrm{e}^{-2 f^3}/f^{0.707}\,,
 \label{eqmyfit}
\end{equation}
which is shown  as the solid curve in each  panel.  The shaded regions
around  each  curve show  the  1  and  2$\sigma$ contours  of  $\Delta
\chi^2$,     where    $\chi^2(\alpha_f)     =     \sum_i    [P_i     -
P(>w_{i,f},\alpha_f)^2]$, is computed for  each mass and redshift bin.
(Equation~\ref{eqmodel1} continues to provide  a good fit if we define
formation  using  $m_1$  rather  than  $m_{\rm  MP}$,  with  the  only
difference   being    that   $\alpha_f   =    0.867\,   \mathrm{e}^{-2
  f^3}/f^{0.8}$.  This $\alpha_f$ is larger than that for $m_{\rm MP}$
by a factor  of $1.06/f^{0.1}$: the median $w_f$  is shifted to higher
redshifts than when formation is defined using the main progenitor.)

\begin{figure*}
\begin{center}
\includegraphics[width=7.5cm]{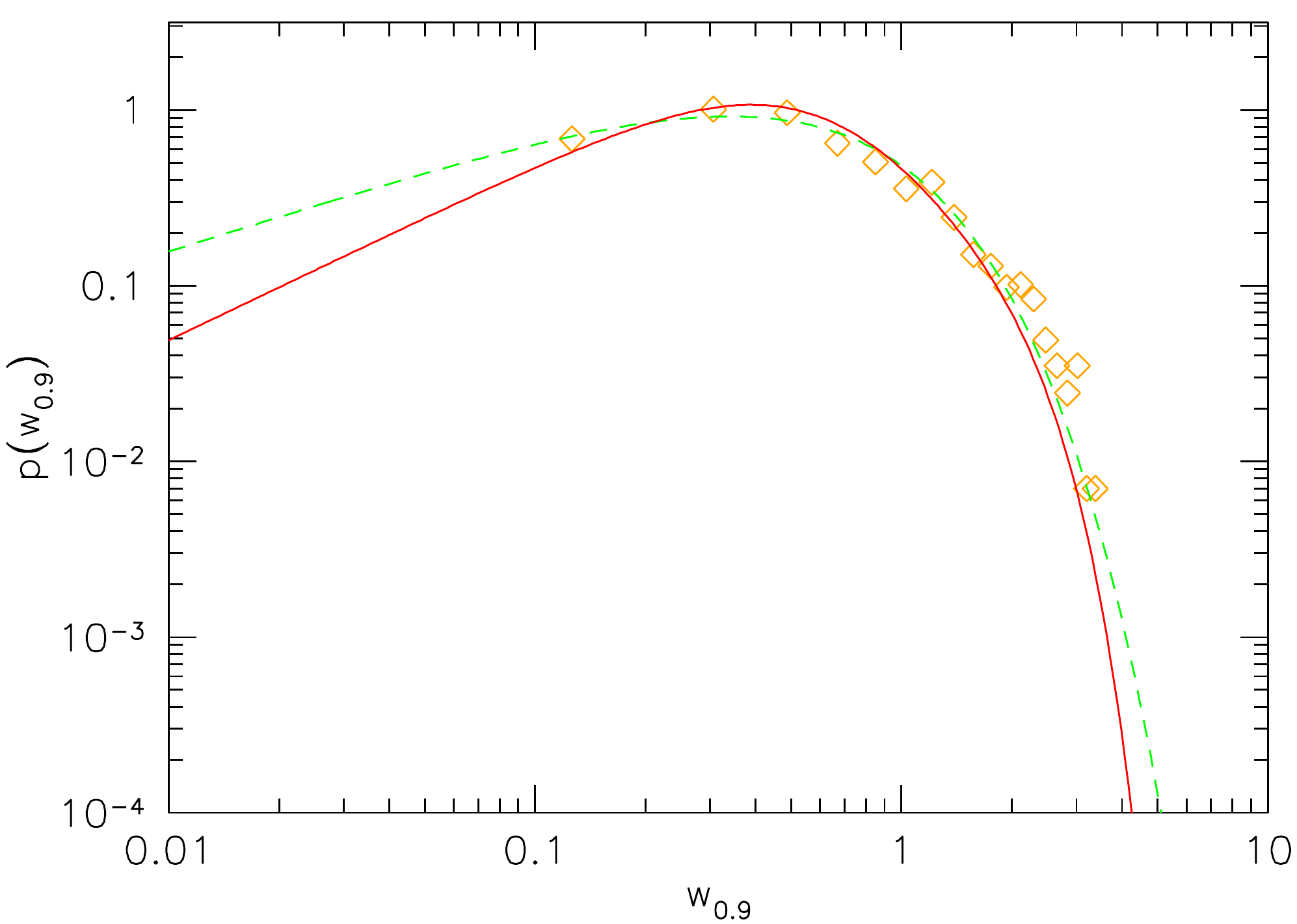}
\includegraphics[width=7.5cm]{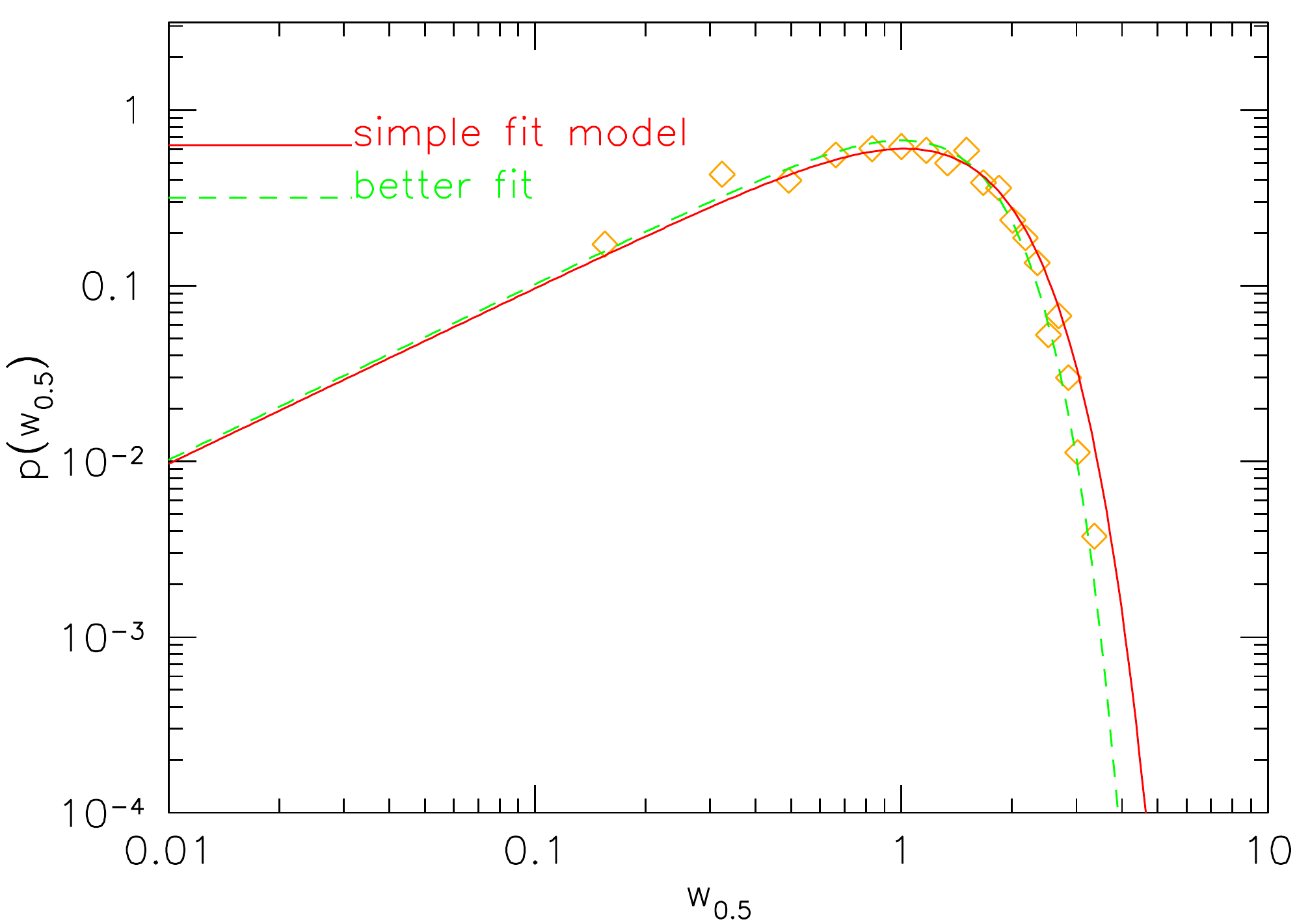}
\includegraphics[width=7.5cm]{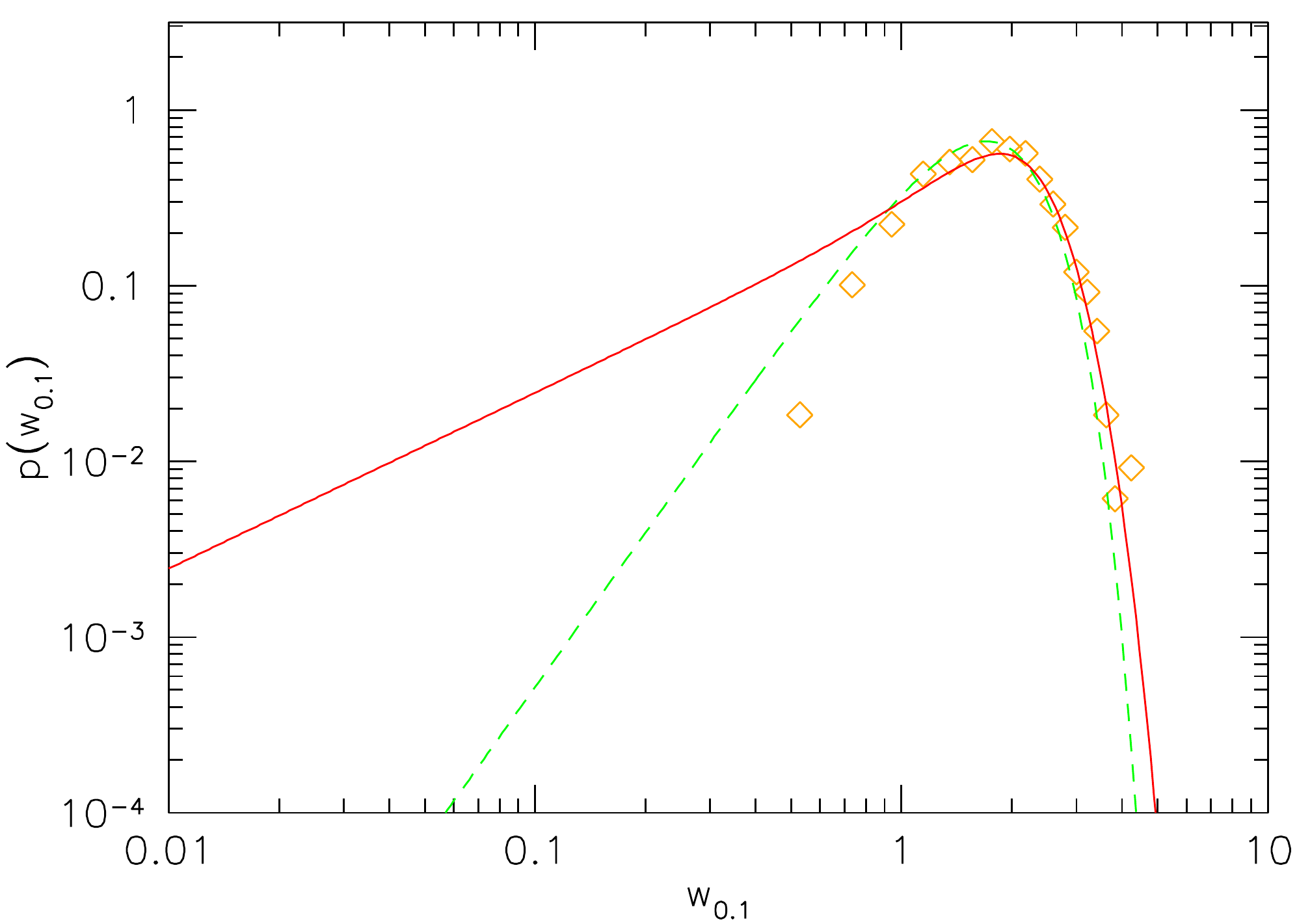}
\includegraphics[width=7.5cm]{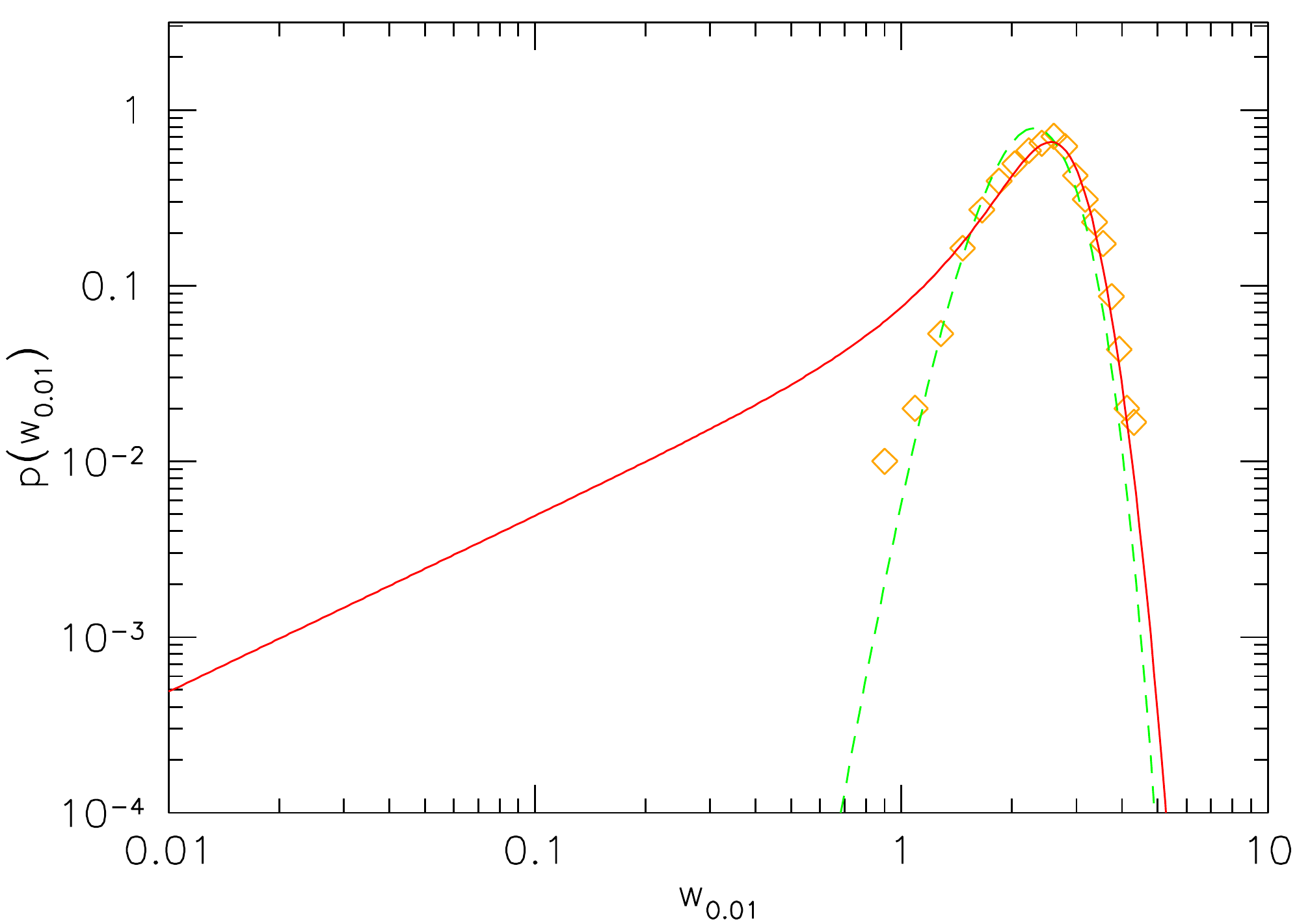}
\caption{\label{fpwmod2}Differential   formation   time  distribution.
  Open diamonds show  the results from the GIF2  simulation for haloes
  more  massive than  $10^{11.5}\,h^{-1}M_{\odot}$ identified  at five
  redshifts $z_0=0$,  $0.52$, $1.05$, $2$  and $4.04$.  The  solid and
  dashed curves show the result of fitting to equations~(\ref{pwmod1})
  and~(\ref{pwmod2}).}
\end{center}
\end{figure*}

\begin{figure*}
\begin{center}
\includegraphics[width=5.8cm]{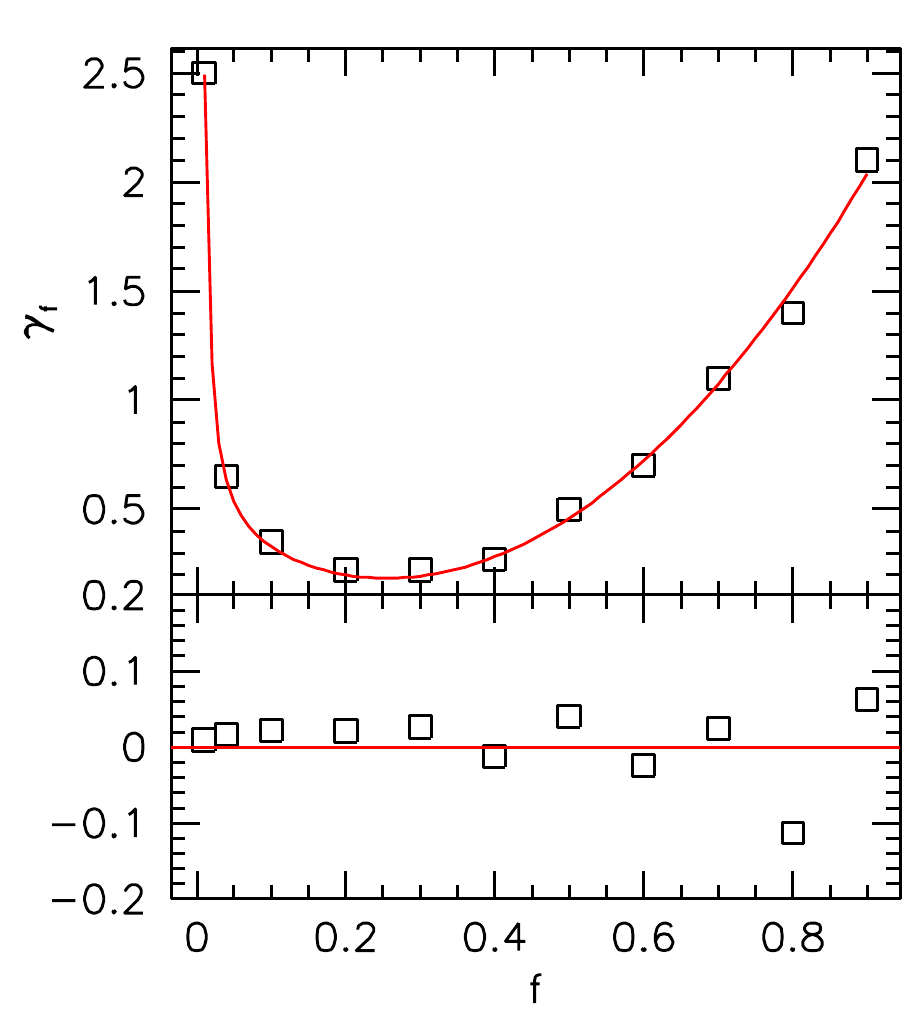}
\includegraphics[width=5.8cm]{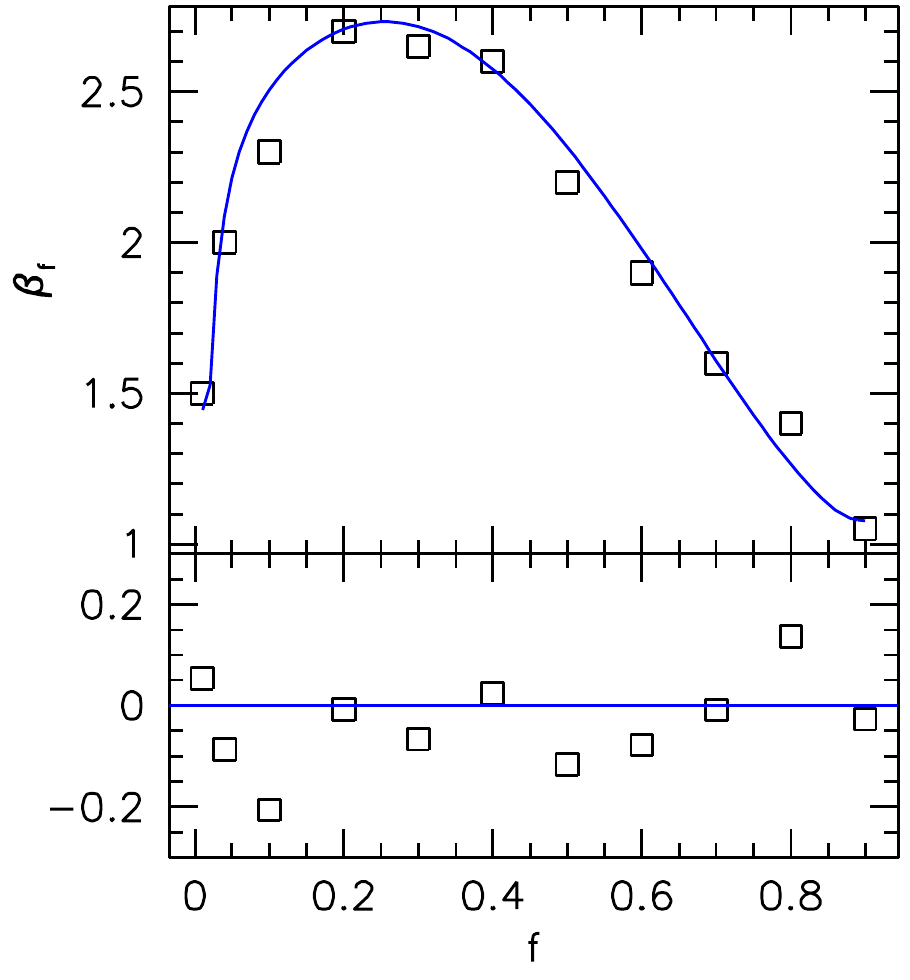}
\includegraphics[width=5.8cm]{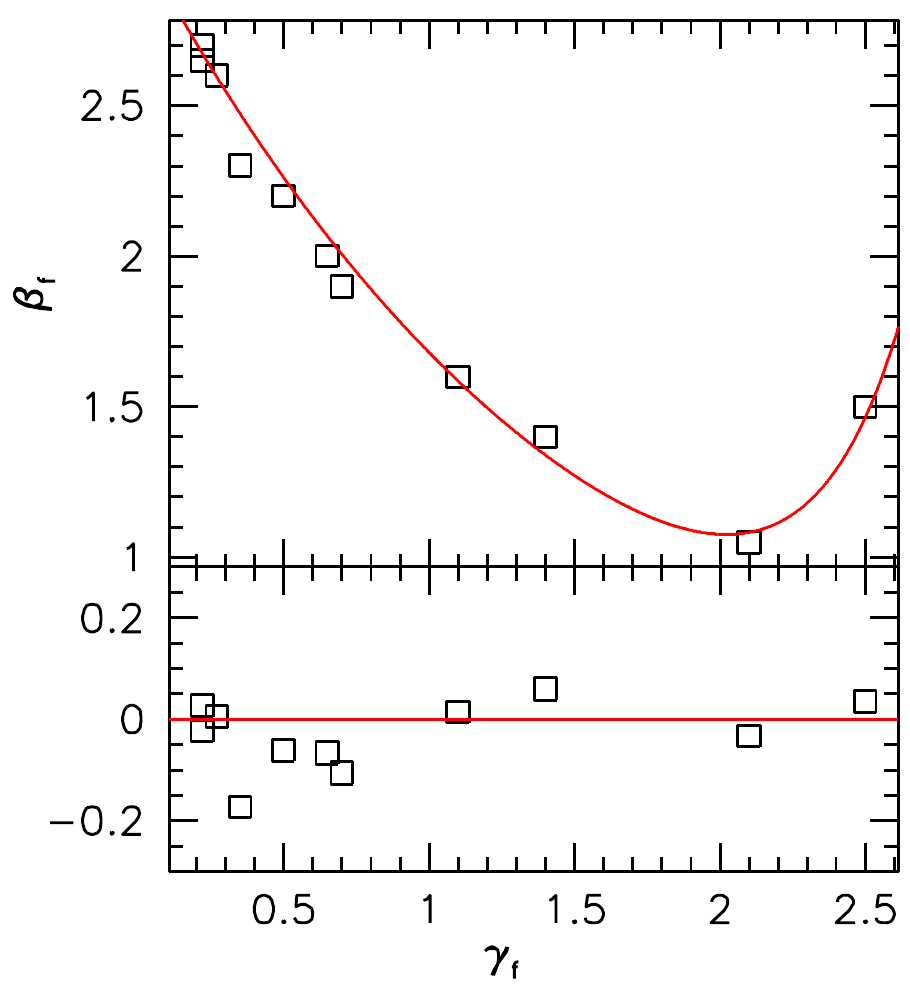}
\caption{Dependence    of   the    best-fit    free   parameters    in
  equation~(\ref{pwmod2}) on  $f$; curves  show the scalings  given by
  equations~(\ref{gf})  and~(\ref{bf}).  The  bottom  panels show  the
  residuals from these relations.  \label{fitbepi}}
\end{center}
\end{figure*}

Figure~\ref{fPformation}   shows  that   equation~(\ref{eqmodel1})  is
reasonably accurate at least between  the first and third quartiles of
the distributions.  The associated differential distribution is
\begin{equation}
 p(w_f) = - \dfrac{\partial P(>w_f)}{\partial w_f}
        = - \dfrac{\alpha_f w \mathrm{e}^{w^2/2}}
                 {\left[ \mathrm{e}^{w^2/2} + \alpha_f -1  \right]^2}.
\label{pwmod1}
\end{equation}
It has median 
\begin{equation}
\tilde{w_f} = \sqrt{2\, \ln \left(\alpha_f + 1 \right)}\,.
 \label{eqmahmod1}
\end{equation}
and the mean, computed via equation~(\ref{meanmmp}), is
\begin{equation}
 \langle w_f\rangle
 = \sqrt{\pi/2}\,\frac{\alpha_f}{1-\alpha_f}\, 
   \sum_{k=1}^\infty \frac{(1-\alpha_f)^k}{k^{1/2}}.
\end{equation}
For large  values of $f$, $\alpha_f\ll 1$,  and $\langle w_f\rangle\to
\sqrt{\alpha_f\pi^2/2}$    whereas    $\tilde{w_f}\to   (2/\pi)\langle
w_f\rangle$.   In  this  limit,  the distribution  is  rather  skewed.
However, it becomes less  skewed as $f$ increases.  When $\alpha_f=1$,
then $\langle w_f\rangle = \sqrt{\pi/2} = 1.25$ whereas $\tilde{w_f} =
\sqrt{2\,\ln(2)}  = 1.18$.   And  when $f=0.04$  then $\alpha_f=8$  so
$\langle w_f\rangle = 2.06$ whereas $\tilde{w_f} = 2.09$.

\begin{figure*}
\begin{center}
\includegraphics[width=7.5cm]{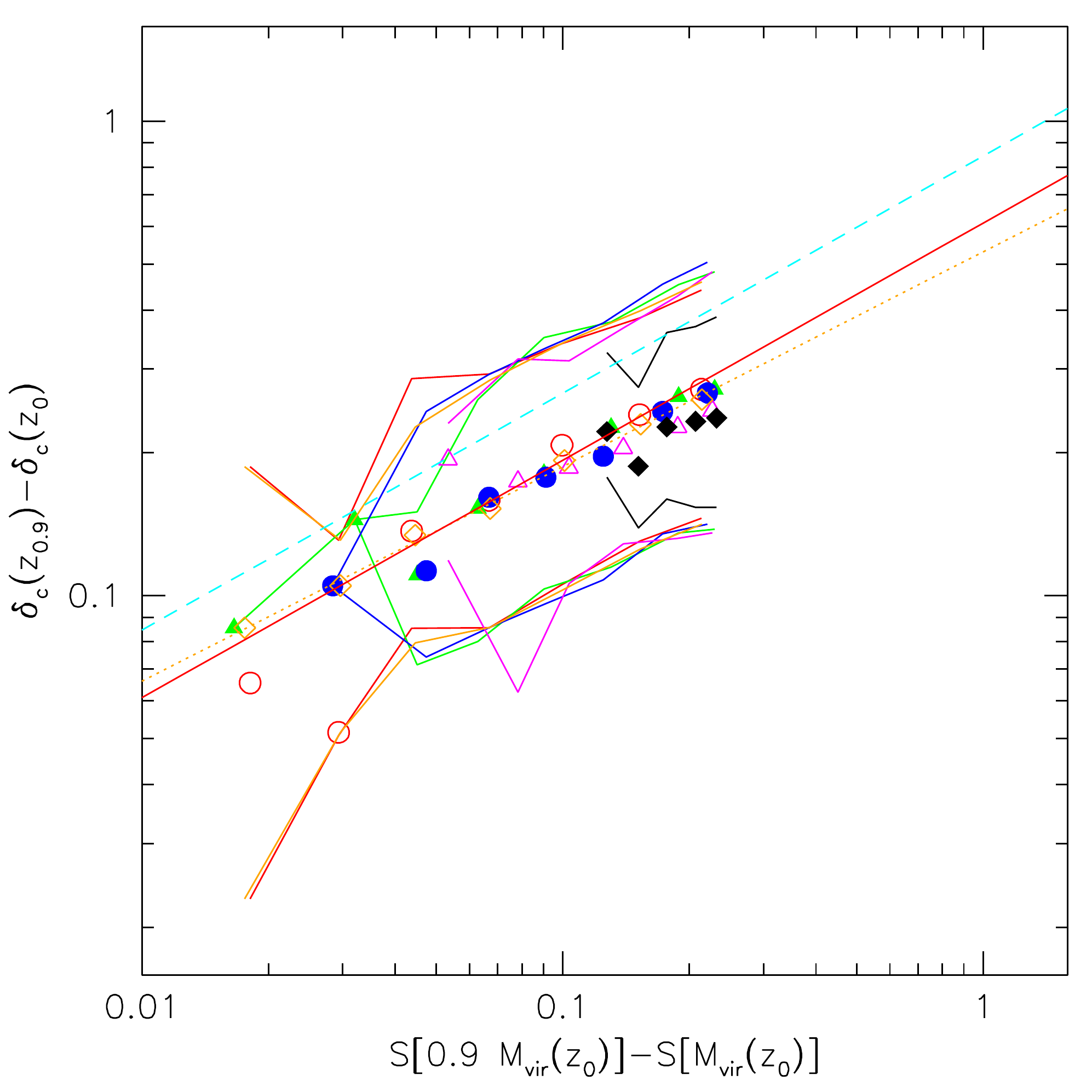}
\includegraphics[width=7.5cm]{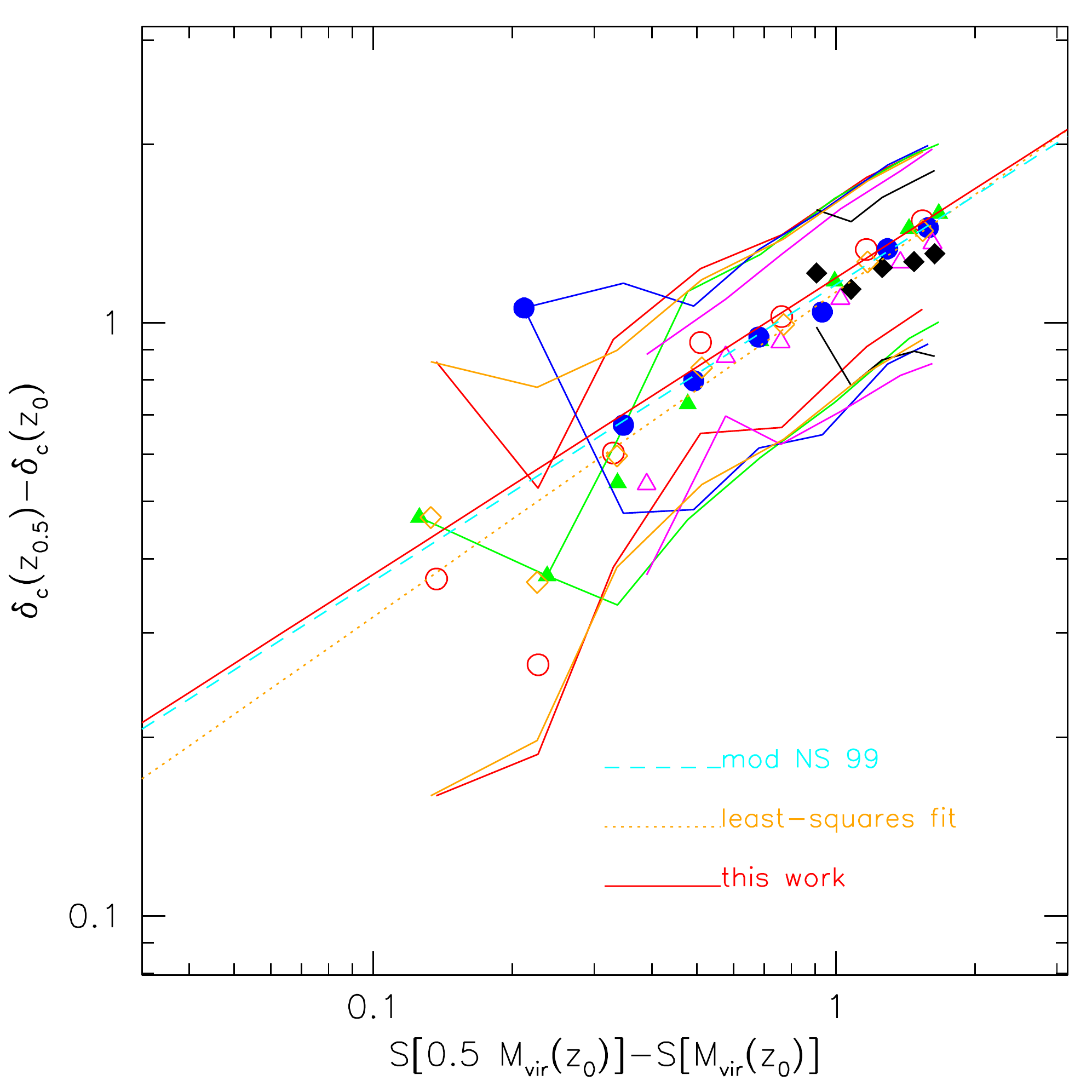}
\includegraphics[width=7.5cm]{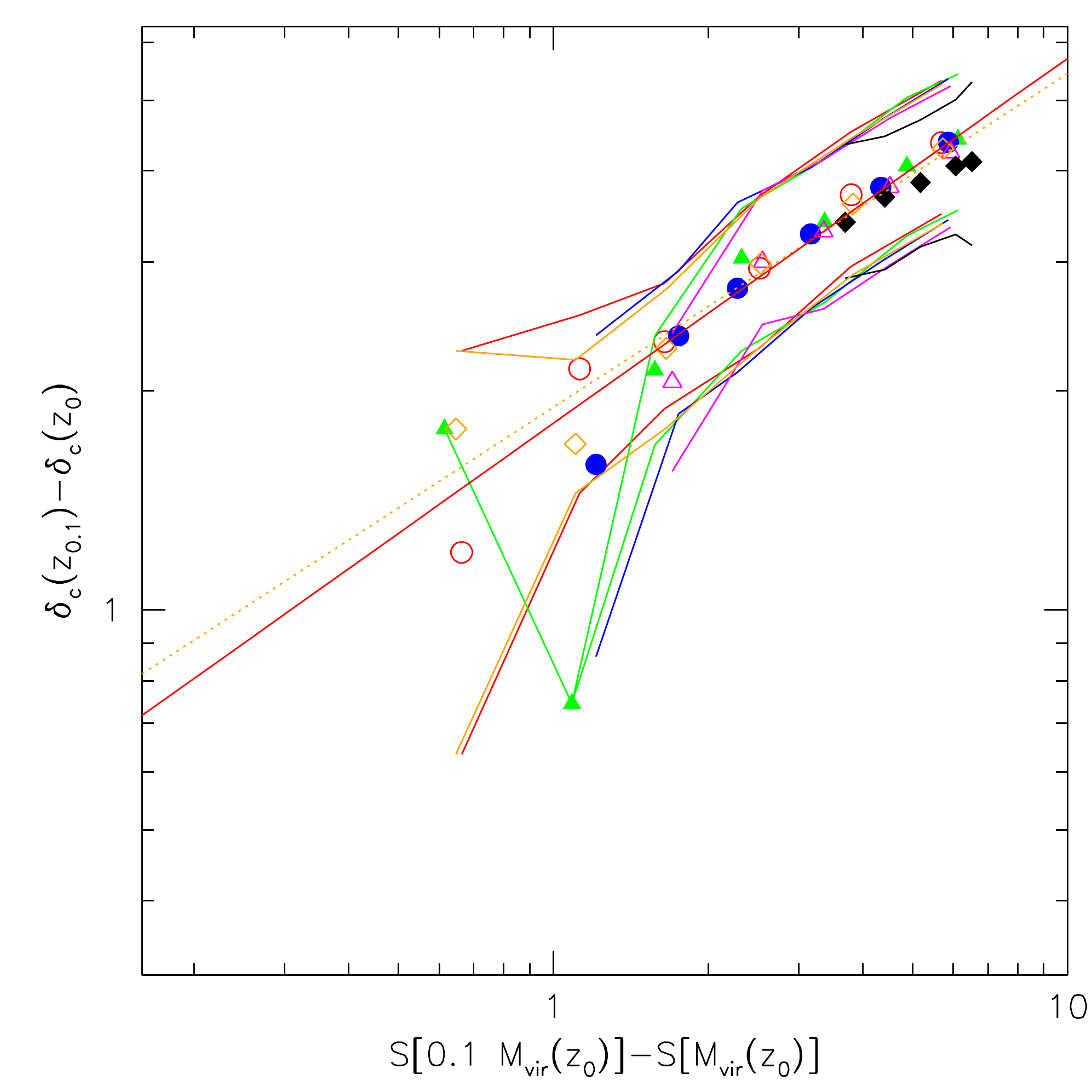}
\includegraphics[width=7.5cm]{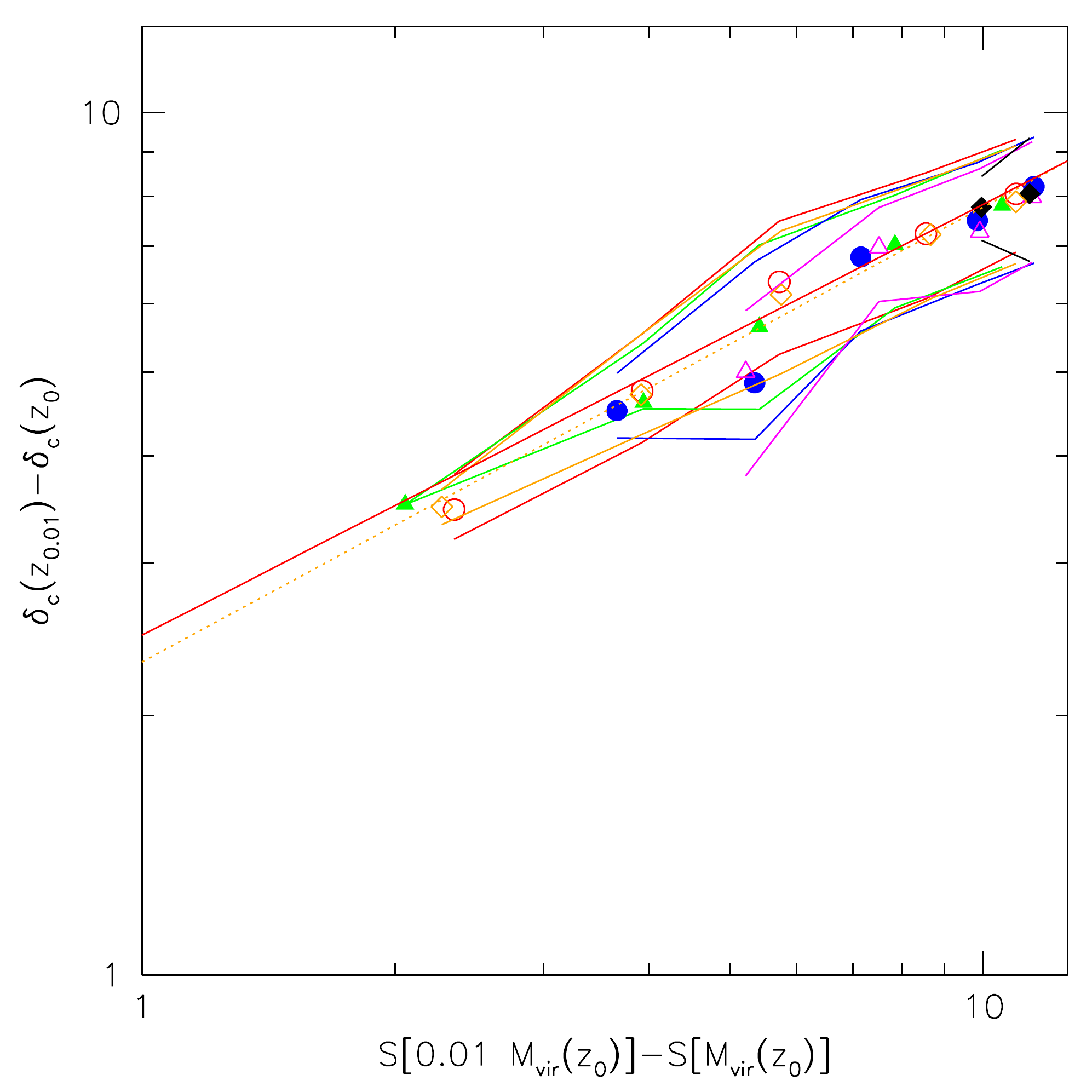}
\caption{Correlation between formation redshift -- defined as the 
  first time that the main progenitor contains a fraction $f$ of 
  the final mass $M_0$ -- and host halo mass, expressed in scaled units.  
  The different panels show results for different choices of $f$ 
  (same as previous figure).  The symbols, also same as previous figure, 
  show results for different $z_0$.   The dotted, dashed, and solid  
  lines show a least-squares to the data at all redshifts (open diamonds), 
  the relation associated with equation~(\ref{eqnusser}), 
  and the prediction of our simple model (equation~\ref{eqmahmod1}).
  In this format, both models predict a line of slope 1/2, so they 
  only differ in the value of the zero-point.
  \label{medwcorr}}
\end{center}
\end{figure*}

Since equation~(\ref{pwmod1}) is not very accurate in the tails of the
distribution, we have also fit $p(w_f)$ to
\begin{equation}
 p(w_f) = A_0\, w_f^{0.63 f^{-2/3}} \,\mathrm{e}^{-\gamma_f w_f^{\beta_f}}\,
\label{pwmod2}
\end{equation}
where  $A_0 = \beta  \,\gamma_f^{B_0}/\Gamma(B_0)$ is  a normalization
factor, with $B_0  = (1+ 0.63 f^{-2/3})/\beta_f$.  For  this model the
mean is
\begin{equation}
 \langle w_f\rangle = \int dw_f\,p(w_f)\,w_f 
  = \gamma_f\, \frac{\Gamma(B_0 + 1/\beta_f)}{\Gamma(B_0)}
\end{equation}
but one must estimate the median numerically.  

Figure~\ref{fpwmod2}  compares  the  differential  formation  redshift
distribution for four  values of the assembled fraction  $f$, with our
two fitting  functions.  The symbols show the  measurements for haloes
more  massive  than  $10^{11.5}\,h^{-1}M_{\odot}$  at  five  different
$z_0$,  as in  Figure~\ref{fPformation}.   For equation~(\ref{pwmod2})
(dashed  curves),  the  best-fit  parameters  $\gamma_f$  and  $\beta$
satisfy
\begin{equation}
 \gamma_f = 0.12 + \dfrac{1}{290\,f^{1.4}} + 4.3 (f-0.24)^2\,,
 \label{gf}
\end{equation}
and
\begin{equation}
 \beta_f = 3.05\, \mathrm{e}^{-0.6 \gamma_f} 
           + \dfrac{\mathrm{e}^{3.2 \gamma_f}}{3800}\,.
 \label{bf}
\end{equation}
Figure~\ref{fitbepi}  shows  these scalings.   Despite  the fact  that
equation~(\ref{pwmod2})  provides a more  accurate description  of the
measurements, we have found  equation~(\ref{pwmod1}) to be more useful
because it comes with a simple expression for the median relation.

\begin{figure*}
\begin{center}
\includegraphics[width=4.25cm]{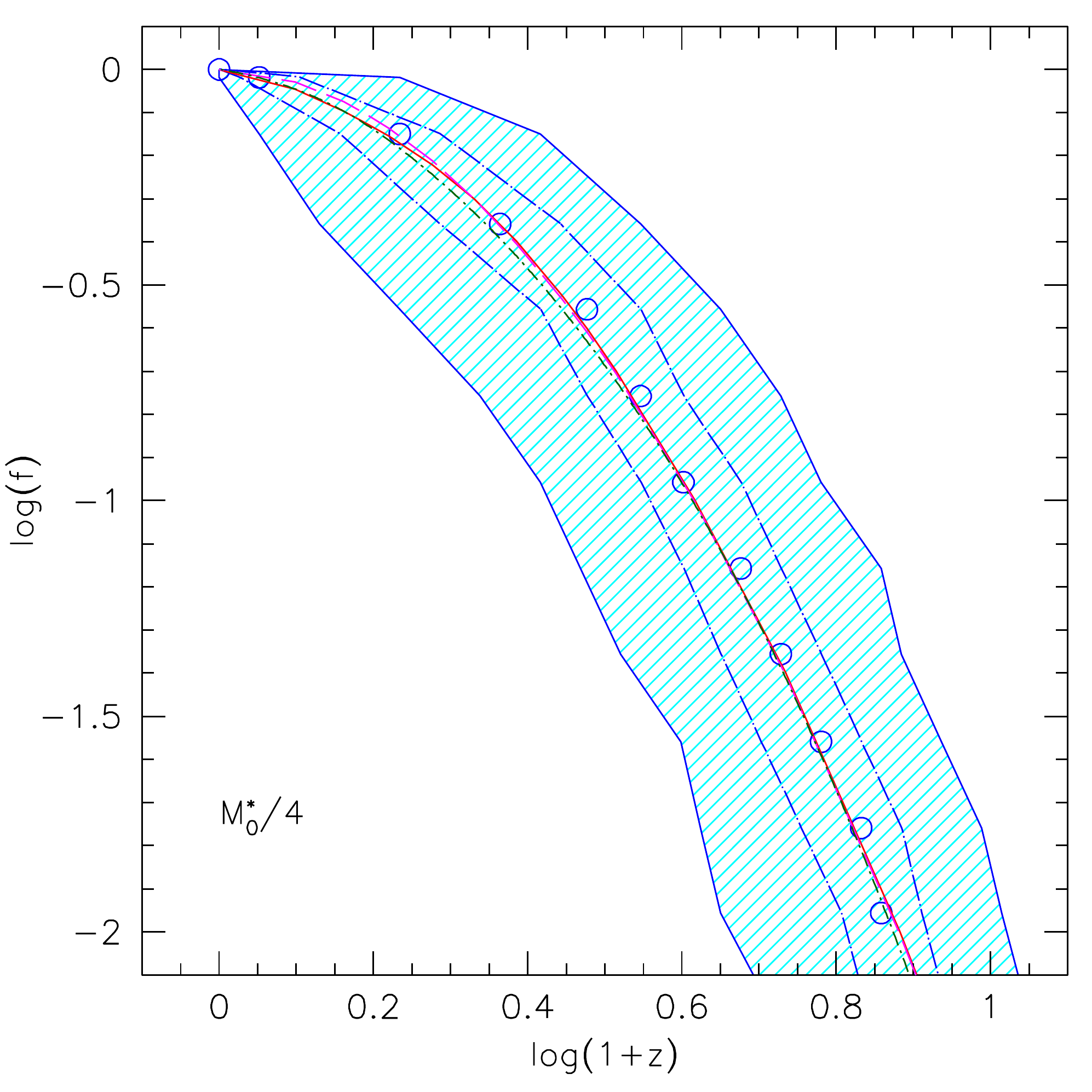}
\includegraphics[width=4.25cm]{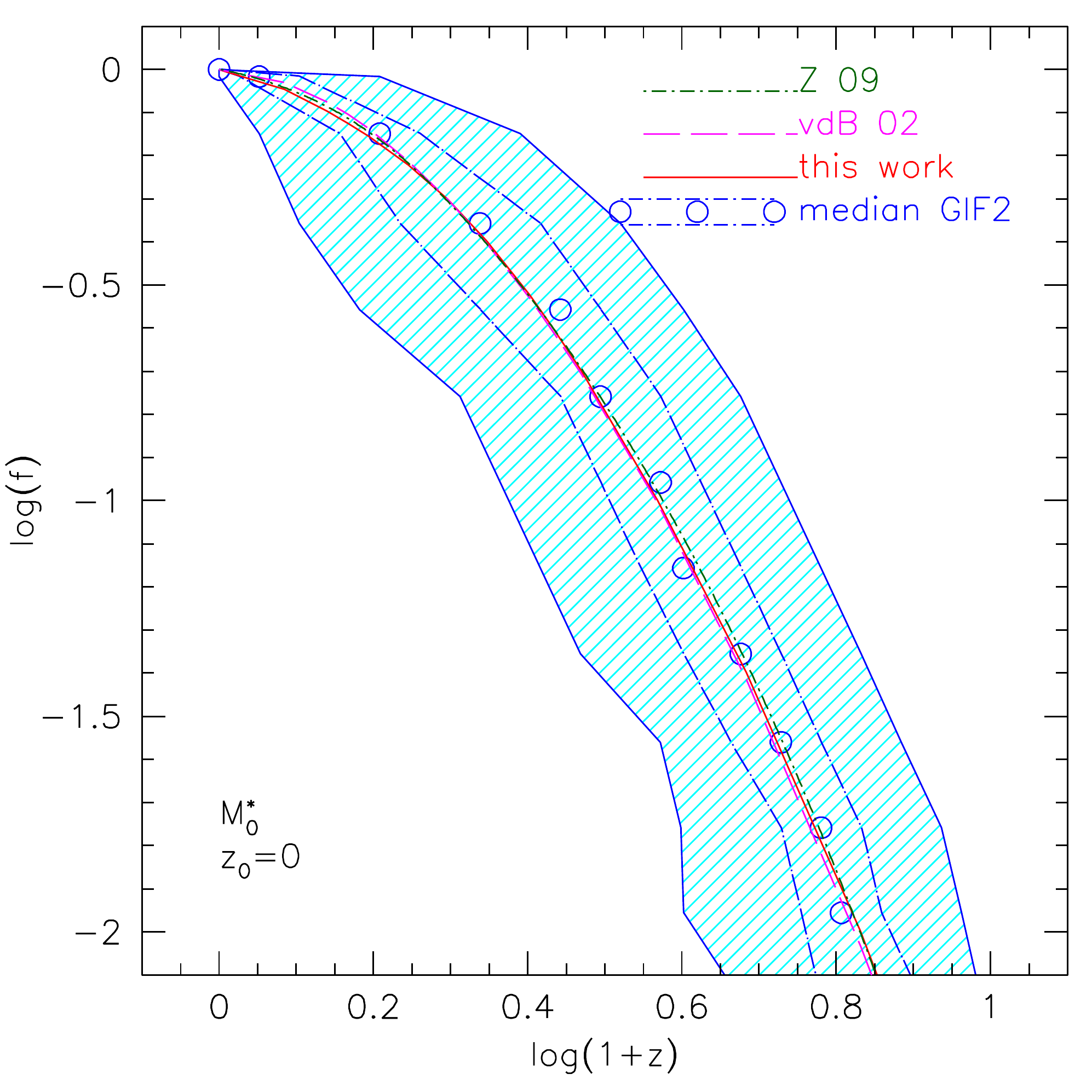}
\includegraphics[width=4.25cm]{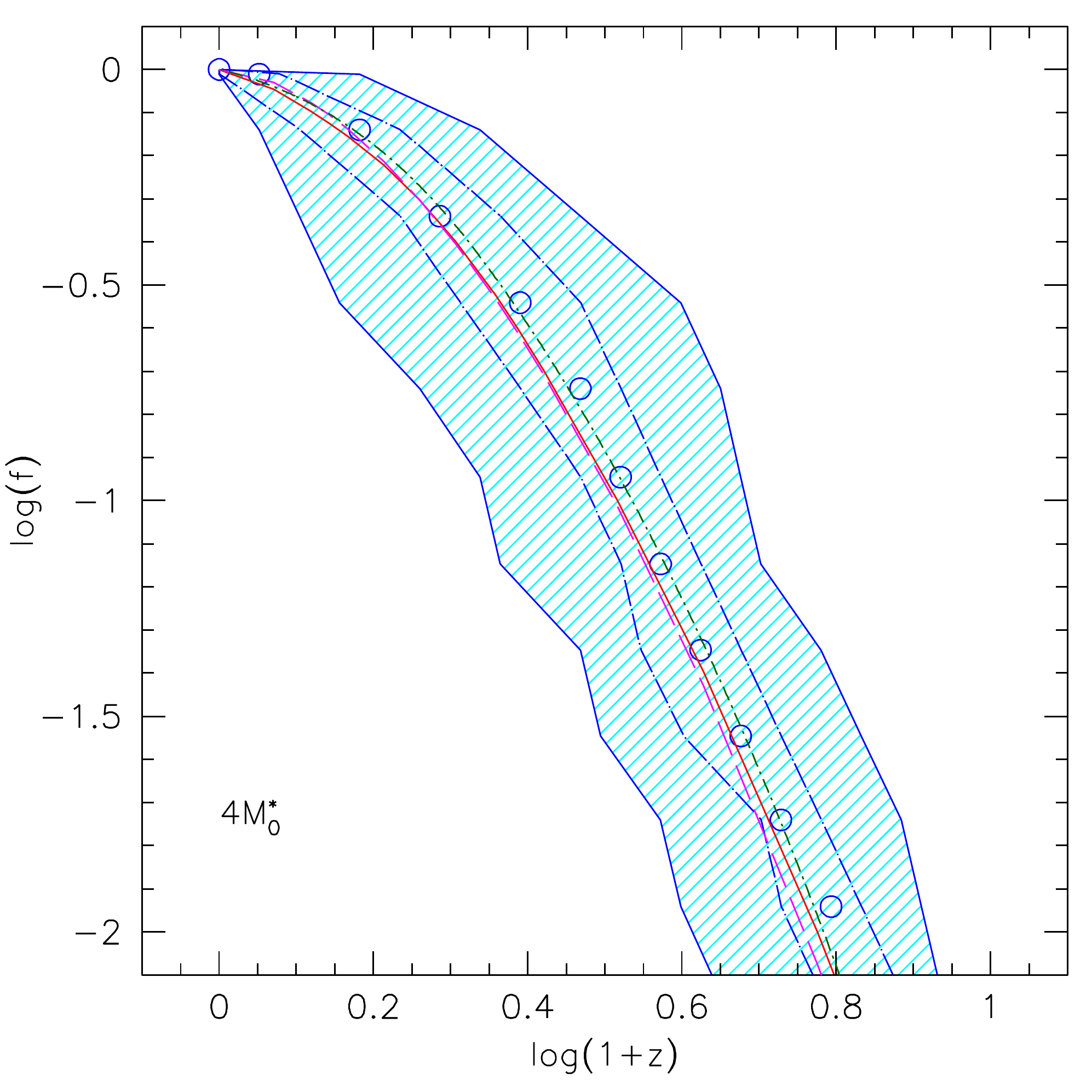}\\
\includegraphics[width=4.25cm]{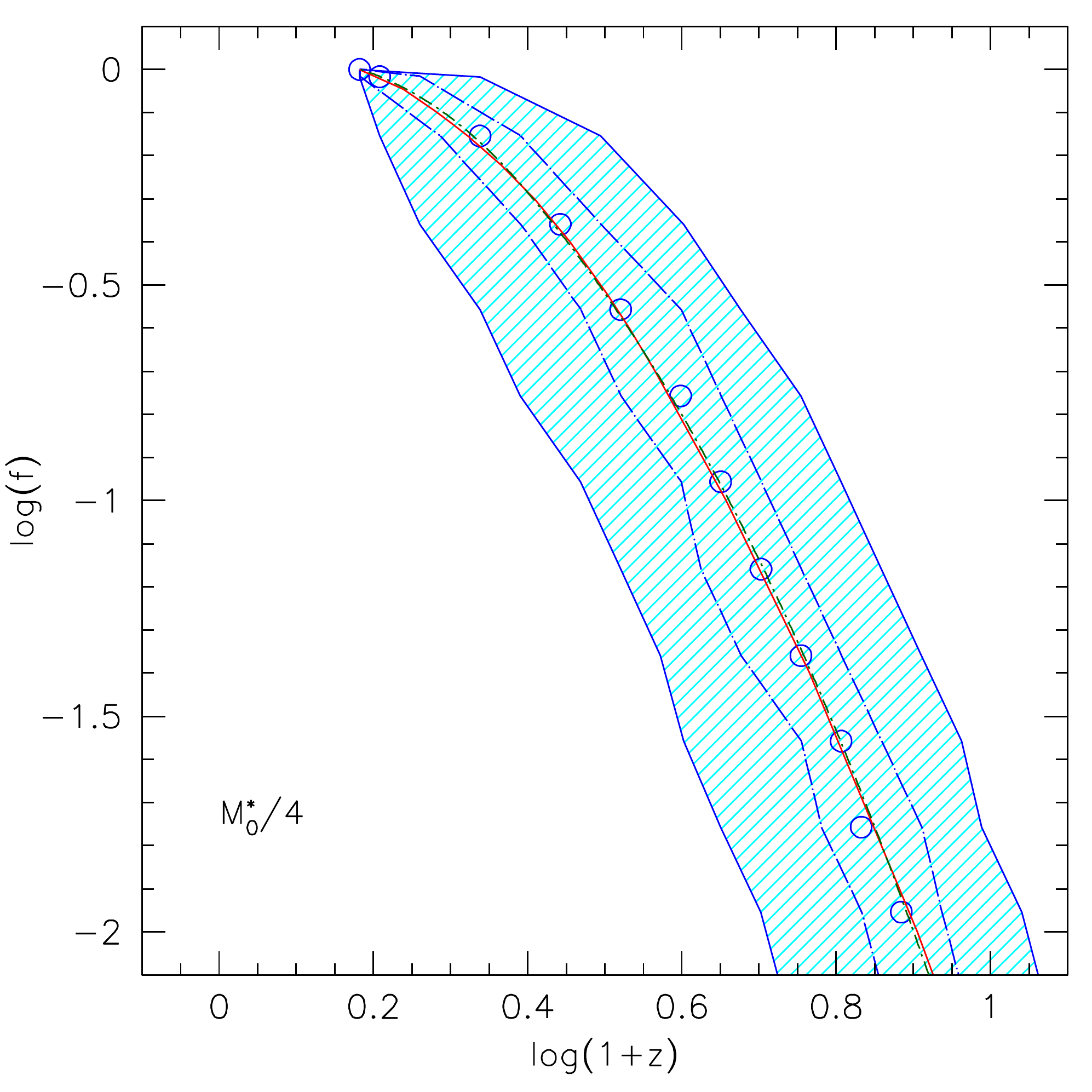}
\includegraphics[width=4.25cm]{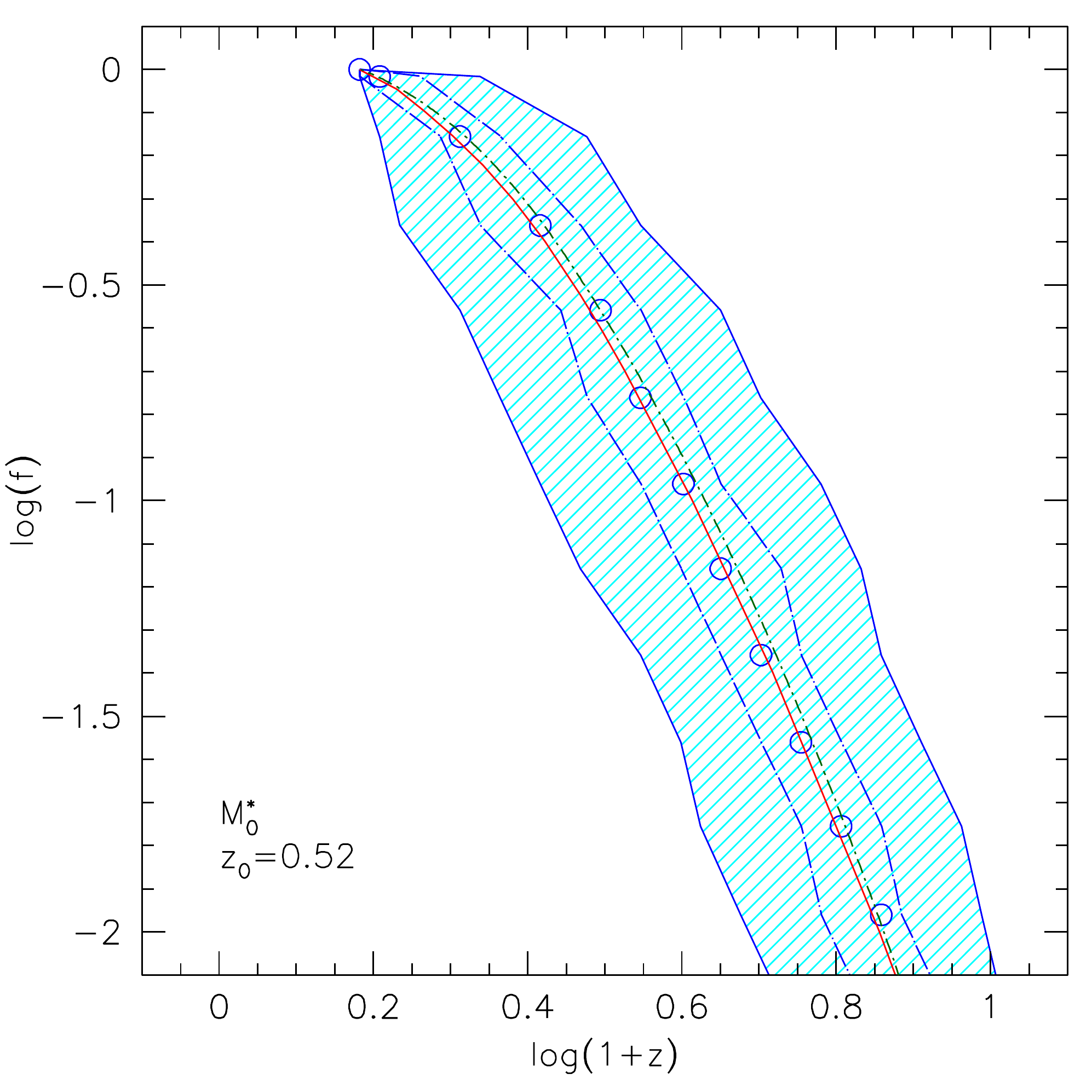}
\includegraphics[width=4.25cm]{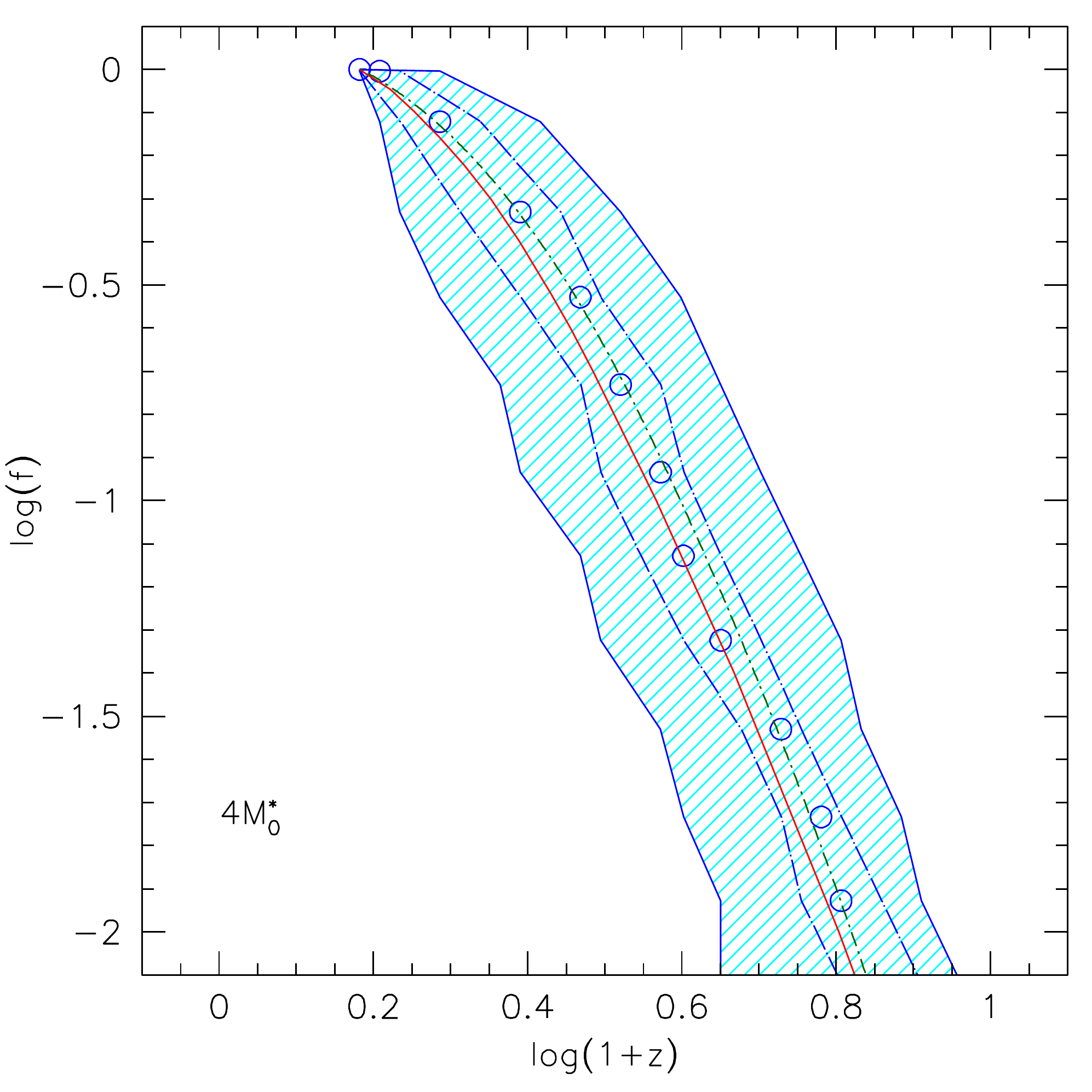} \\ 
\includegraphics[width=4.25cm]{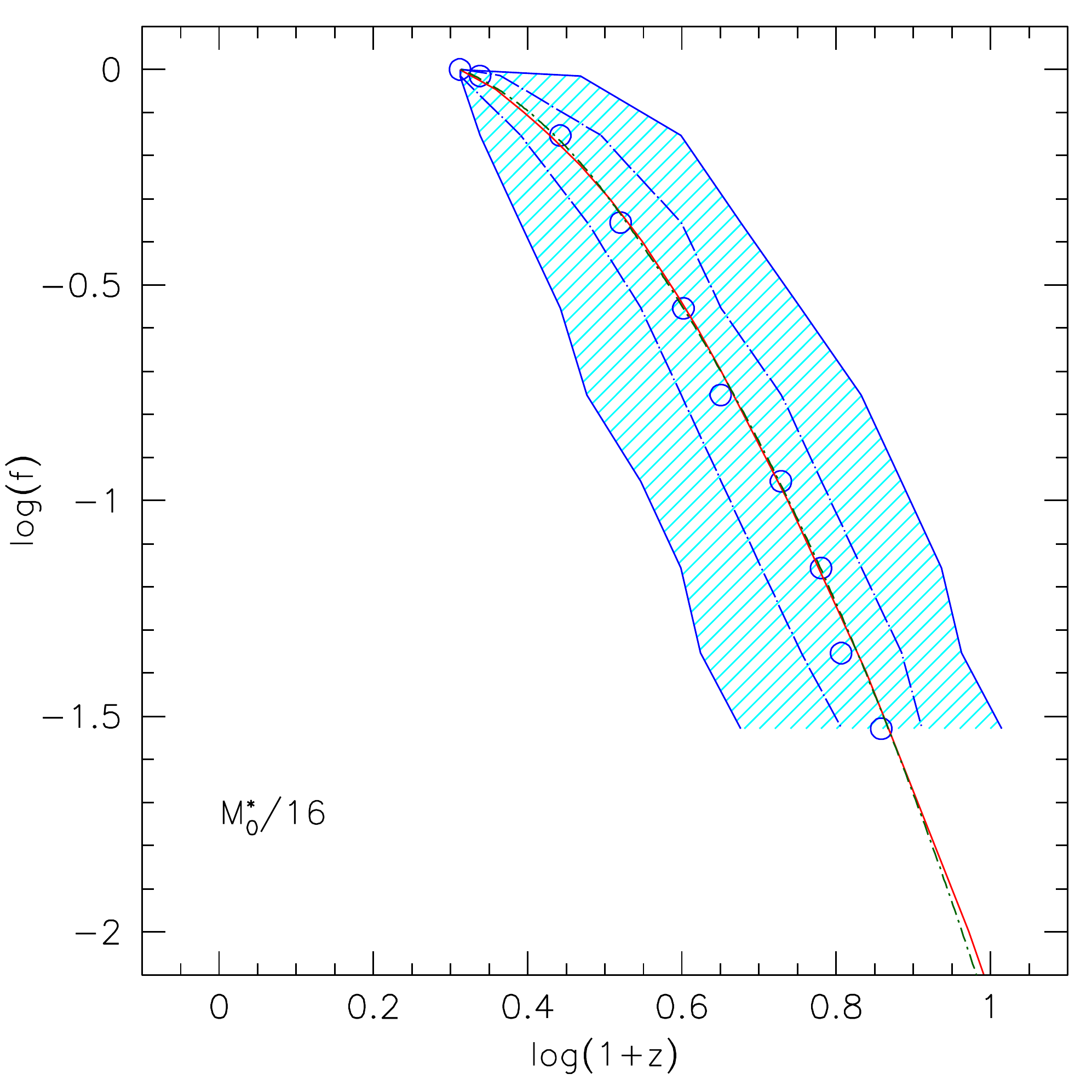}
\includegraphics[width=4.25cm]{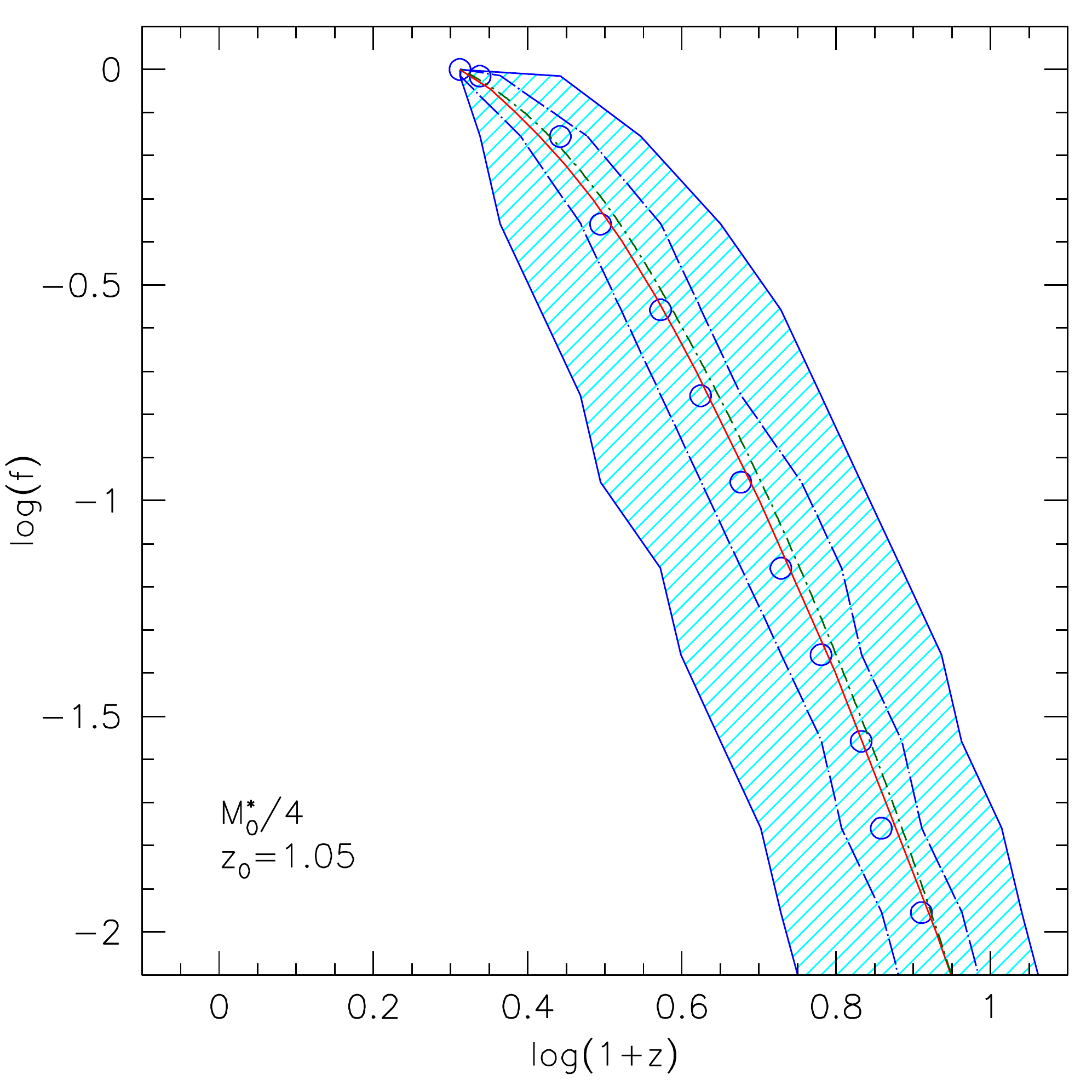}
\includegraphics[width=4.25cm]{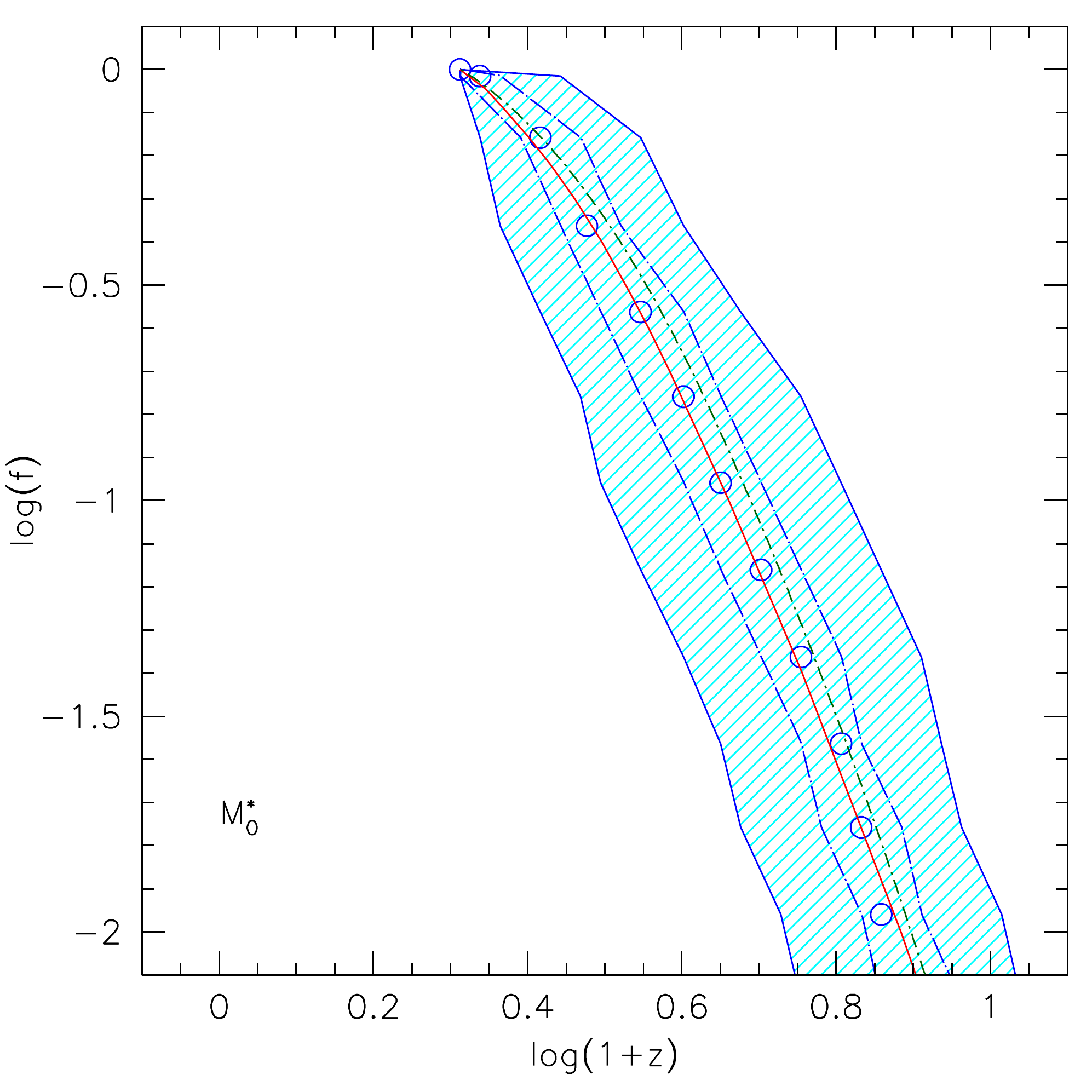} \\
\includegraphics[width=4.25cm]{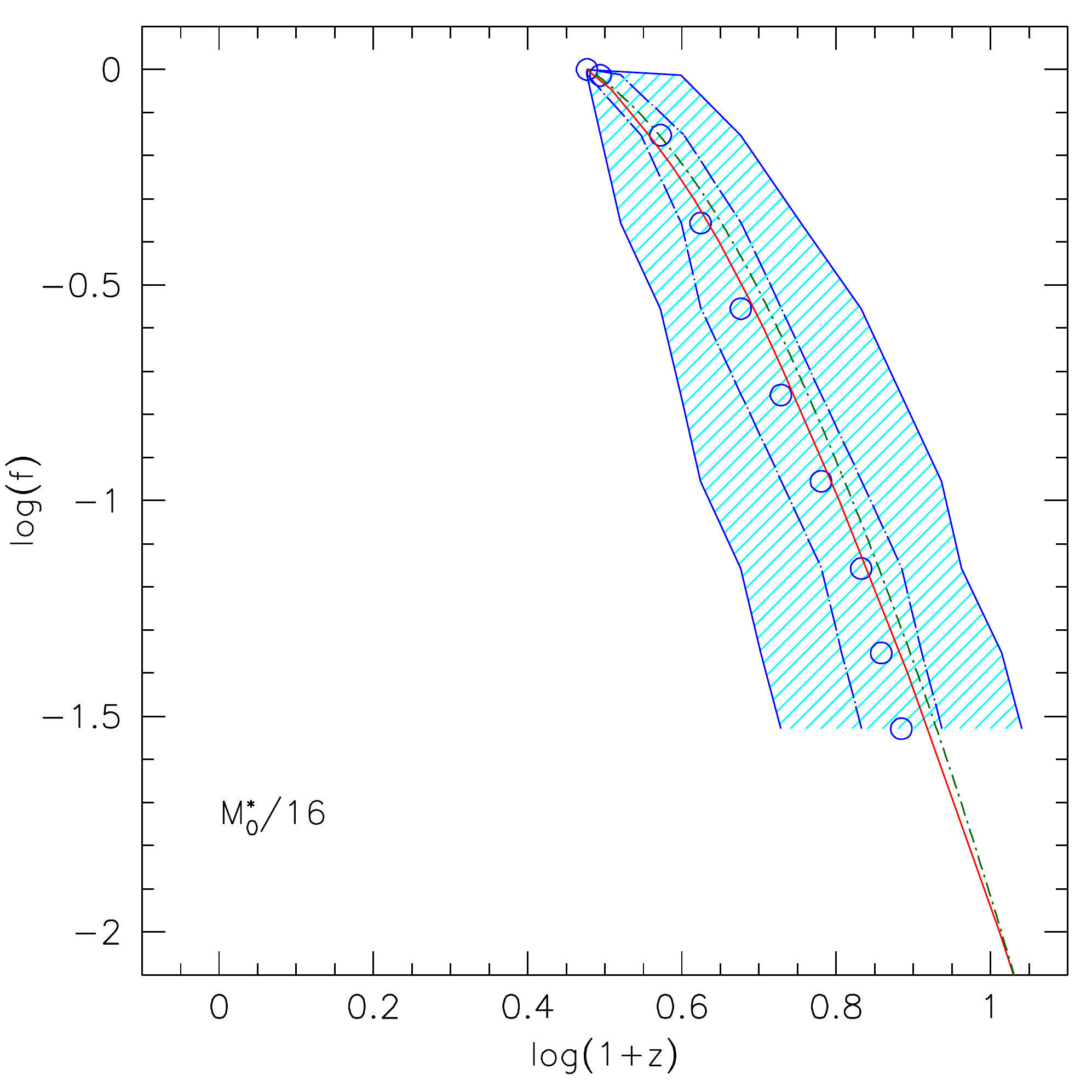}
\includegraphics[width=4.25cm]{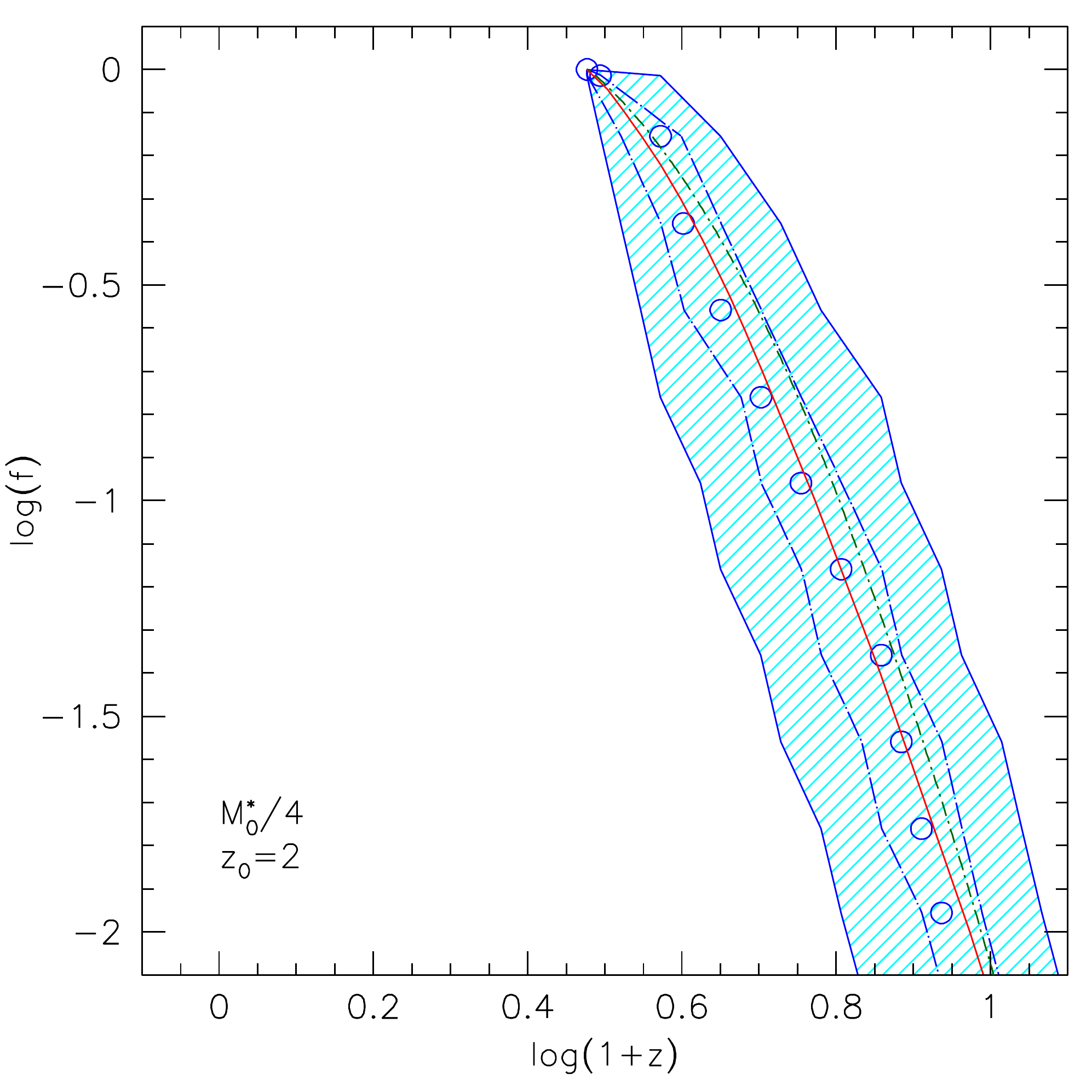}
\includegraphics[width=4.25cm]{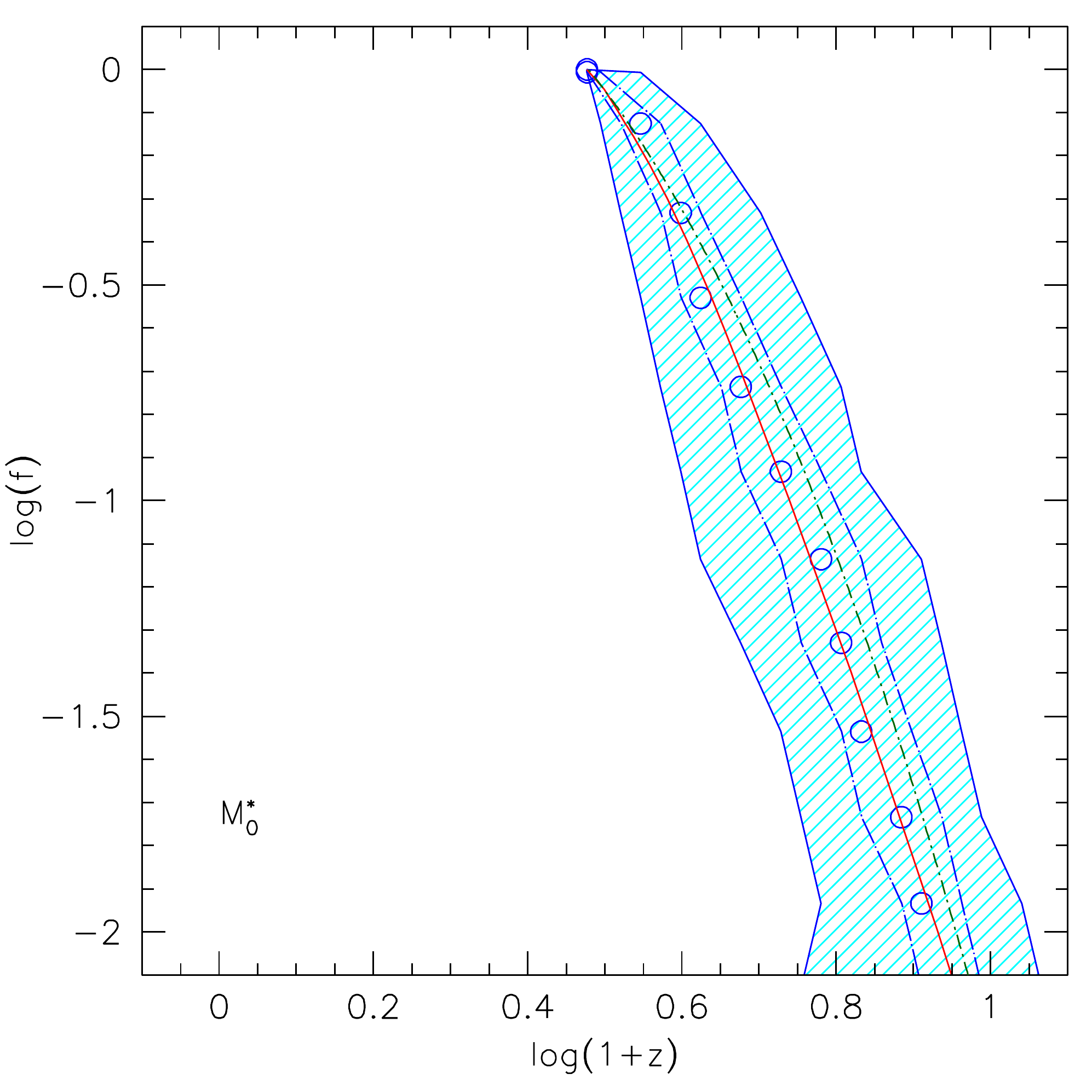} \\
\includegraphics[width=4.25cm]{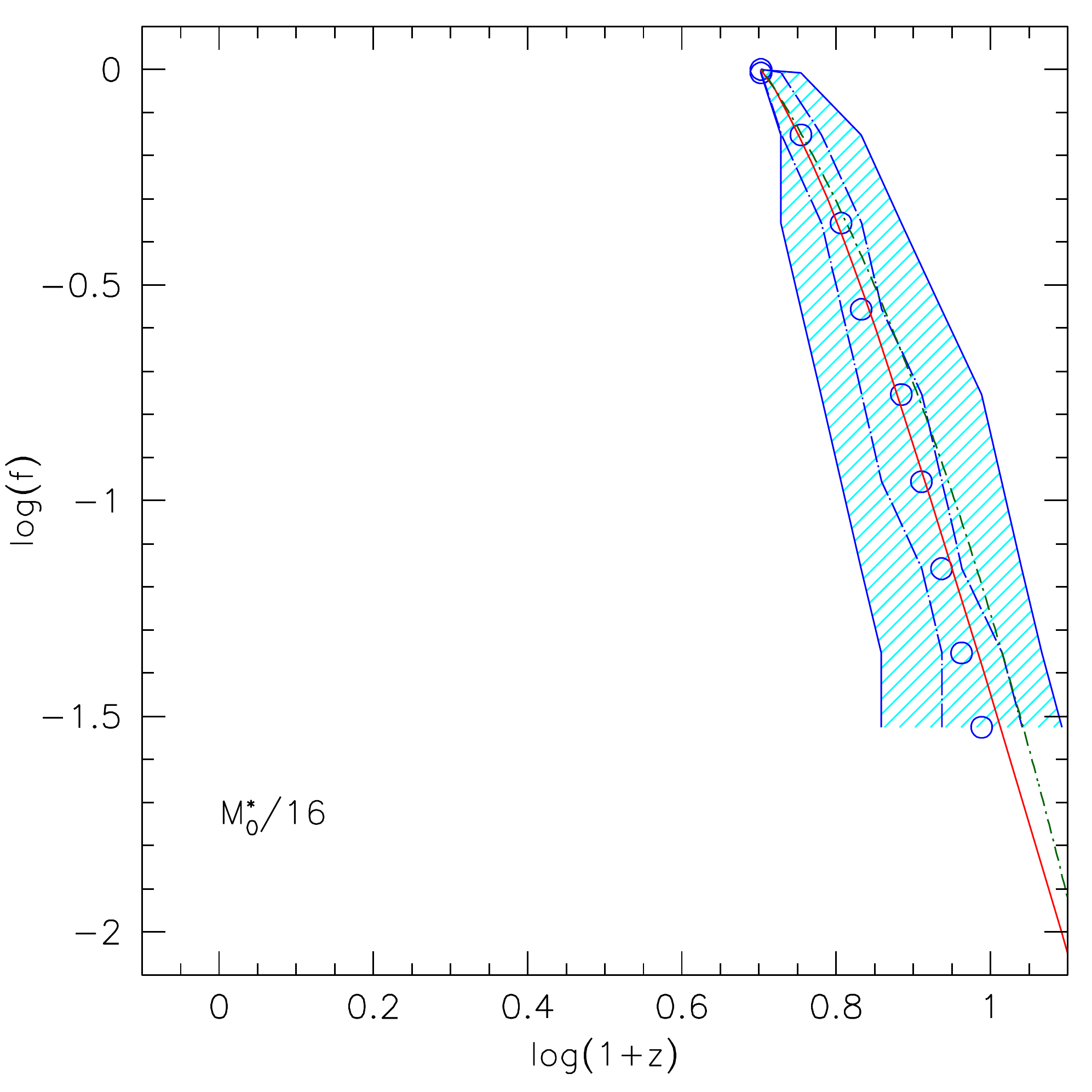}
\includegraphics[width=4.25cm]{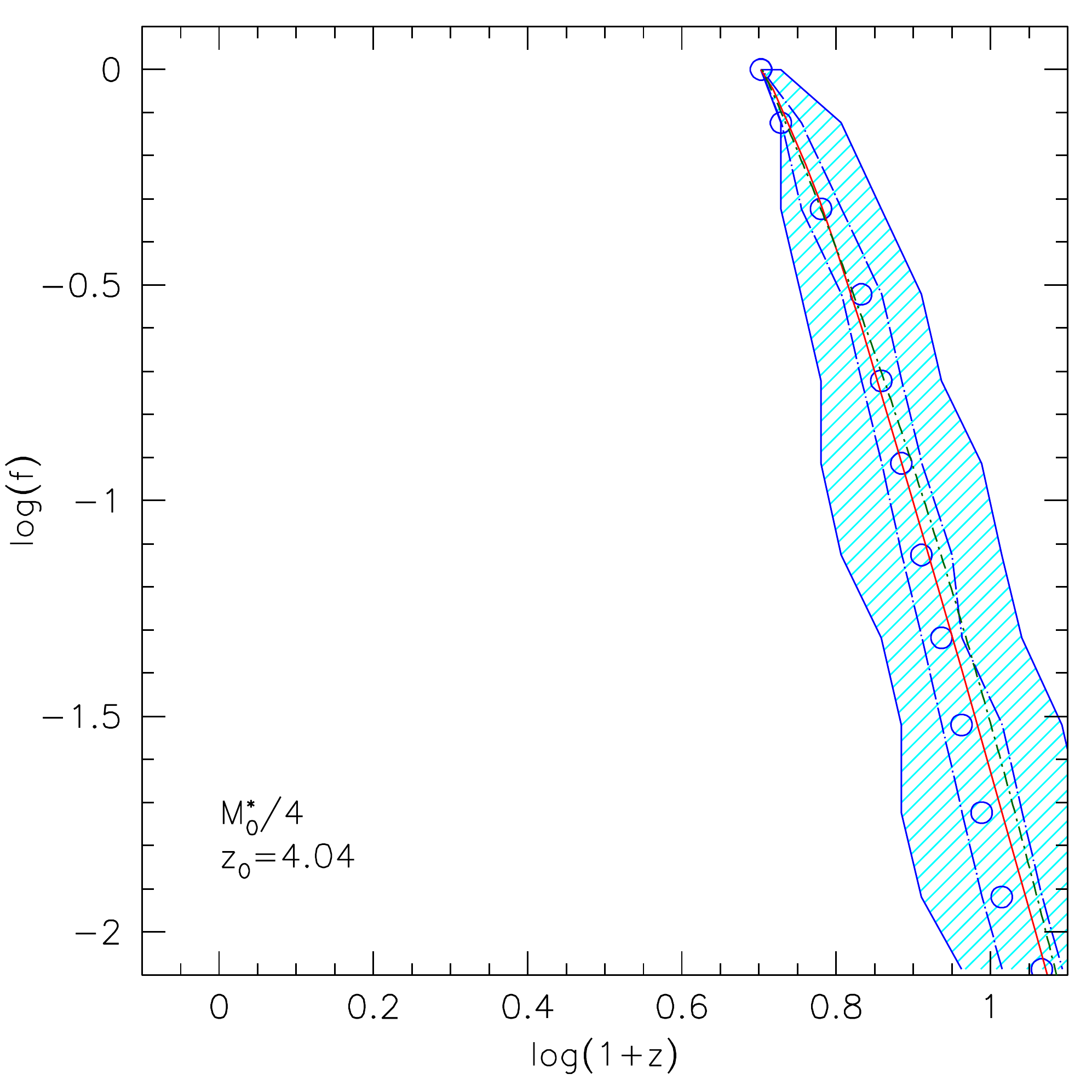} 
\caption{Mass   growth  history   of   dark  matter   haloes  in   the
  $\mathrm{\Lambda CDM}$ GIF2 cosmology. The different panels refer to
  various redshifts and host halo  masses, as labeled.  In each panel,
  open circles show the median $z$ at fixed $m/M$, and the curves that
  enclose  them  show  the  first  and  the  third  quartiles  of  the
  distribution defined by the GIF2  sample.  These curves are the same
  as  those  for the  median  $m/M$ at  fixed  $z$:  as expected  from
  equation~(\ref{zfmmp}).   The dot-dashed,  dashed  and solid  curves
  show the model  of \citet{zhao09}, \citet{vandenbosch02} (valid only
  for      haloes     starting      from     $z_0=0$)      and     our
  equation~(\ref{medzf}).\label{fmah}}
\end{center}
\end{figure*}

\subsection{Test of scaled units}
Figure  \ref{medwcorr}  shows the  correlation  between the  formation
redshift (for $f=0.9$, $0.5$, $0.1$ and $0.01$) and the host halo mass
for the same  choices of $z_0$ as before, in  scaled units.  I.e., the
redshift is expressed as $\delta_c(z_f)-\delta_c(z_0)$ and the mass as
$S(fM_0)-S(M_0)$.   The different  symbols  represent different  $z_0$
(same  as  Figure~\ref{fPformation});  they  show the  median  of  the
correlation, i.e., the median of $\delta_c(z_f)-\delta_c(z_0)$ in bins
of $S(fM_0)-S(M_0)$, while the solid  lines of the same color show the
first and third  quartiles.  The dotted and solid  lines in each panel
show a least-squares fit to the correlation, and the prediction of our
model  (equation~\ref{eqmahmod1}).  The  dashed  line in  the two  top
panels  shows the  prediction, $N(>m,z_f|M_0,z_0)  =  1/2$, associated
with  equation~(\ref{eqnusser}).  Note  that,  in the  format we  have
chosen to present our results, the  prediction boils down to a line of
slope 1/2, with zero-point equal to $\log_{10} \tilde w_f$.

\subsection{The median mass accretion history}
As described above, we have  obtained the merger history tree for each
halo more massive than $10^{11.5}\,h^{-1}M_{\odot}$ at $z_0$, for four
different choices  of $z_0$.  Figure~\ref{fmah} shows the  mass of the
main halo progenitor, in units of  the mass at $z_0$, as a function of
$z$.  The different panels show  results for parent halos of different
masses, here expressed as  fractions of $M^*_0$ (where $S(M^*_0)\equiv
\delta^2_c(0)$,    so    $M^*_0$    corresponds   to    $8.9    \times
10^{12}\,M_{\odot}/h$).  From top to bottom we show the halo accretion
histories starting  from different redshifts $z_0$.   The open circles
and dashed  curve which enclose  them represent the median,  the first
and the third quartile of  the tree extracted from the simulation. The
solid curves enclose $95\%$ of the trees.

Equation~(\ref{zfmmp})  shows that,  for these  median  quantities, it
should not matter if  we plot the median $z_f$ at fixed  $m/M = f$, or
the  median $f$ at  fixed $z$.   This is  indeed the  case, and  is in
contrast to  what would  have happened  if we had  chosen to  show the
corresponding   mean   values:  the   curve   which  traces   $\langle
z_f|f\rangle$ as  a function  of $f$ is  slightly different  from that
which  traces  $\langle  f|z_f\rangle$  as  a function  of  $z$.   Our
equation~(\ref{zfmmp}) shows  that this is  a consequence of  the mean
not being the same as the median: the distributions are skewed.

The  dot-dashed   curve  in  each   panel  shows  the   prediction  of
\citet{zhao09}: it  describes our  measurements very well.  The dashed
curve (for  $z_0=0$ only),  shows the model  of \citet{vandenbosch02}.
The solid curve  in each panel shows our  model, evaluated as follows:
given $M_0$  at redshift $z_0$, we  compute $S(M_0)$, $\delta_c(z_0)$,
and   a   table   of   $S(f\,M_0)$   for   any   $0<f<1$.    Inverting
equation~(\ref{eqmahmod1})       then       gives      $z_f$       via
equation~(\ref{medzf}).   This  agrees very  well  with the  numerical
simulation   results   and   therefore   also  with   the   model   of
\citet{zhao09}.  But  note that our prescription is  {\em much} easier
to implement.

\subsection{Correlation between formation times}
\begin{figure}
\begin{center}
\includegraphics[width=7.5cm]{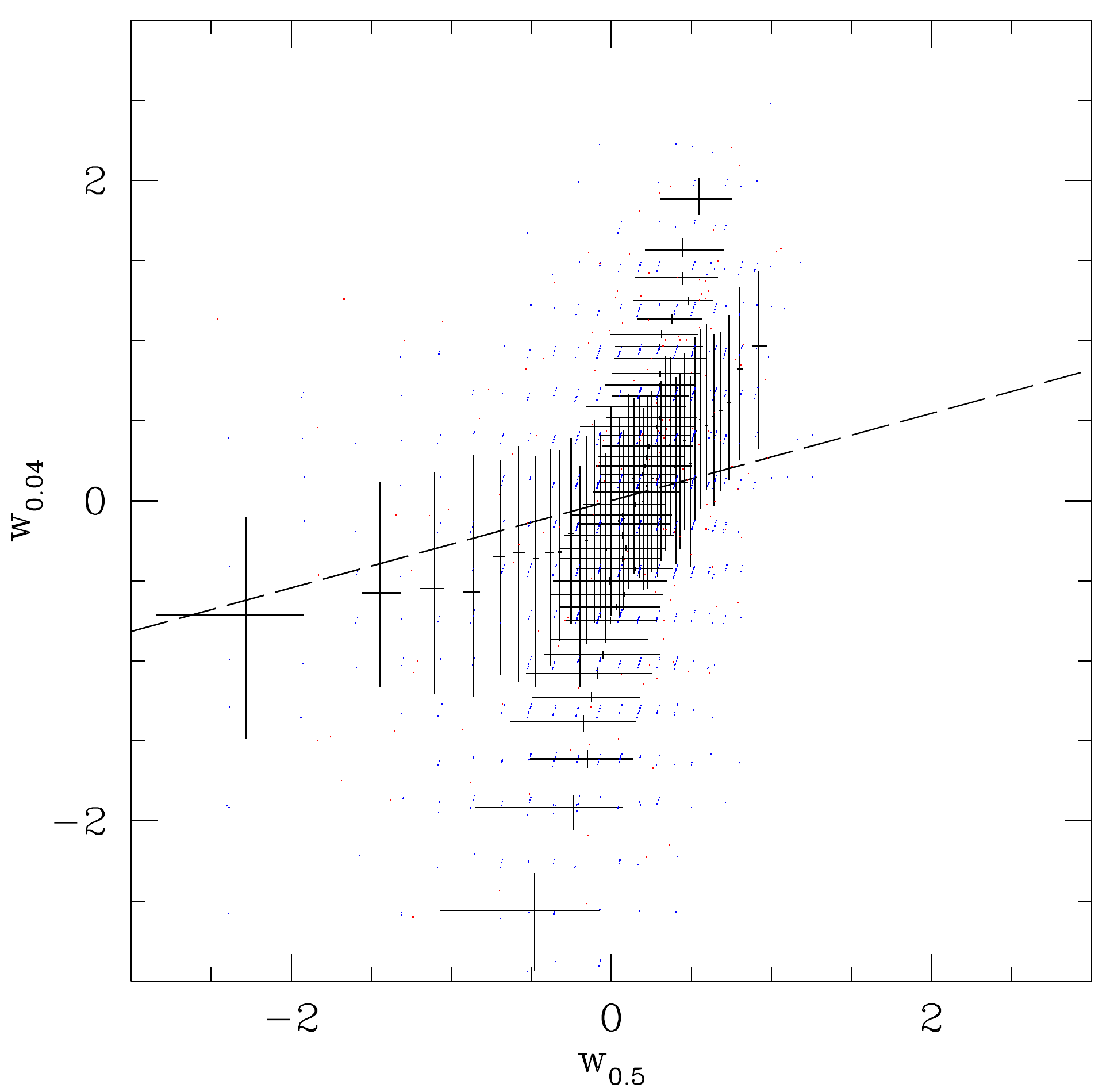}
\caption{\label{whw04}Correlation   between   formation  times,   when
  formation  is defined  as the  first time  that the  main progenitor
  $m_{\rm  MP}$ contains  0.5 and  0.04 of  the total  mass.   We have
  scaled  the variables  by their  rms values,  so the  fact  that the
  median  $w_{0.5}/\sigma_{0.5}$ at fixed  $w_{0.04}/\sigma_{0.04}$ is
  rather different  from the median  $w_{0.04}/\sigma_{0.04}$ at fixed
  $w_{0.5}/\sigma_{0.5}$ indicates that  the correlation between these
  two is weak.}
\end{center}
\end{figure}

So  far, we  have mainly  tested  if $w_f$  is indeed  a good  scaling
variable  for  the  entire range  of  $f$.   In  what follows,  it  is
interesting to  know if the  different formation time  definitions are
correlated.  E.g.,  if a halo  is above the median  formation redshift
for  one value  of $f$,  is it  also likely  to have  formed  at above
average  $w_f$ for  another $f$?   Figure~\ref{whw04} shows  that this
correlation is  weak, at least for  $f=1/2$ and 0.04:  When $w_{0.5} -
\langle w_{0.5}\rangle$  and $w_{0.04} -  \langle w_{0.04}\rangle$ are
normalized by  their rms values,  then the correlation  coefficient is
$\sim 0.25$.

\subsection{Halo concentrations from masses and formation times}
\begin{figure*}
\begin{center}
\includegraphics[width=5.8cm]{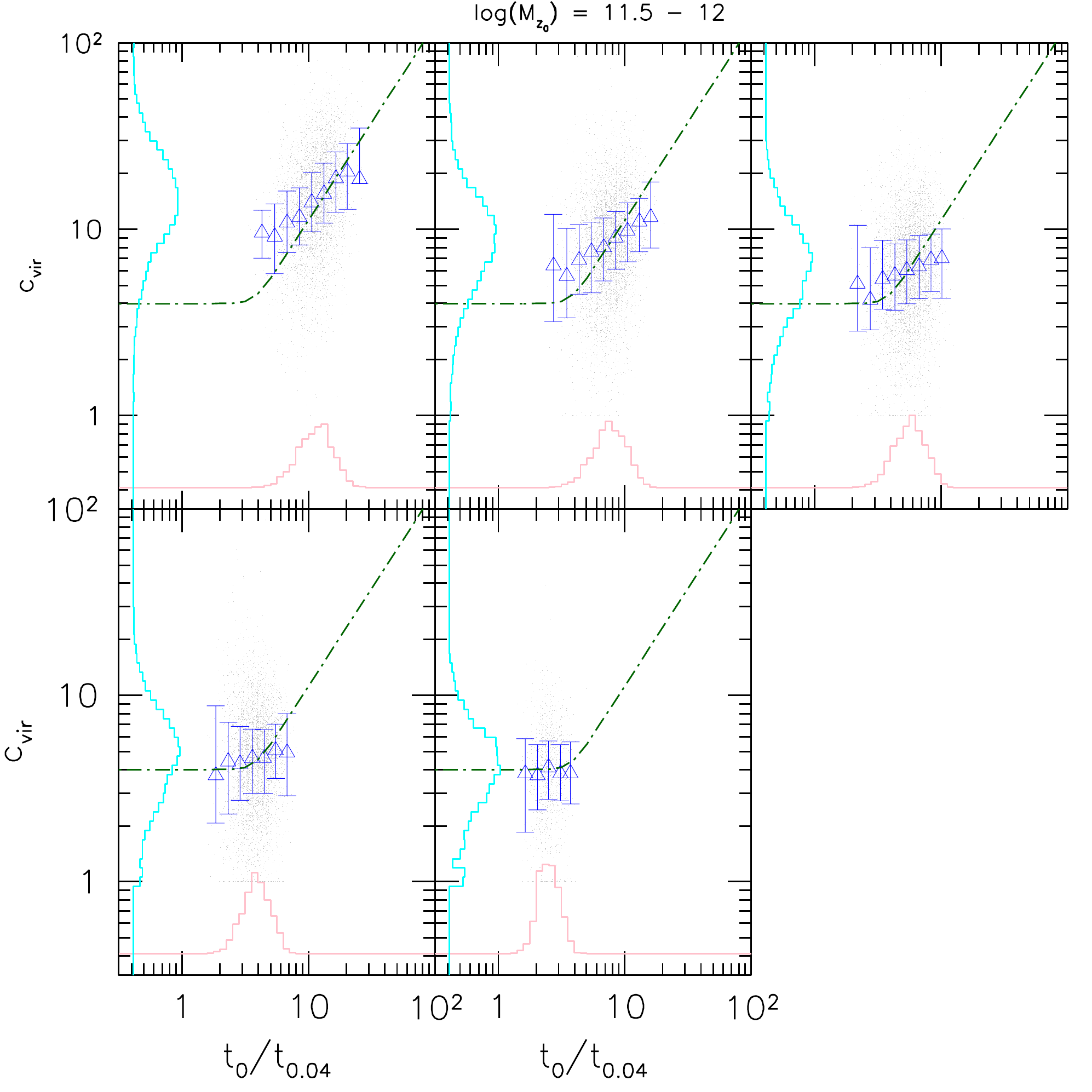}
\includegraphics[width=5.8cm]{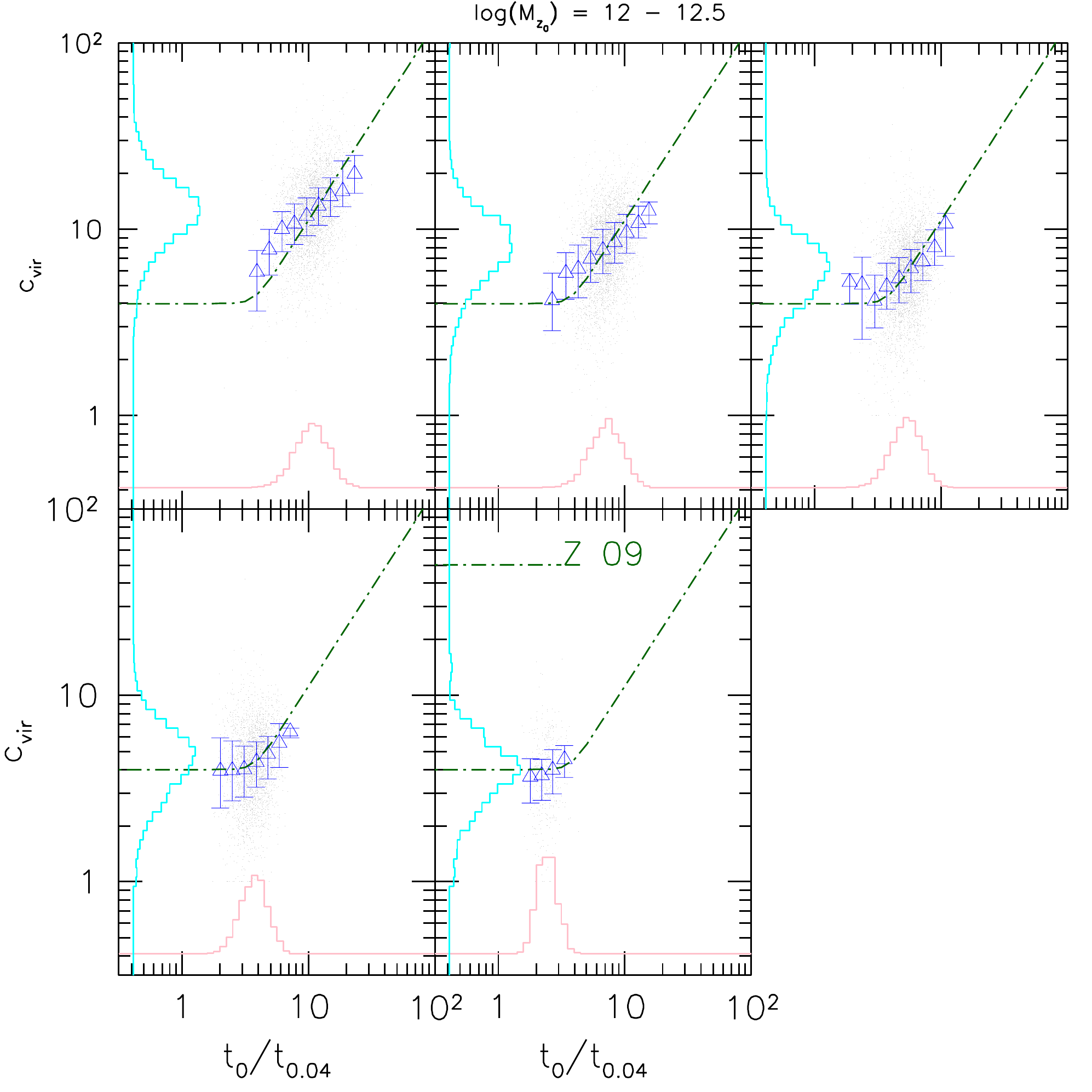}
\includegraphics[width=5.8cm]{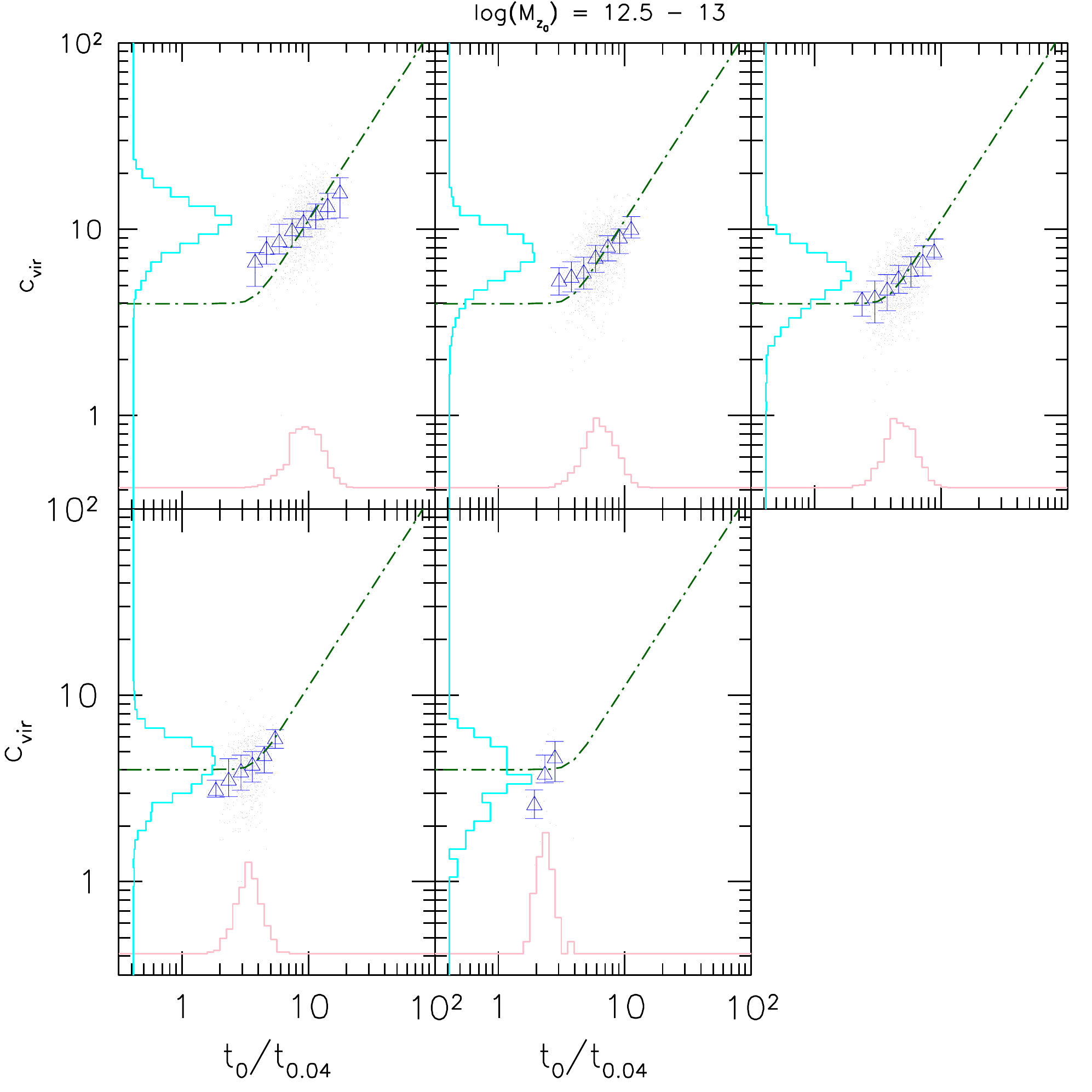}
\caption{Correlation between  host halo concentration and  the time at
  which the system assembled $4\%$  of its mass.  The different panels
  (left,  central and  right)  show results  for  different host  halo
  masses $M_0$, while the different windows show results for different
  $z_0$  (same  as  Figure~\ref{fmcrel}).   The  histograms  show  the
  distributions of $c_{vir}$ and  $t_0/t_{0.04}$, and the symbols with
  error bars show  the correlation between them: the  median and first
  and  third  quartiles  of  the  distribution of  $c_{vir}$  at  each
  $t_0/t_{0.04}$.        The        dot-dashed       curve       shows
  equation~(\ref{czhao09}).  \label{fCT}}
\end{center}
\end{figure*}

Equation~(\ref{czhao09}) states that  the formation time $t_{0.04}$ of
a halo  can be  converted into an  estimate of its  concentration.  To
check this,  we estimated $c_{vir}$ for  each halo in  GIF2 by fitting
the matter density profile to the NFW functional form,
\begin{equation}
  \rho(r|M_{vir}) = \frac{\rho_s}{r/r_s(1+r/r_s)^2}\,,
\end{equation}
where 
\begin{equation}
  \rho_s = \frac{M_{vir}}{4 \pi (R_{vir}/c_{vir})^3}
  \left[ \ln(1+c_{vir}) - \frac{c_{vir}}{1+c_{vir}}\right]^{-1}\,
\end{equation}
and $c_{vir}\equiv r_s/R_{vir}$.  
Figure~\ref{fCT} shows the correlation between $c_{vir}$ and 
$t_0/t_{0.04}$, for various $M_{vir}$ and $z_0$.  
The dot-dashed curve (same in each panel) shows equation(~\ref{czhao09}):  
if $c_{vir}$ were completely predicted by $t_0/t_{0.04}$ then there 
should be no scatter around this relation.  

In each panel, dots show the halos, and triangles with error bars show
the median and first  and third quartiles in ${\rm log}_{10}(c_{vir})$
for  narrow  bins  in  $t_0/t_{0.04}$.  The  actual  distributions  of
$\log_{10}(c_{vir})$ and $\log_{10}(t_0/t_{0.04})$ are shown along the
left axis and across the bottom.   (Note that, for each $M_0$ bin, the
peak  of the distribution  of $t_0/t_{0.04}$  shifts towards  unity as
$z_0$ increases, for the reason given following equation~\ref{medzf}.)
Clearly,  although   equation~(\ref{czhao09})  provides  a  reasonable
description  of the  triangles, the  scatter around  this  relation is
considerable.   Quantitatively, the  scatter in  $\log_{10}c_{vir}$ is
$0.13$~dex for $z_0=0$ halos with $M_{vir} \ge 10^{13}h^{-1}M_\odot$.

The previous subsection showed that $t_{0.5}$ and $t_{0.04}$ of a halo
are only weakly correlated.  So it is reasonable to ask if some of the
scatter in $c_{vir}$ can be  removed if one includes information about
$t_{0.5}$.  E.g.,  one might  reasonably expect the  most concentrated
halos to have small values of both $t_{0.5}$ {\em and} $t_{0.04}$.  We
have found that the typical concentration scales as
\begin{equation}
 \log_{10} c_{vir} = \log_{10} 0.45 \,
      \left[4.23 + \left(\frac{t_1}{t_{0.04}}\right)^{1.15} 
                 + \left(\frac{t_1}{t_{0.5}}\right)^{2.3} \right].
 \label{cznew}
\end{equation}
For halos  more massive  than $10^{13}h^{-1}M_\odot$, the  rms scatter
around this relation  (0.1~dex) is indeed reduced with  respect to the
case  in which only  $t_{0.04}$ is  known (0.12~dex).   Evidently, the
remaining  scatter is due  to other  processes.  Table~\ref{cvirtable}
summarizes how this scatter decreases as more information is added.

\begin{table}
\caption{ \label{cvirtable} Precision of concentration estimates 
  for halos with $M_{vir}>10^{13}\,h^{-1}M_{\odot}$.}
\begin{center}
\begin{tabular}{ l | r |}
  \hline  \hline
 rms$(\log_{10}c_{vir}|M_{vir})$ & $0.124$ \\ 
 rms$(\log_{10}c_{vir}|\tau_{0.04},M_{vir})$ & $0.120$ \\ 
 rms$(\log_{10}c_{vir}|\tau_{0.5},\tau_{0.04},M_{vir})$ & $0.100$ \\ \hline
\hline
\end{tabular}
\end{center}
\end{table}

\begin{figure}
\begin{center}
\includegraphics[width=\hsize]{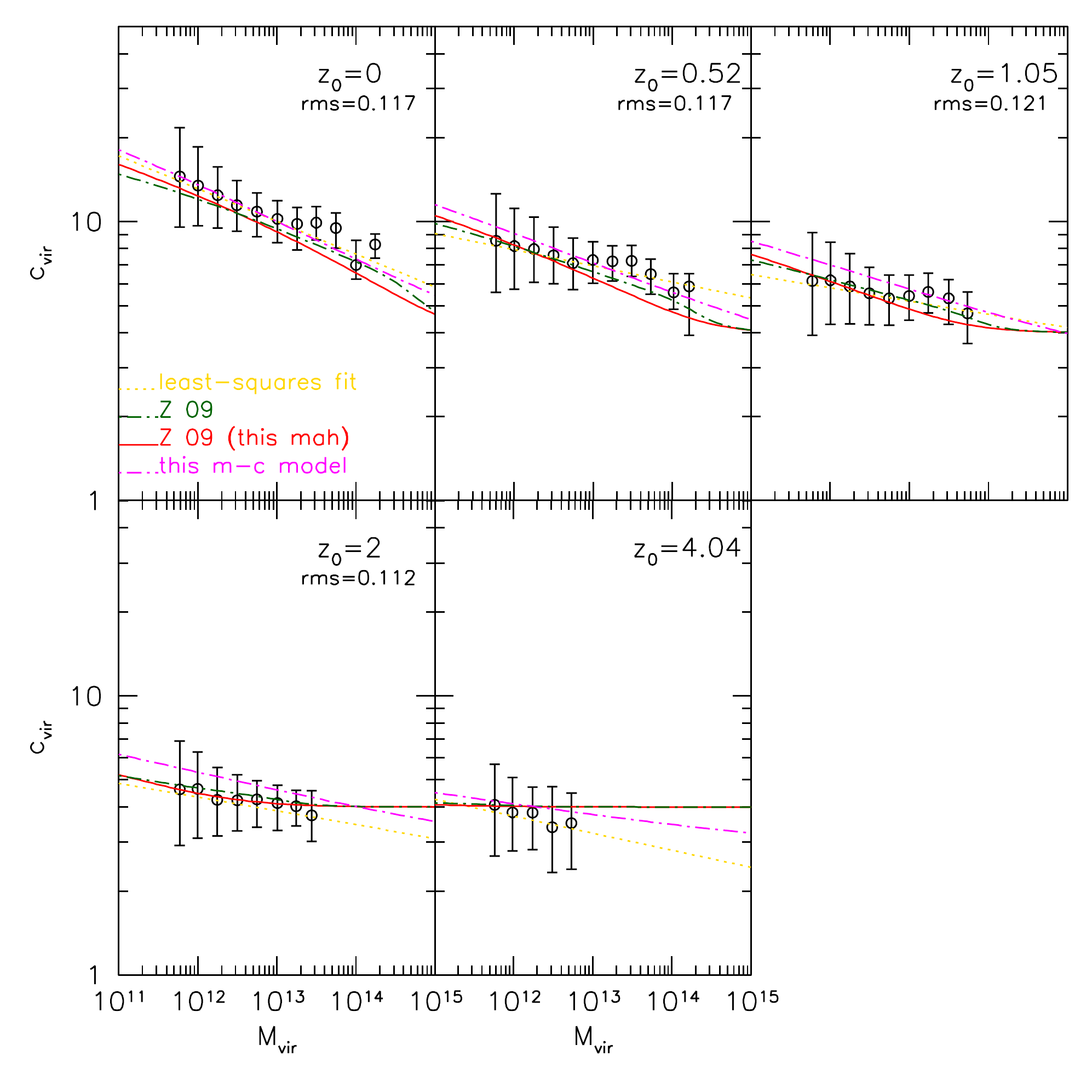}
\caption{Median  mass concentration relation  in the  GIF2 simulation.
  Open  circles  show  haloes  in  the simulation  at  five  different
  redshifts  $z_0$, as  labeled in  the panels.   Error bars  show the
  first and the third quartiles of the distribution for each mass bin.
  The dotted  line shows a  least-squares fit to the  data; dot-dashed
  and the solid  curves show the predictions based  on the median mass
  growth history of \citet{zhao09} and our equation~(\ref{medzf}).  In
  each panel  we show the  rms in $\log_{10}(c_{vir})$ at  fixed mass,
  for           haloes           more           massive           than
  $10^{13}\,M_{\odot}/h$.  \label{fmcrel}}
\end{center}
\end{figure}

One of the  main goals of this paper is to  provide a simple algorithm
for  generating the  $M-c$ relation.   Figure~\ref{fmcrel}  shows this
relation  in  GIF2.    The  open  circles  show  the   median  of  the
distribution at fixed  mass, and the error bar  encloses the first and
third  quartiles.    Different  panels  show   results  for  different
redshifts.  The dotted  line shows  a least-squares  fit to  the data,
while the  dot-dashed and solid  curves show the relations  which come
from inserting the mass  accretion histories of \citet{zhao09} and our
equation~(\ref{medzf}) into equation~(\ref{czhao09}).  Both approaches
capture the  mass dependence of the concentration,  and its flattening
with redshift, very well.  But note  that our estimate of the value of
$t_{0.04}/t_1$ that  is appropriate  for a given  mass bin  (e.g., our
equation~\ref{medzf}), is much more straightforward to implement.

\begin{figure}
\begin{center}
\includegraphics[width=\hsize]{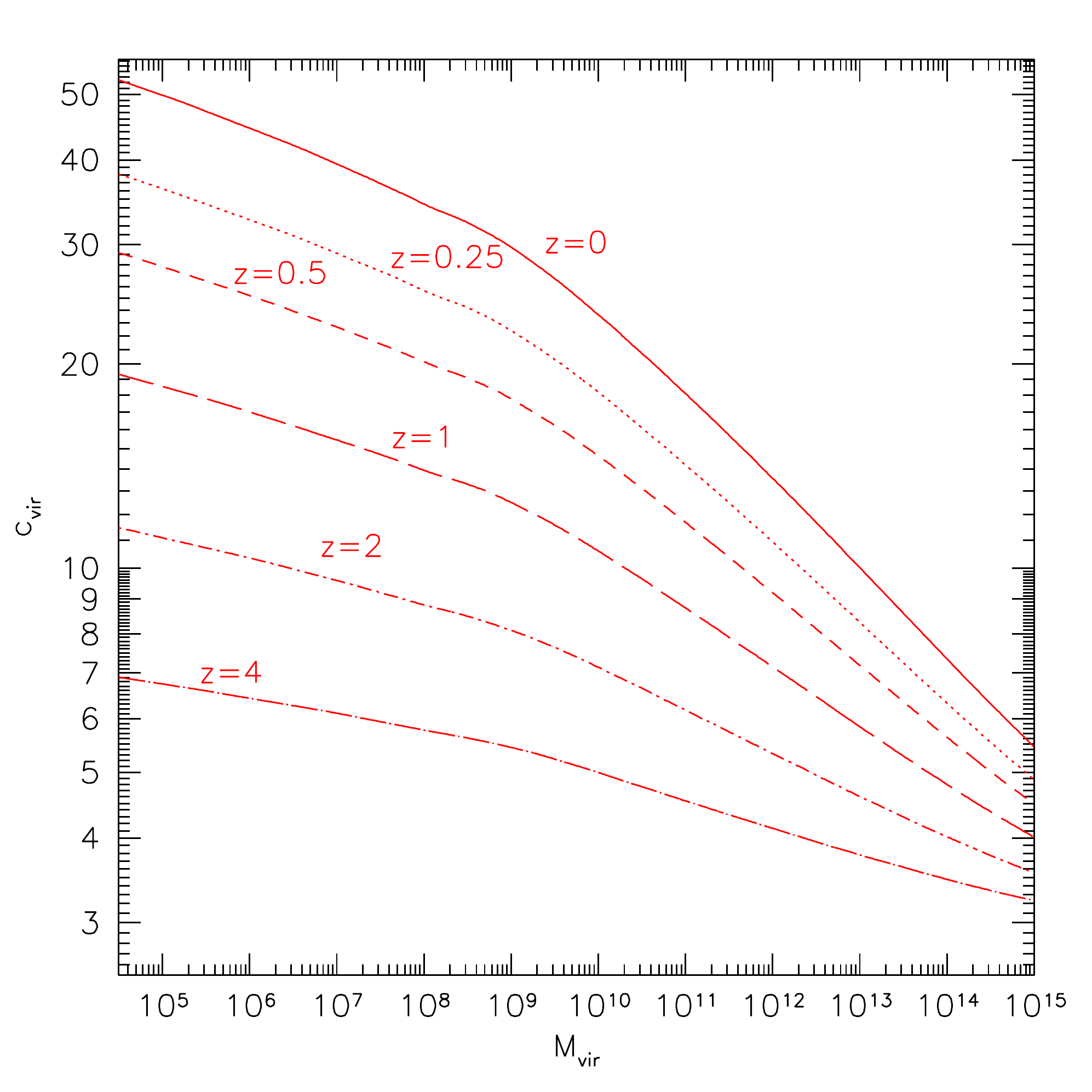}
\caption{Mass concentration relation at six different redshifts predicted
    using equation~(\ref{medzf}). The concentration prediction has been extended
    down to small masses: $3 \times 10^4 M_{\odot}/h$.\label{fmcrelsmall}}
\end{center}
\end{figure}

The $M-c$ relation for masses of order $10^6 M_{\odot}$ is needed 
for modeling the milli-lensing effects of galaxy-size halo substructures 
\citep{keeton03,keeton03b,sluse03,metcalf01,metcalf12}.  The small mass 
regime is also important for estimating the dark matter annihilation 
signal coming from the integrated effect of substructures and field 
haloes at masses smaller than $10^6 M_{\odot}$ 
\citep{pieri05,giocoli08a,giocoli09a}.  However, this is below the 
mass resolution of current numerical simulations.  
Therefore, Figure~\ref{fmcrelsmall} shows the result of using our model 
(equation~(\ref{medzf}, where $t_{0.5}$ and $t_{0.04}$ are obtained using 
our mass accretion history model) to estimate the $M-c$ relation at six 
different redshifts down to $3\times 10^{4}h^{-1}M_{\odot}$.

\subsection{Formation times from masses and concentrations}
In real datasets,  while we may hope to have  a reasonable estimate of
the mass  of the host halo,  and perhaps of its  concentration, we are
unlikely  to  know  either  $t_{0.04}$  or $t_{0.5}$.   So  one  might
reasonably  ask how  well $t_f$  can be  predicted given  knowledge of
either $M_{vir}$, $c_{vir}$ or both.

\begin{table}
\caption{ \label{tformtable} Precision of formation time estimates 
  for halos with $M_{vir}>10^{13}\,h^{-1}M_{\odot}$.}
\begin{center}
\begin{tabular}{ l | r |}
  \hline  \hline
 rms$(\tau_{0.5}|M_{vir})$ & $0.123$ \\ 
 rms$(\tau_{0.5}|c_{vir})$ & $0.088$ \\ 
 rms$(\tau_{0.5}|c_{vir},M_{vir})$ & $0.087$ \\ \hline
 rms$(\tau_{0.04}|M_{vir})$ & $0.176$ \\ 
 rms$(\tau_{0.04}|c_{vir})$ & $0.111$ \\ 
 rms$(\tau_{0.04}|c_{vir},M_{vir})$ & $0.109$ \\ \hline
\hline
\end{tabular}
\end{center}
\begin{flushleft}
$^{\star}$ $\tau_f \equiv \log_{10}(t_f/t_1)$ \\
\end{flushleft}
\end{table}

Since equation~(\ref{eqmodel1}) gives  the distribution of $z_f$ given
$M_{vir}$ and $z_0$, it can be manipulated to give the distribution of
$t_f/t_1$ at fixed $M_{vir}$.  In practice,  we can get a good idea of
the rms scatter  around the mean by estimating how  far on either side
of the median  value one must go before 68\%  of the total probability
has been enclosed.  These correspond to values of
$w_f 
= \sqrt{2 \ln(1 + 0.188\,\alpha_f)}$ and
$w_f 
=  \sqrt{2   \ln(1  +   5.303\,\alpha_f)}$.   For  $f=0.04$   we  have
$\alpha_f=8$  so $w_f  = 1.35$  and $2.75$  respectively.  For  $M_0 =
8.9\times  10^{12}h^{-1}M_\odot$,  $S(M_0)/\delta_c^2(z_0)  =  1$,  so
$\delta_c(z_f)/\delta_c(z_0)  = 2.81$  and 4.68,  making  $(t_f/t_0) =
0.18$ and  $0.08$.  Thus, the  rms value of $\log(t_f/t_0)$  should be
about 0.16~dex.  This  is indeed close to the  value 0.17~dex measured
directly in  the simulations.  A  similar analyis for  $f=0.5$ returns
similarly good agreement.

\begin{figure*}
\begin{center}
\includegraphics[width=7.5cm]{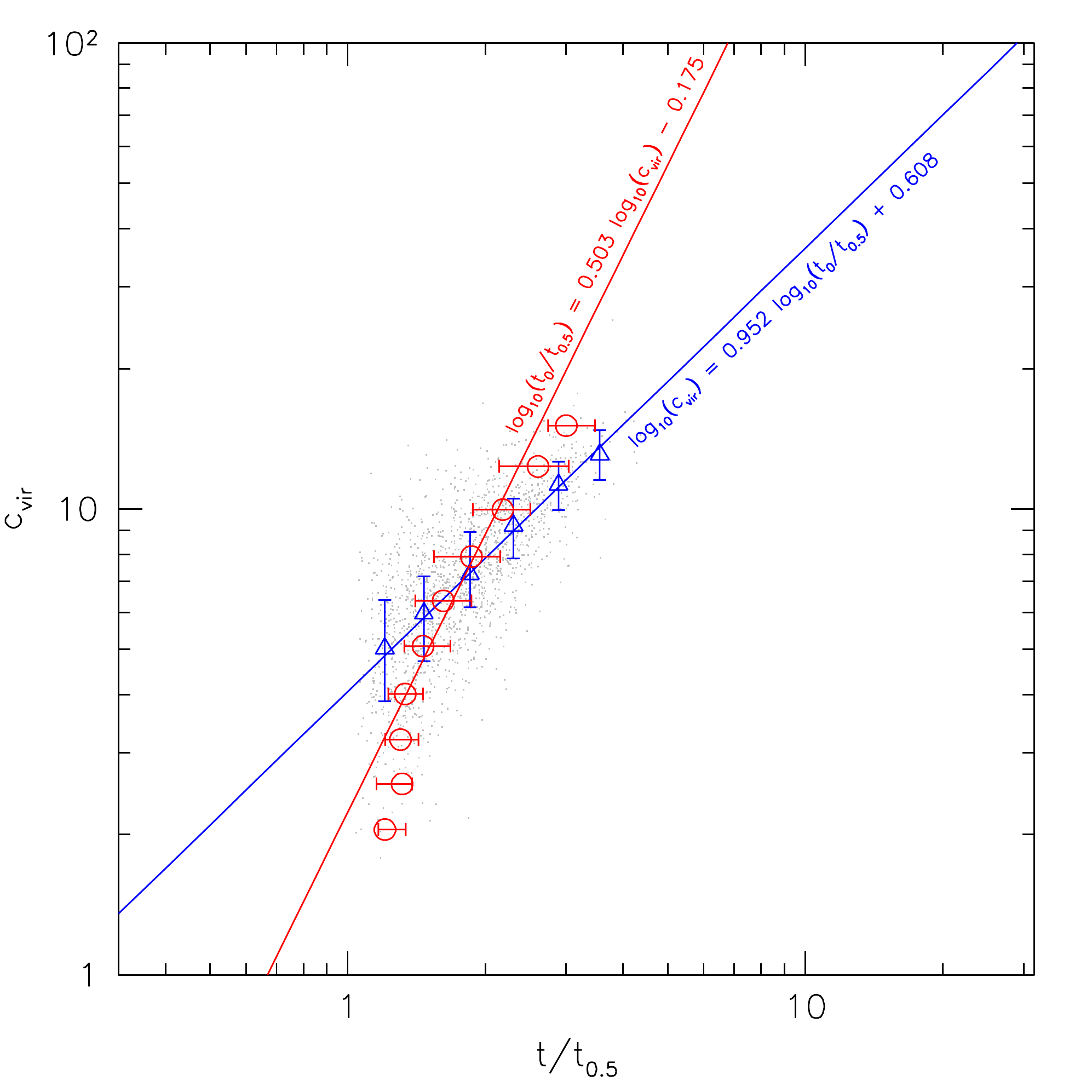}
\includegraphics[width=7.5cm]{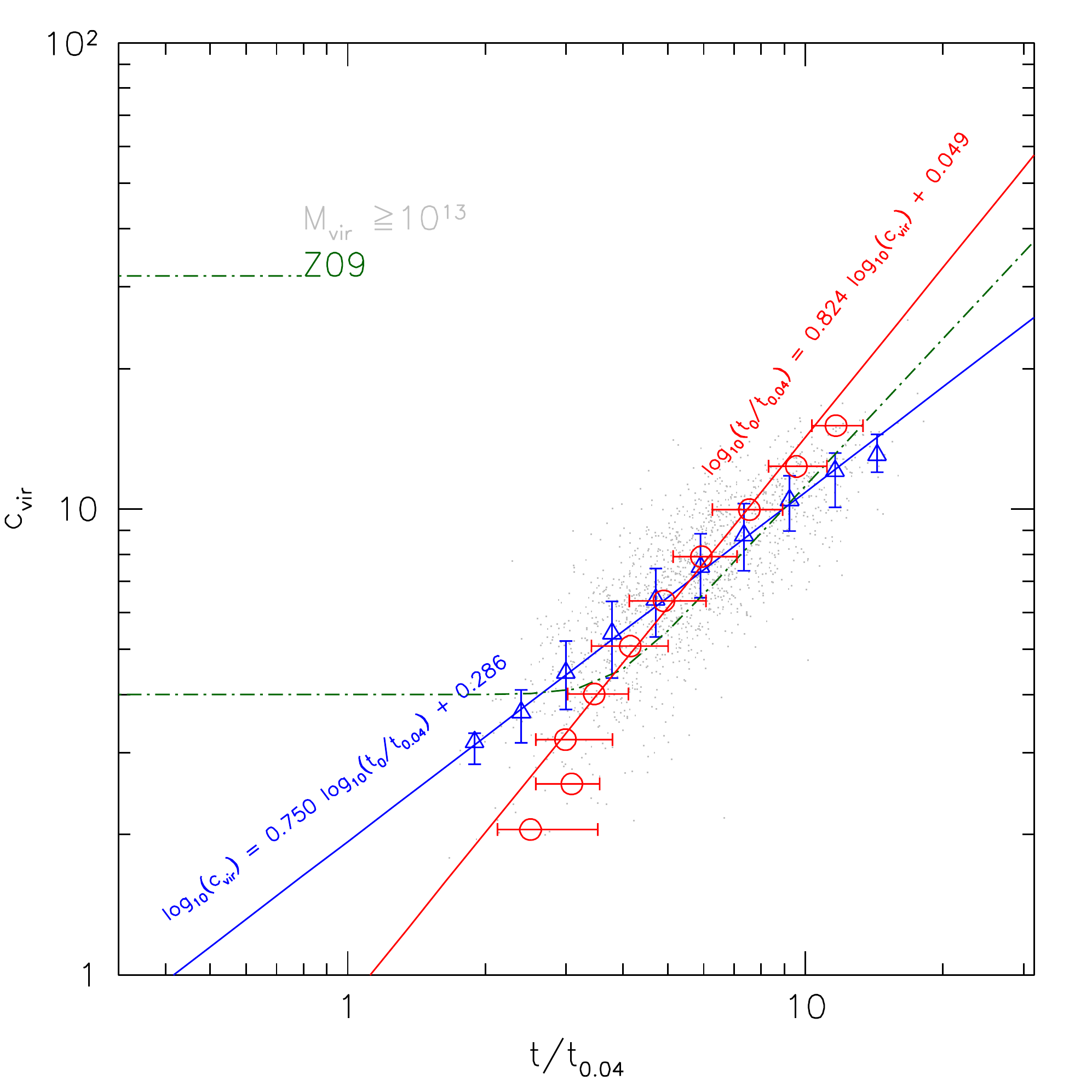}
\caption{\label{timeandc}Correlation    between    concentration   and
  $\tau_f$ ($f=0.5$ on the left  and $f=0.04$ on the right panel). The
  data show simulation  measurements at any $z_0$ for  all haloes more
  massive than $10^{13}\,M_{\odot}/h$. Open triangles and circles show
  the  median with the  quartiles of  respectively $c_{vir}$  at fixed
  $\tau_f$ and vice versa. The solid lines represent the least-squares
  fit to the simulation data points in both axsis.}
\end{center}
\end{figure*}

The  $t_f-c_{vir}$  relation  is   more  complicated.   If  we  define
$\tau_f\equiv  \log_{10}(t_f/t_0)$,   then  at  any   $z_0$,  $\langle
\tau_{0.5}|c_{vir}\rangle  \approx  0.175  -  0.503  \log_{10}c_{vir}$
whereas  $\langle \tau_{0.04}|c_{vir}\rangle  \approx  -0.049 -  0.825
\log_{10}c_{vir}$.  In both  cases, the rms scatter around  the fit is
about  0.1~dex.    Adding  the  mass  does  not   reduce  the  scatter
significantly,   so   we   do   not   report  fits   to   this   case.
Table~\ref{tformtable}  summarizes the precision  with which  a halo's
formation time  can be predicted  from its mass and  concentration. In
Figure  \ref{timeandc}  we  show  the  correlation  between  the  halo
concentration and $\tau_f$, where $f=0.5$  on the left and $f=0.04$ on
the right panel, for haloes with $M_{vir}\ge 10^{13}\,M_{\odot}/h$ and
at any  considered $z_0$.   The data points  with error bars  show the
median, with the  quartiles, of both the concentration  at fixed $t_f$
and  vice versa.  The  solid lines  show the  leas-squares fit  to the
simulation data in both axsis.

\section{Monte-Carlo realizations of mass accretion histories}
\label{mcs}
We have  used the distribution  of formation times discussed  above to
generate  Monte  Carlo realizations  of  mass  accretion histories  as
follows.  In reality, $m_{\rm MP}$ changes continuously; the amount by
which  it changes  at any  given  time is  a random  variable that  is
different for each halo.  Most merger tree algorithms approximate this
by taking (typically small) discrete steps in redshift, and then allow
the mass  to vary stochastically  \citep[e.g.]{nusser99}; others allow
large  steps  in  time   \citep{moreno09}.   Our  approach  is  rather
different: we predetermine  discrete steps in mass, and  at each step,
we draw a redshift from $p_{\rm MP}$ for that mass.

The choice of mass-step is critical.   Two steps in mass $f_1 M_0$ and
$f_2 M_0$  (with $f_1>f_2$) would  be strongly correlated  if $f_1\sim
f_2$,  but  mass   scales  separated  by  $\Delta  f/f   \sim  1$  are
approximately        uncorrelated       \citep{peacock90,paranjape11b}.
Figure~\ref{whw04} shows  that, at least  for large $\Delta  f/f$, the
distributions are indeed reasonably independent.

\begin{figure*}
\begin{center}
\includegraphics[width=5.8cm]{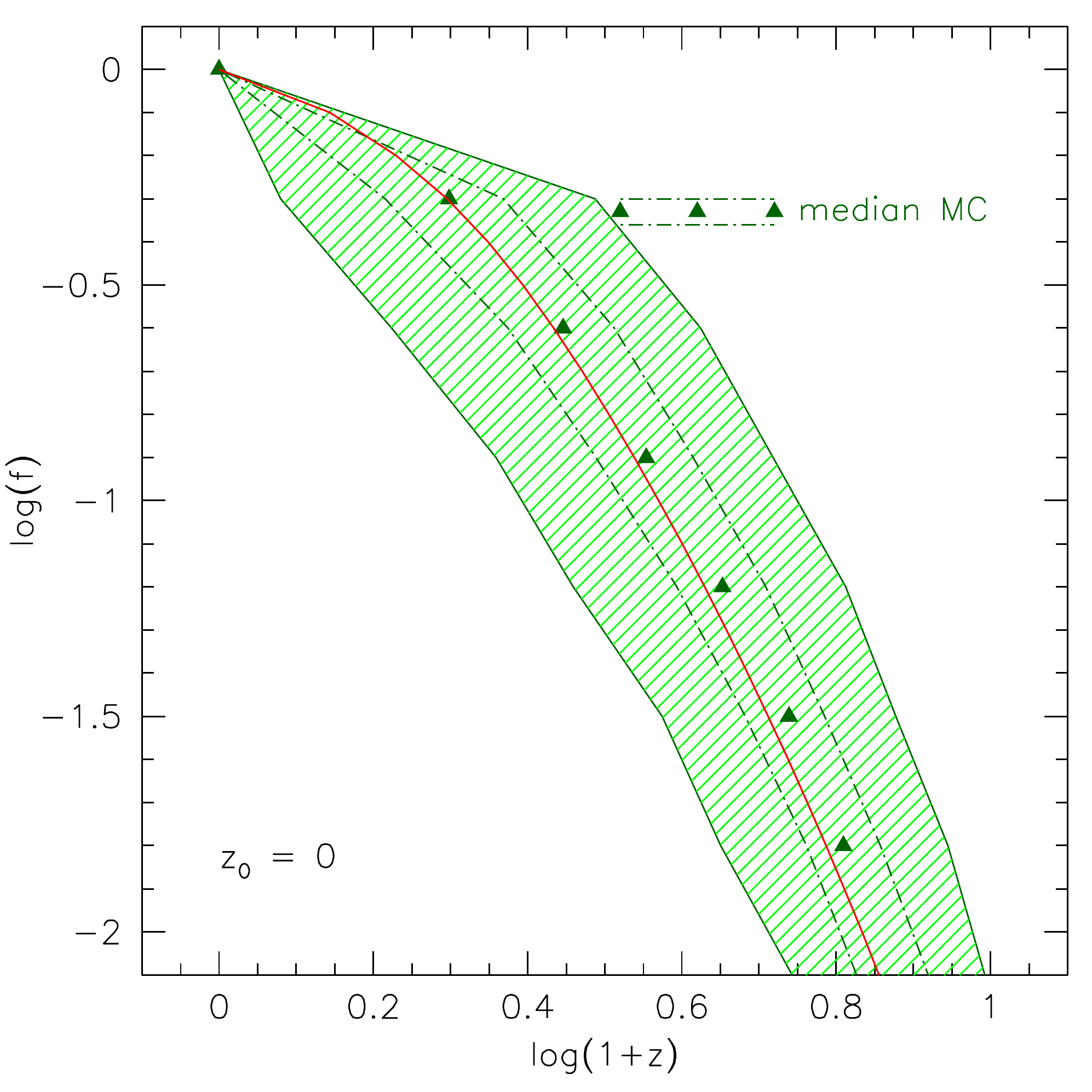}
\includegraphics[width=5.8cm]{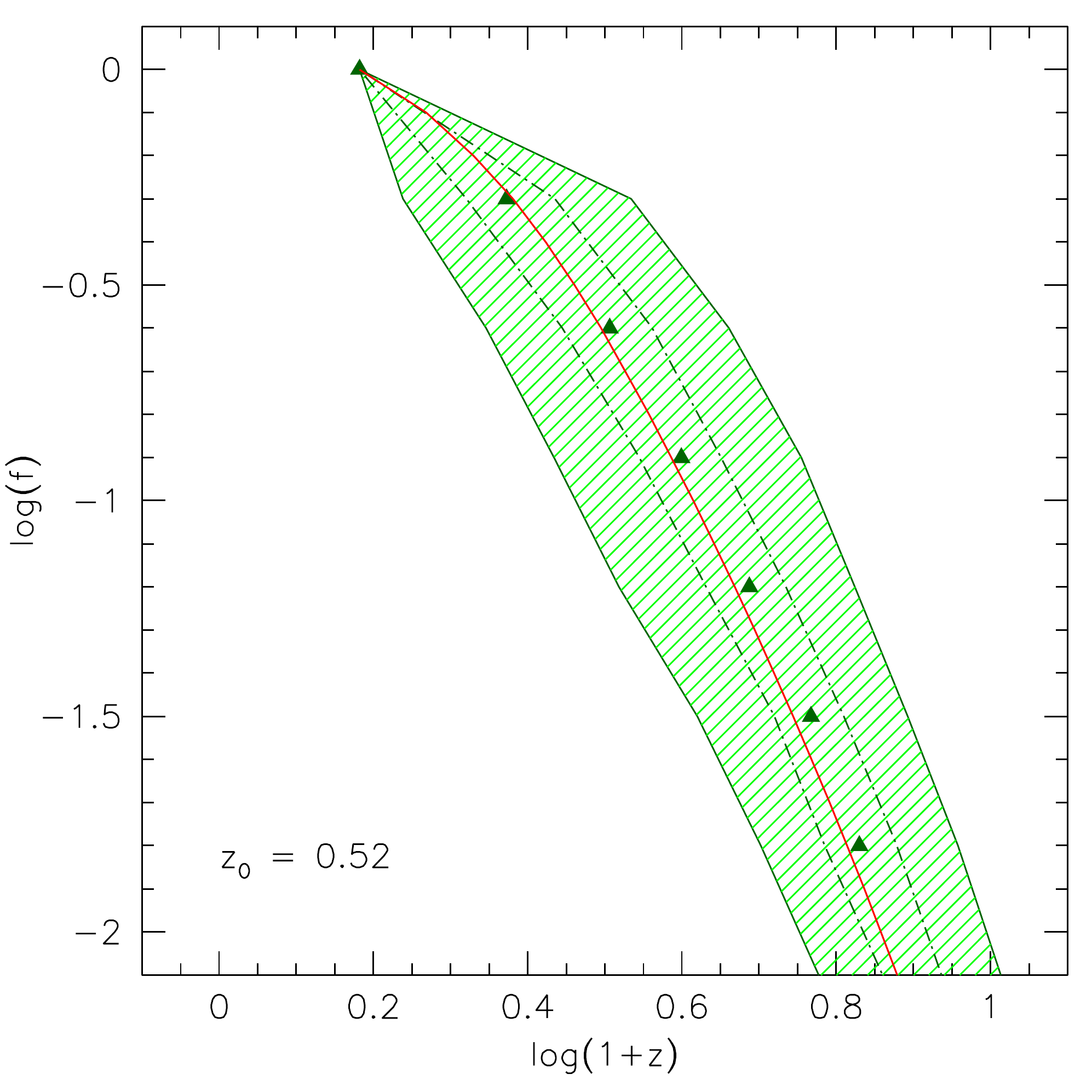}
\includegraphics[width=5.8cm]{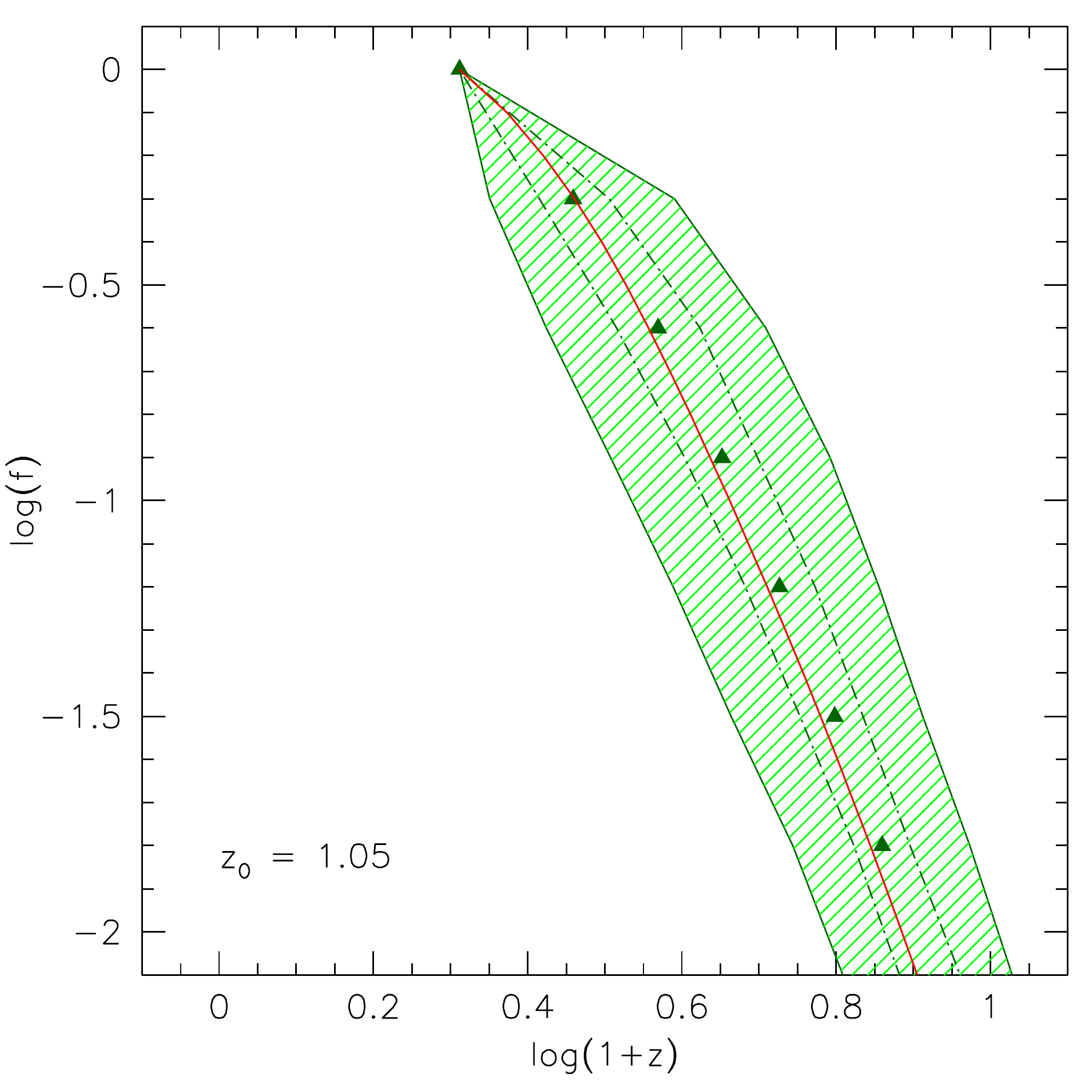}
\includegraphics[width=5.8cm]{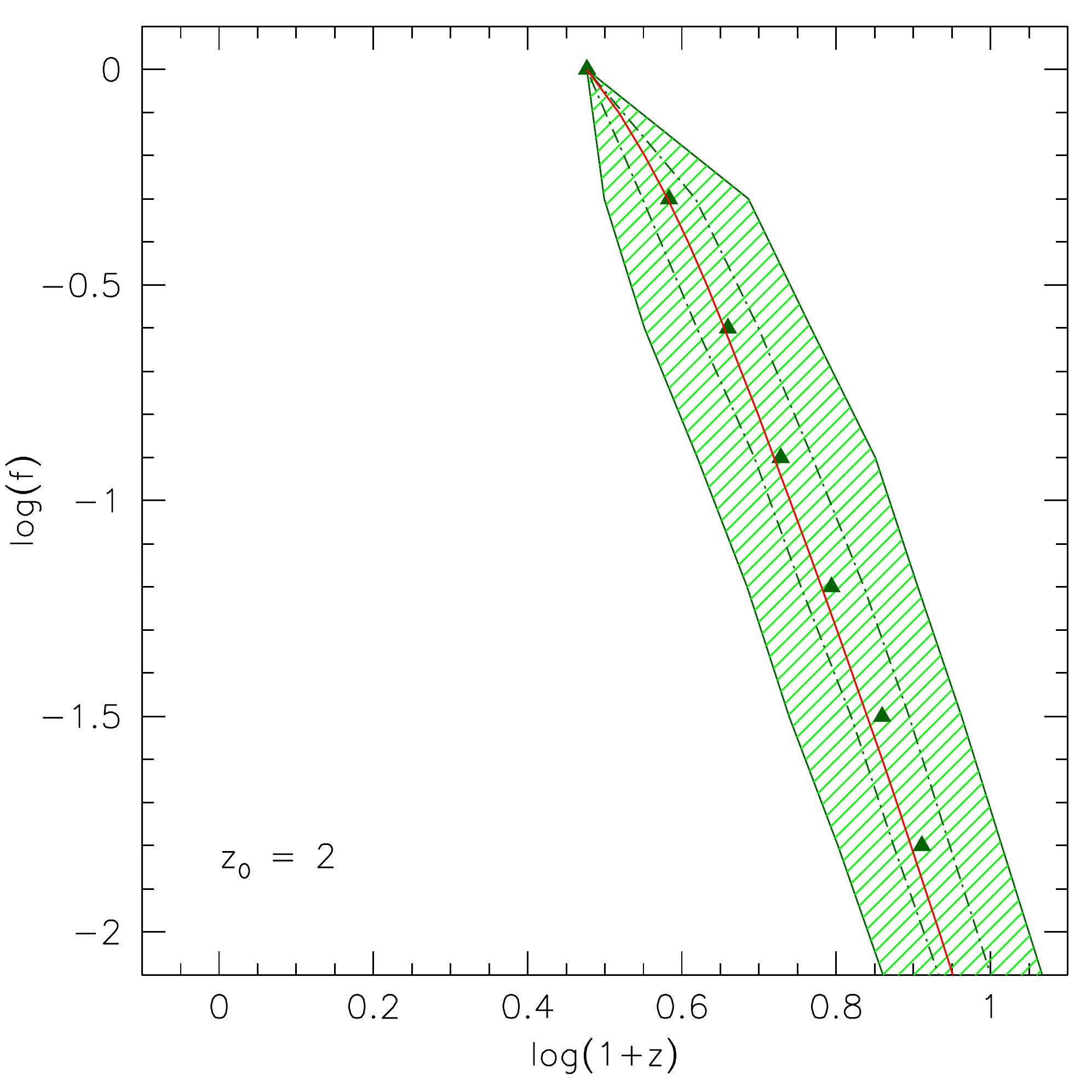}
\includegraphics[width=5.8cm]{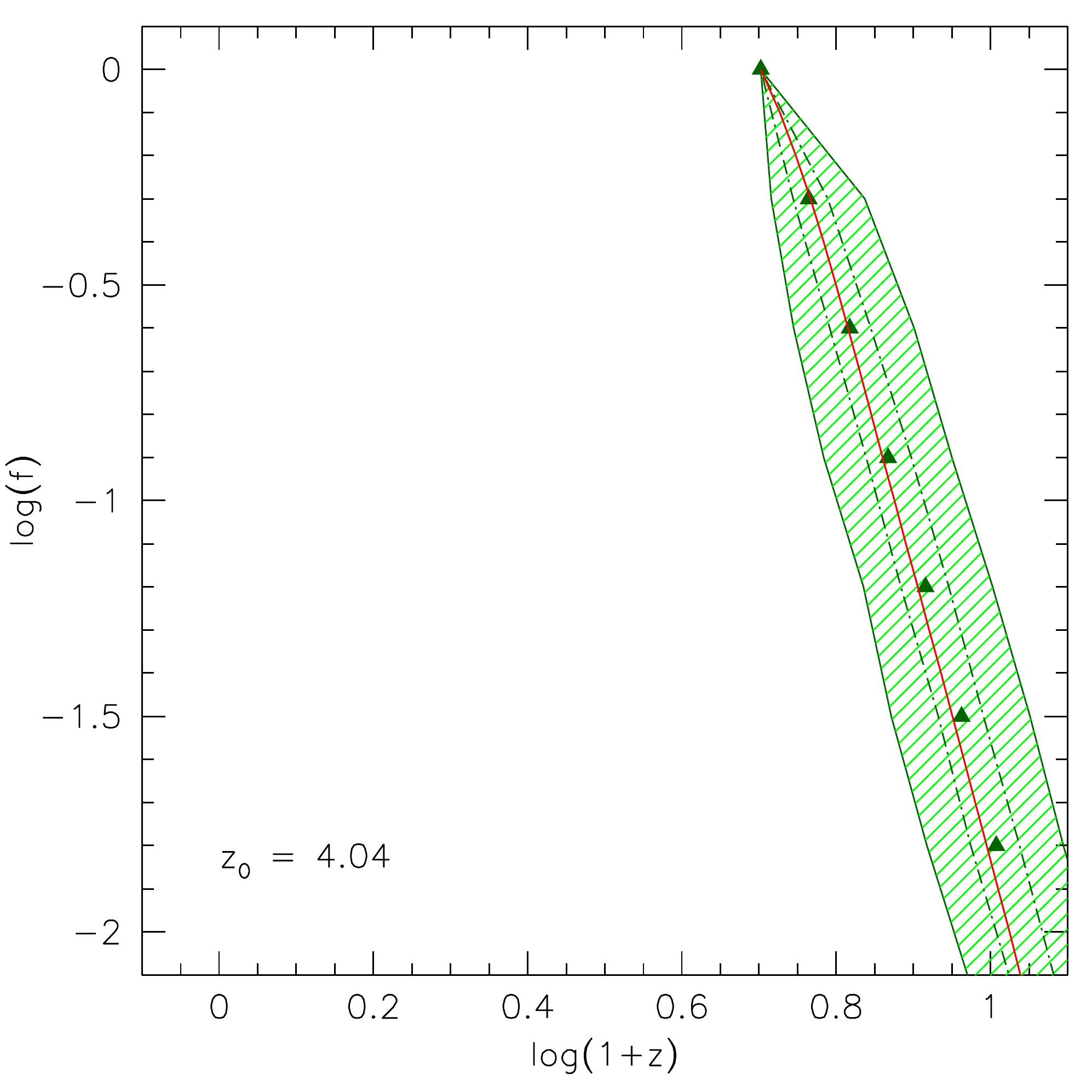}
\caption{Mass growth  history of a  sample of $M_0^*$  haloes starting
  from different redshifts $z_0$.  The sample as been built sampling a
  discrete number of mass steps with $\Delta \log_{10} M=-0.3$ for the
  rescaled formation redshit  distributions between $z_0$ and $z>z_0$.
  The filled triangles  and the curves that enclose  them indicate the
  median and the first and the third quartiles.  The solid line is our
  mass growth history model (equation~\ref{medzf}).\label{fmahMC}}
\end{center}
\end{figure*}

We have  found that if we  set $\Delta \log_{10} M  \approx -0.3$ then
the  distributions $p(w_{f_1})$  and $p(w_{f_2})$,  when  expressed as
functions  of  redshift,  do   not  overlap  much.   Since  this  mass
difference  corresponds  approximately  to  $\Delta f/f\sim  0.5$,  we
assume  that  we  can  take  independent  picks  from  these  redshift
distributions, with the condition that for the small fraction of picks
where  $z_2<z_1$ occurs  we should  set $z_2=z_1$.   If there  were no
overlap at  all, our approach  would correspond to  having independent
jumps in redshift, always having the correct distribution of redshifts
at the predetermined  discrete mass scales.  Figure~\ref{fmahMC} shows
the results.  The filled triangles  and the dashed curve that encloses
them  indicate the  median and  the first  and third  quartiles, solid
curves  enclose  $95\%$  of  the  trees.  The  solid  line  shows  our
analytical prediction (equation~\ref{medzf}).
 
\section{Applications} 
\label{forecasts}
\begin{figure*}
\begin{center}
\includegraphics[width=7.0cm]{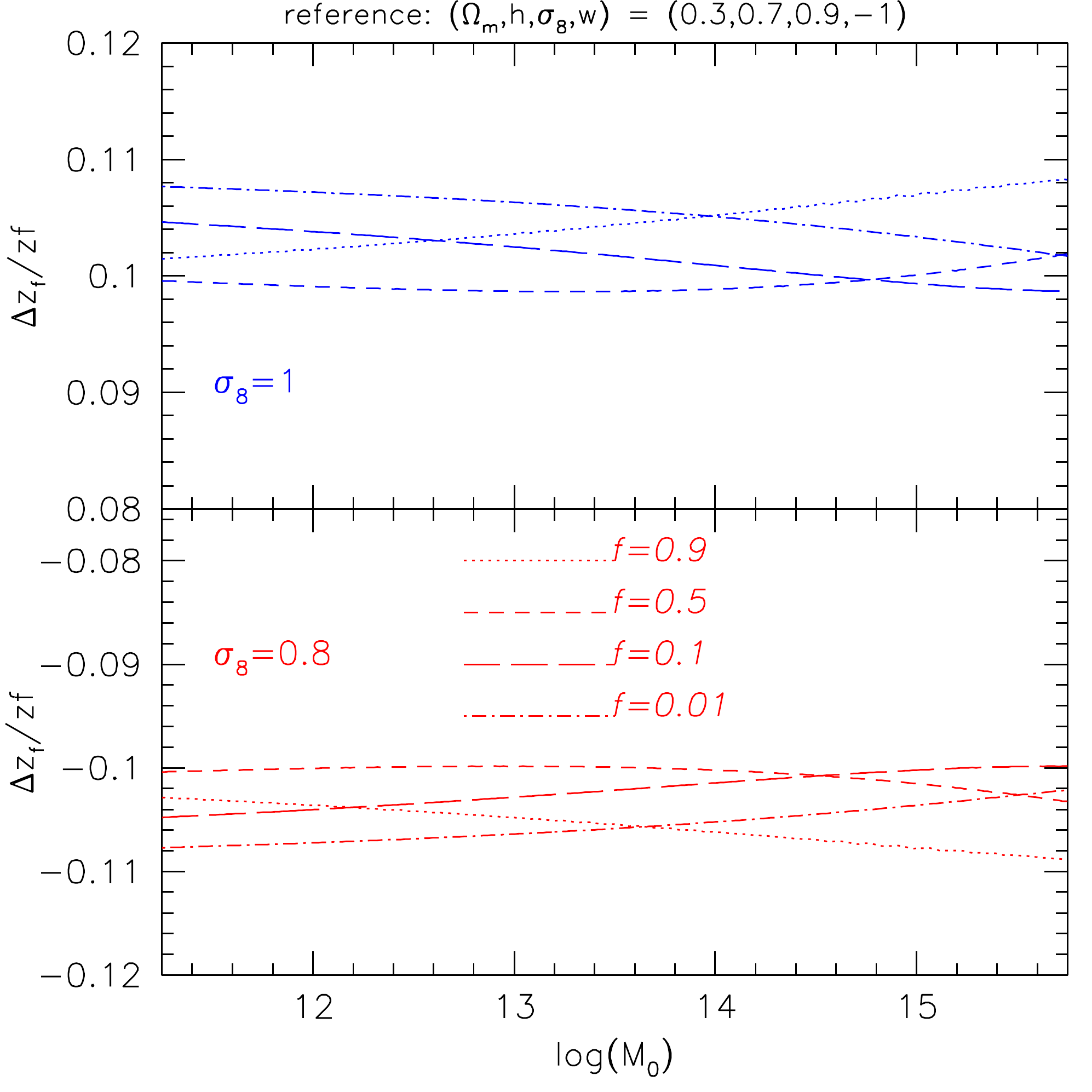}
\includegraphics[width=7.0cm]{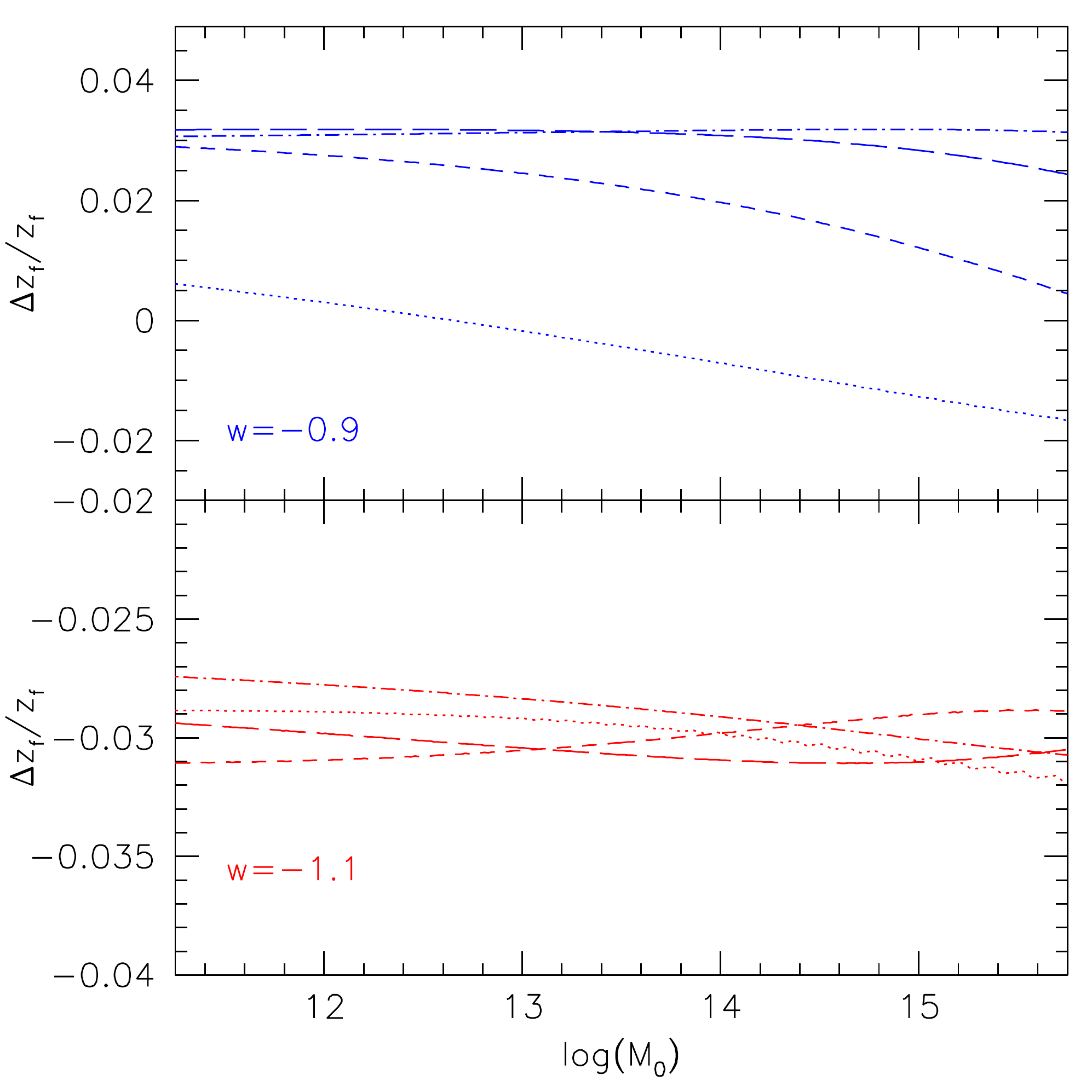}
\caption{Correlation  between the  formation redshift  and  the virial
  host  halo  mass at  $z_0=0$,  for  different  choices of  the  mass
  fraction   required  for   formation  ($f=0.9$,   $0.5$   $0.1$  and
  $0.01$). Left panel shows  the relative change in formation redshift
  (and  so  the   main  halo  growth  and  the   halo  abundance)  for
  $\sigma_8=0.8$ and  $\sigma_8=1$ relative to  $\sigma_8=0.9$.  Right
  panel  shows  the dependence  on  the  equation  of state  parameter
  $w$.\label{fmzf1}}
\end{center}
\end{figure*}

Extended-\citet{press74} approach gives  the possibility of estimating
the  main halo  growth  as function  of  the growth  factor, the  mass
variance and  the overdensity threshold for collapse.   Once fixed the
cosmological  model  these  quantities  can  be  translated  in  mass,
redshift and  time.  For  this reason the  halo formation,  growth and
concentration   depend   on   the  cosmological   parameters   adopted
\citep{harker06,power11}.  In the left panel of the Figure \ref{fmzf1}
we show  how much the median  redshift of haloes at  the present time,
with $z_0$=0, differs with respect  to a reference model in assembling
$90\%$,  $50\%$, $10\%$  and $1\%$  of their  mass.  We  have compared
$\sigma_8=0.8$  (bottom)  and   $\sigma_8=1$  (top)  with  respect  to
$\sigma_8=0.9$.   $\sigma_8$ represents  the amplitude  of  the linear
power spectrum  on a scale of  $8$ Mcp$/h$, from the  figure we notice
that universes  with a lower value  of $\sigma_8$ tend  to form haloes
more  recently. The  same  is valid  if  we consider  the dark  energy
equation  of  state parameter  $w$.   In  the  right panel  of  Figure
\ref{fmzf1} we show the relative change in formation redshifts between
$w=-1.1$ and $w=-0.9$ with respect to $w=-1$.

\begin{figure*}
\begin{center}
\includegraphics[width=7.0cm]{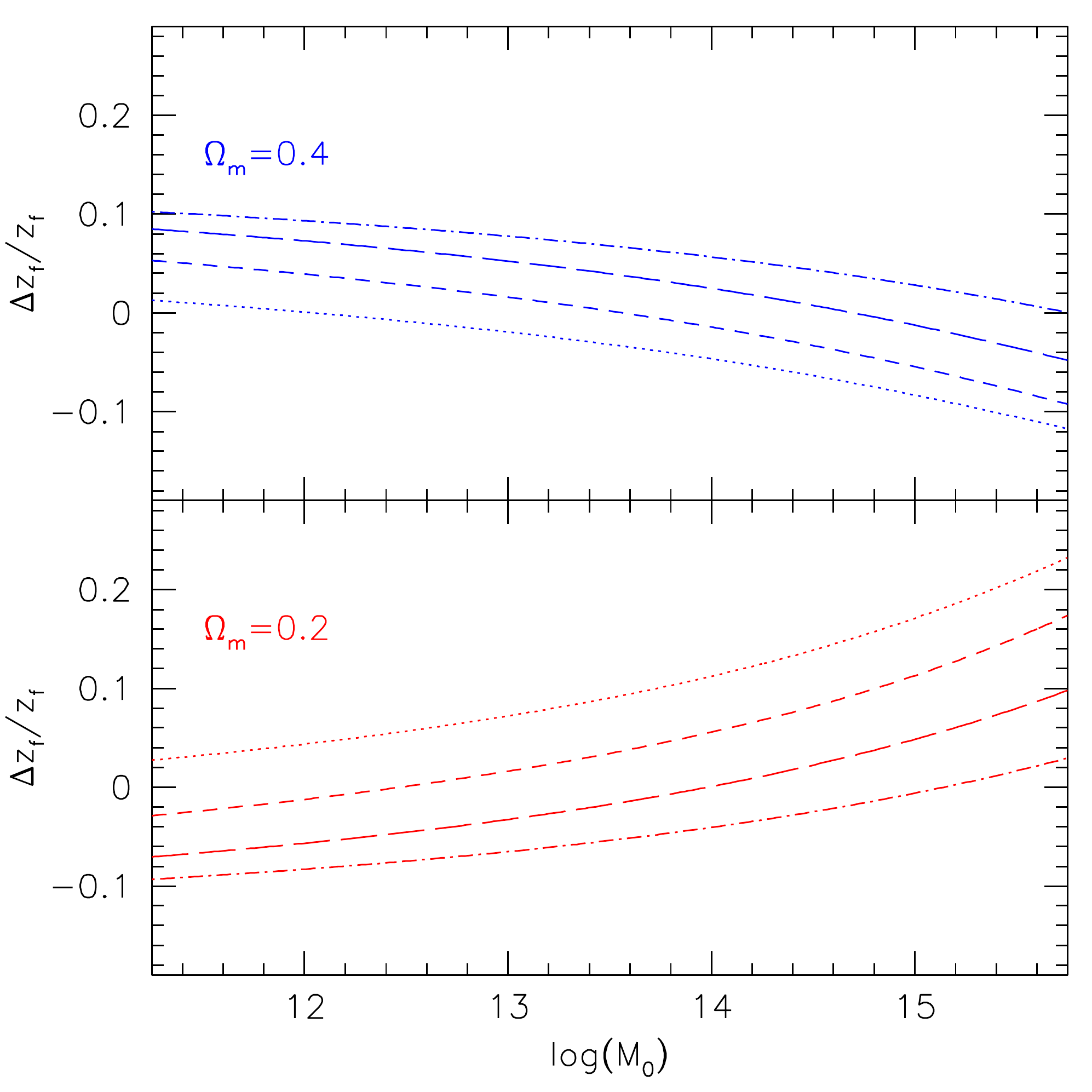}
\includegraphics[width=7.0cm]{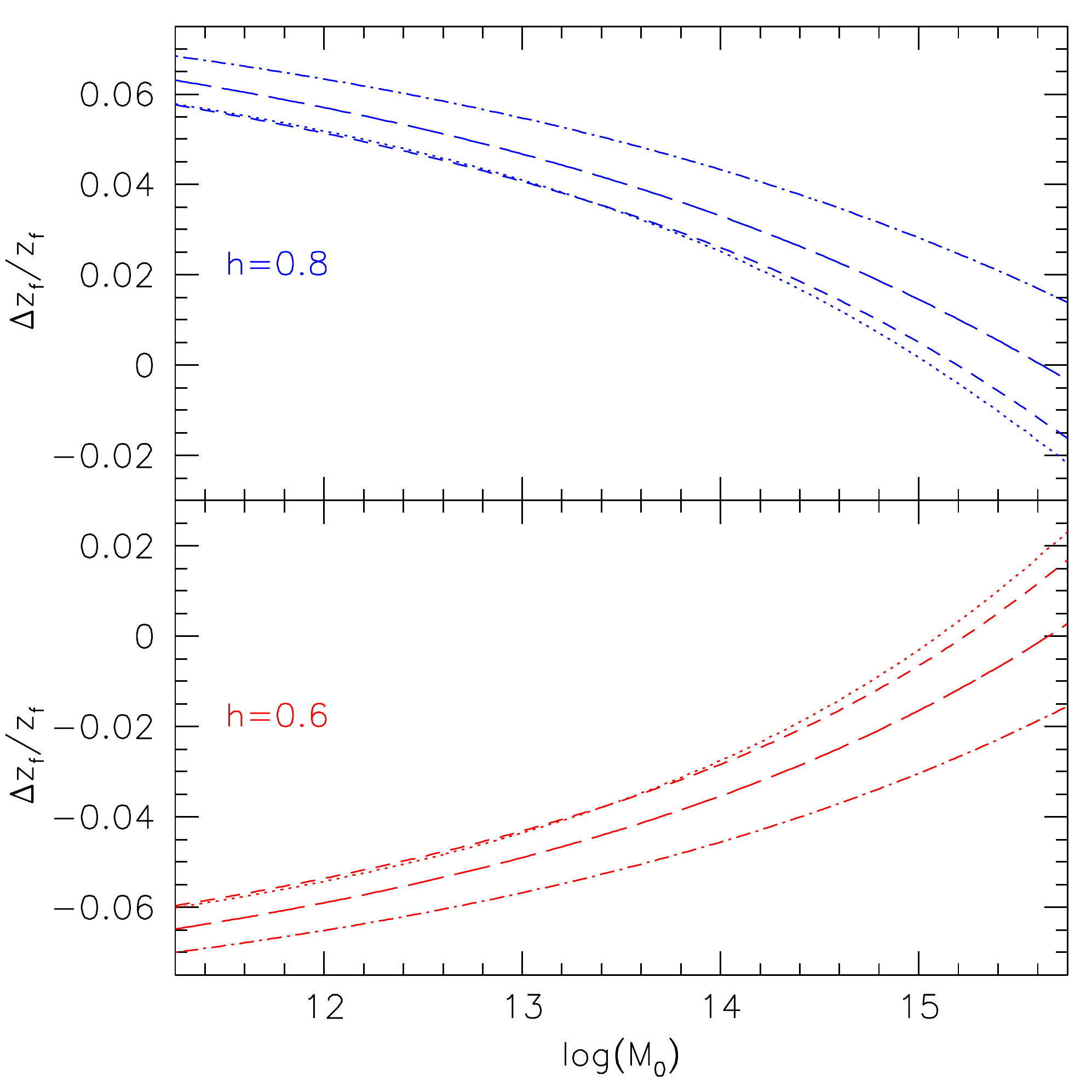}
\caption{The same as  Figure \ref{fmzf1}, in this case  on the left we
  show the  relative difference  in formation redshifts  for universes
  with lower  and higher  density parameter with  respect to  the GIF2
  cosmology, the right with different Hubble constant. \label{fmzf2}}
\end{center}
\end{figure*}

Similarly to  the previous figure,  in Figure \ref{fmzf2} we  show the
relative difference in formation redshifts in universes with lower and
higher matter density parameter with  respect to the model of the GIF2
simulation (we have assumed  that $\Omega_m+\Omega_{\Lambda} = 1$). In
the  same figure,  the right  panel quantify  the change  in formation
redshift due to the change in Hubble constant.

From the figures above we notice that a bigger difference in formation
redshifts  can be  mainly  seen, for  masses  of the  order of  galaxy
clusters, in  universes with a different  power spectrum normalization
$\sigma_8$ and matter density  parameter $\Omega_m$, a change of $0.1$
in  $w$  and  $h$  do  not  influence  too  much  the  formation  time
definitions.

\section{Discussion and Conclusions}
If the formation time of a halo  is defined as the first time that its
main progenitor  exceeds a fraction $f$  of the final  mass $M$, then,
for  halos of a  given $M$,  there will  be distribution  of formation
times which depends on $f$.   We argued that the median formation time
as a  function of $f$  should be the  same as the  median of $f$  as a
function  of time  (equation~\ref{zfmmp}),  and showed  that this  was
indeed the case (Figure~\ref{fmah}).

We  also argued  that, to  a good  approximation, the  distribution of
formation times  should be  a function of  the scaling  variable $w_f$
defined in  equation~(\ref{wf}), even when  $f<1/2$.  The distribution
of scaled formation times $p(w_f)$ is indeed approximately independent
of  $M$ and  $z_0$ (Figure~\ref{fPformation}).   This  distribution is
well-described by equation~(\ref{eqmodel1}).   Its median has a simple
form.  When combined  with a relation which relates  formation time to
halo  concentration (equation~\ref{czhao09}),  this provides  a simple
way  to  estimate  the   concentration  of  a  halo.   Including  more
information      about       the      mass      accretion      history
(e.g.  equation~\ref{cznew})  helps  predict  the  concentration  more
precisely (Table~\ref{cvirtable}).

We also  studied how well a  halo's formation time  could be predicted
from knowledge of its mass and concentration.  If the concentration is
known, then the  mass adds little new information  about the formation
time (Table~\ref{tformtable}).   When the formation  time is expressed
in units of the time $t_1$  at which the halo was identified, then the
concentration  estimates $\log_{10}(t_{0.5}/t_1)$  with  rms 0.09~dex,
and $\log_{10}(t_{0.04}/t_1)$ with rms 0.11~dex.  It may be that these
scalings will  find use in studies  which compare the mean  age of the
stars in the  central galaxy of a cluster with the  time when the mass
in the central core was first assembled.

The  formation times for  sufficiently different  $f$ are  only weakly
correlated (Figure~\ref{whw04}).  We argued why this is not unexpected
(Section~\ref{mcs})  and  then  used  this  fact to  devise  a  simple
Monte-Carlo  algorithm for generating  mass accretion  histories.  The
algorithm is rather accurate  (Figure~\ref{fmahMC}), and allows one to
generate  realistic mass  accretion histories  over a  large  range of
masses and  redshifts.  We expect it  to be useful for  studies of how
the  $M-c$  relation depends  on  the  background cosmological  model.
Algorithms for estimating the  median main halo progenitor mass growth
history  and  the  concentration-mass  relation  are  available  here:
\href{http://cgiocoli.wordpress.com/research-interests/concentrationmah}{http://cgiocoli.wordpress.com/research-interests/concentrationmah}.

\section*{Acknowledgements} 
We are grateful to the anonymous referee for comments 
that improved how we present our results.
Thanks  to Lauro  Moscardini and Massimo Meneghetti  for  reading the
manuscript and  providing useful comments, also  for future projects.
Many  thanks to  Stefano Ettori, Federico Marulli  and 
Cristiano  De Boni  for helpful discussions.   CG  is supported  by 
the  ASI contracts:  I/009/10/0, EUCLID- IC fase A/B1 and PRIN-INAF 2009.  
RKS is supported in part by NSF-AST 0908241.

\appendix
\section{Appendix}
The main text defined halo formation using the main progenitor $m_{\rm
  MP}$, rather than the  most massive progenitor $m_1$.  In principle,
predicting the distribution  of $m_1(z)$, the most massive  of the set
of progenitors  at $z$, is  just an extreme value  statistics problem.
In practice, calculating this  distribution is complicated by the fact
that the partition of $M_0$ into $m_1,\ldots,m_n$ does not involve $n$
independent picks from the  same distribution.  (E.g., for white-noise
initial conditions, the  joint distribution of the $n$  picks is known
in closed  form -- see  \citet{sheth99a} -- and  it is not  simply the
product  of  $n$ identical  distributions.)   Similarly,  there is  no
simple expression for the distribution of $m_{\rm MP}(z)$.

\begin{figure*}
\begin{center}
\includegraphics[width=7.5cm]{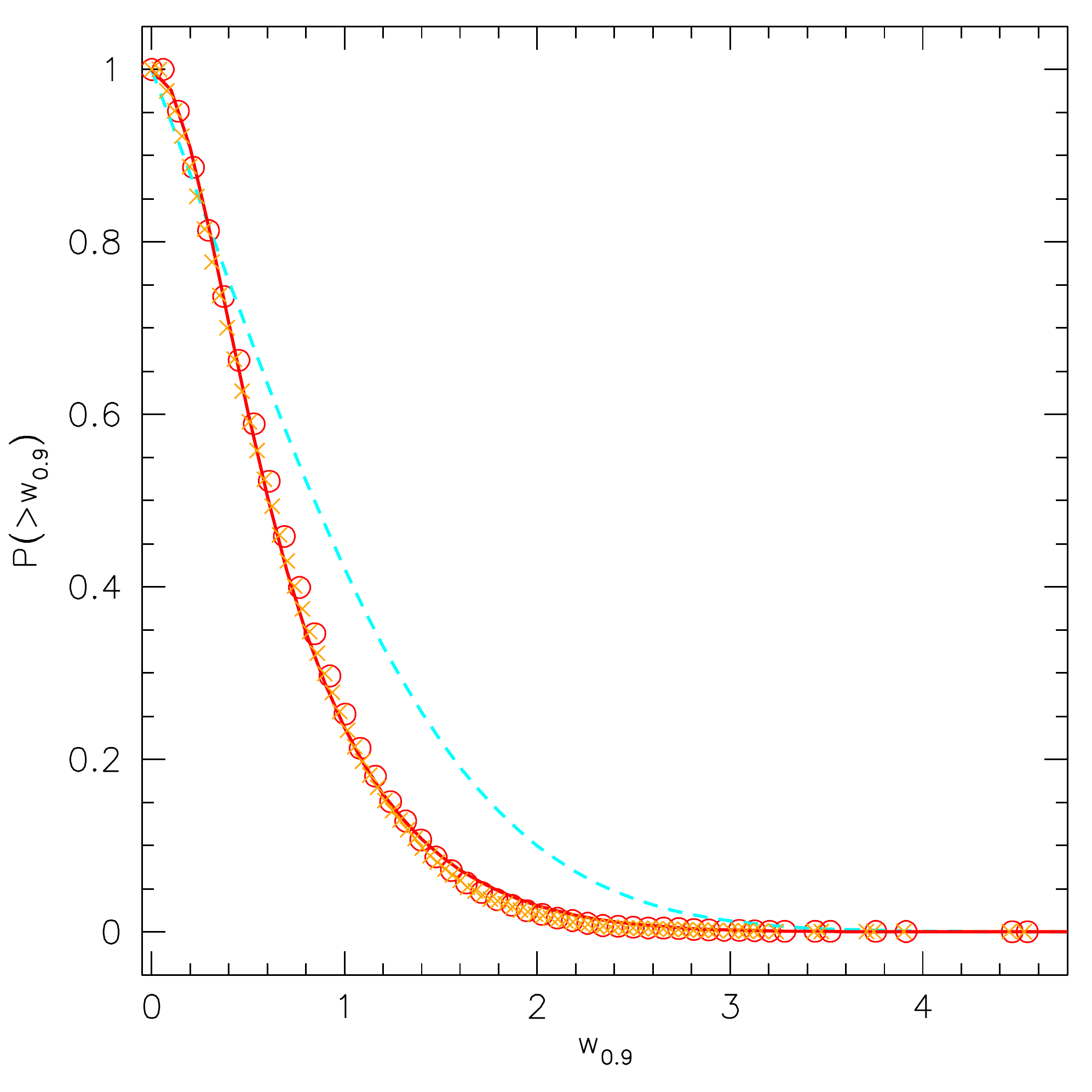}
\includegraphics[width=7.5cm]{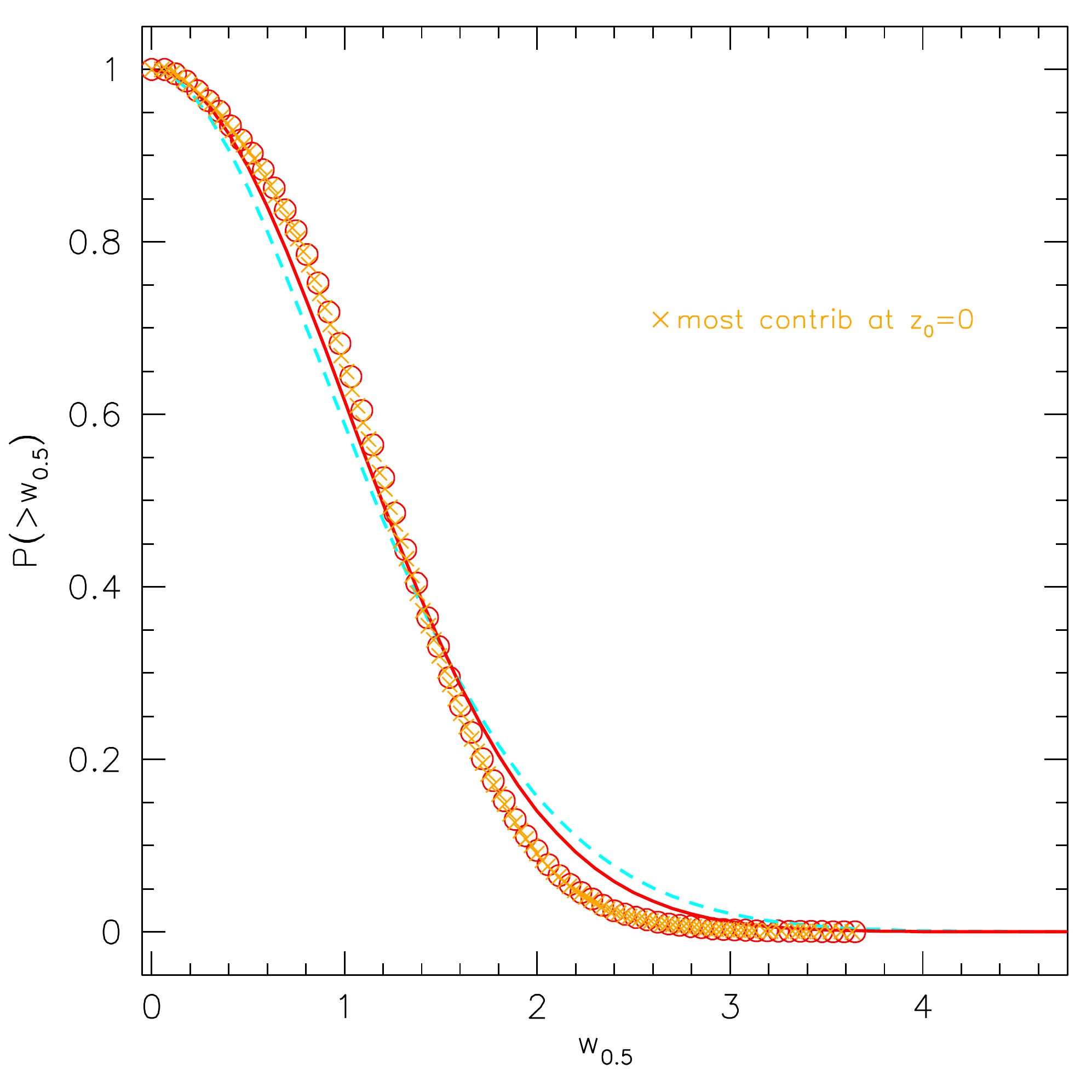}
\includegraphics[width=7.5cm]{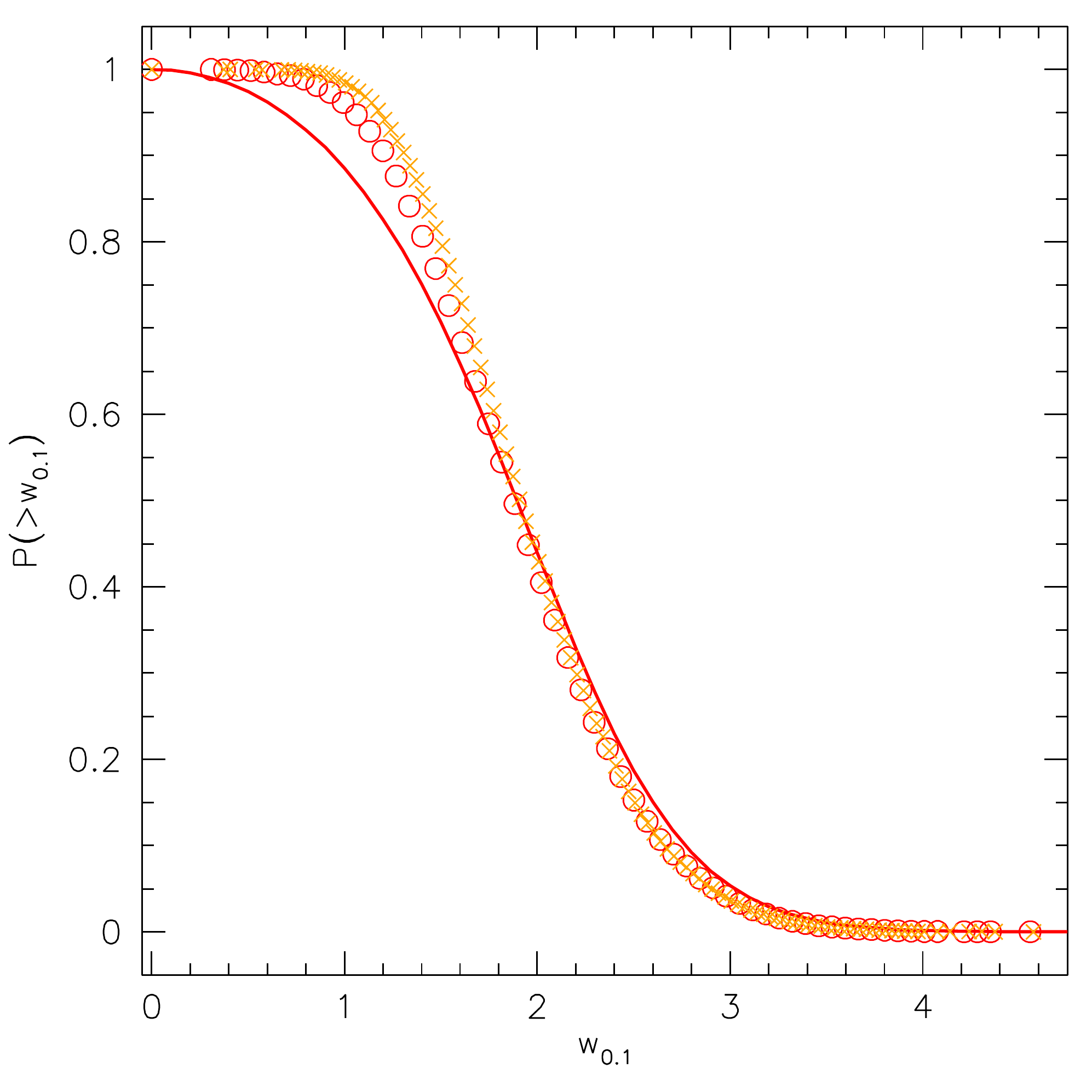}
\includegraphics[width=7.5cm]{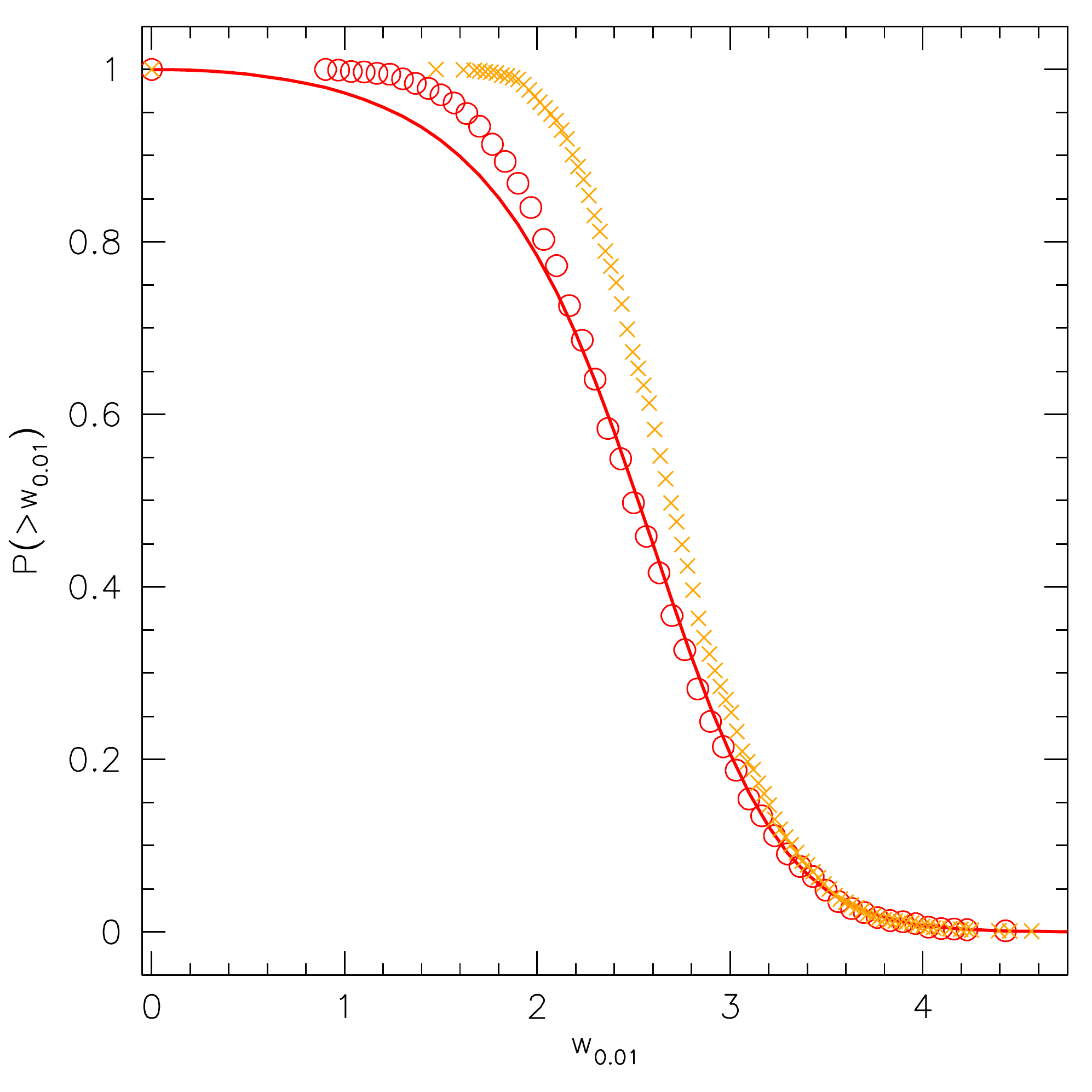}
\caption{Distribution of scaled formation times for dark matter haloes 
  identified at $z_0=0$, when formation is defined as the earliest time 
  that $m_{\rm MP} \ge fM_0$ (circles) or as $m_1 \ge fM_0$ (crosses) for 
  different values of $f$. The solid curve shows our simple fitting 
  function (equation~\ref{eqmodel1}) with $\alpha_f$ given by 
  equation~(\ref{eqfit2}) rather than~(\ref{eqmyfit}).  
  Note that the two definitions coincide for $f\ge 1/2$, but that 
  otherwise $m_1$ yields higher formation redshifts.
  \label{fPformation2}}
\end{center}
\end{figure*}

Figure~\ref{fPformation2}   shows   the   cumulative  formation   time
distribution of the  sample of haloes identified at  $z_0=0$. The open
circles (same  as Figure  \ref{fPformation}) show the  measurements on
the merger-tree where the main  halo progenitor has been defined among
the most  contributing between two  consecutive time steps,  while for
the crossed  data we  defined the main  halo as the  most contributing
progenitor at $z_0=0$. The dashed and  the solid curve are the same as
in  Figure \ref{fPformation} and  represent our  fit and  the modified
model  by  \citet{nusser99}.    As  expected,  the  distributions  are
essentially  identical when  $f=0.9$ and  $0.5$.  However,  at smaller
values of $f$, the difference between the two definitions increases.

We  can   quantify  this  difference   by  noting  that,  if   we  fit
equation~(\ref{eqmodel1})  to  this  distribution, then  the  required
scaling of $\alpha_f$ with $f$ becomes
\begin{equation}
 \alpha_f = 0.867\, \mathrm{e}^{-2 f^3}/f^{0.8}\,.
 \label{eqfit2}
\end{equation}
Since this $\alpha_f$ is larger than that in the main text by a factor
of  $1.06/f^{0.1}$, equation~(\ref{eqmahmod1})  shows that  the median
$w_f$ is  shifted to higher  redshifts than when formation  is defined
using the main progenitor.

\bibliographystyle{mn2e}
\bibliography{gstv1.0}
\label{lastpage}
\end{document}